\documentclass{article}
\usepackage{graphicx} 
\usepackage{amsmath}
\usepackage{amssymb}
\usepackage{amsthm}
\usepackage{physics}
\usepackage{xcolor}
\usepackage{appendix}
\usepackage{hyperref}
\usepackage{caption}
\usepackage{bm}                           
\usepackage{float}                        

\usepackage{makecell}                     
\usepackage{multirow}                     

\usepackage[backend=biber, style=numeric, sorting=none]{biblatex}
\DeclareMathOperator\Det{det}

\usepackage{geometry}
\hypersetup{
    colorlinks=true,
    linkcolor=blue,
    filecolor=magenta,
    urlcolor=red,
    citecolor=green,
}

\labelformat{section}{Section #1}
\labelformat{subsection}{Section #1}
\labelformat{equation}{Eq.~(#1)}
\labelformat{figure}{Fig.~#1}
\labelformat{subfigure}{Fig.~#1}
\labelformat{table}{Tab.~#1}

\title{Bipartite temporal Bell inequality for squeezed coherent state of inflationary perturbations}
\author{Aurindam Mondal \footnote{aurindammondal99@gmail.com, \, aurindam\_r$@$mail.isical.ac.in } \\ 
\small{$^{*}$ Physics and Applied Mathematica Unit}, \\
\small{$^{*}$ Indian Statistical Institute, 203, Barrackpore Trunk Road, Kolkata - 700108, India}}

\date{\vspace{-0.8 cm}}

\begin{document}

\maketitle
\begin{abstract}
    We investigate the role of the bipartite temporal Bell inequality, an analogue of the spatial Bell inequality, in probing the quantum imprints of primordial perturbations when the initially chosen Bunch–Davies vacuum is replaced by a coherent state. Although it is based on the same principles of locality and realism, its primary advantage lies in the fact that it does not require two distinct set of observables for its construction. Instead, measurements performed on a single component of the pseudo-spin operator at different times are sufficient. Consequently, it is particularly well suited for cosmological scenarios, where observational constraints typically allow access to only one component of the pseudo-spin operator. 

    Assuming a coherent state as the initial condition, we derive an analytical expression for the expectation value of the bipartite temporal Bell operator and demonstrate the absence of temporal Bell violation in such a scenario. Interestingly, the results for squeezed coherent state is found to differ—albeit slightly—from those of squeezed vacuum state for large values of the squeezing parameter. This suggests that the ability to distinguish among different initial states of primordial perturbations does not rely on the violation of temporal Bell inequality. Furthermore, the dependence of the temporal Bell inequality on a purely imaginary phase factor of the wave function appears to be an unique feature, which is entirely absent in the context of spatial Bell inequalities.

\end{abstract}

\tableofcontents
\section{Introduction}\label{sec:1}

Inflation, a period of rapidly accelerated expansion in the early universe, plays a crucial role in explaining the temperature anisotropies observed in the Cosmic Microwave Background (CMB) as well as the origin of the large-scale structures seen today \cite{PhysRevD.23.347,Linde:1981mu,Martin:2018ycu,Baumann:2009ds}. The primordial density perturbations that seed the observable signatures of inflation, arise from vacuum fluctuations of the quantum fields present during the inflationary epoch. These fluctuations are then amplified and stretched over cosmological scales during the inflationary epoch and make the formation of large scale structure become feasible. It establishes the quantum origin of inflationary perturbations and provides a primary motivation for the investigation of quantum measures within a cosmological framework. 

In the context of early universe cosmology, several studies have provided compelling evidence for the emergence of a highly entangled quantum state, known as the two-mode squeezed vacuum state \cite{Albrecht:1992kf, Martin:2012ua}. The inflationary dynamics of the early universe plays a central role in generating the non-local properties of such quantum states. This, in turn, motivates the quantification of entanglement using various quantum measures, such as entanglement entropy \cite{Brahma:2020zpk}, cosmological complexity \cite{Bhattacharyya:2022rhm}, quantum discord \cite{Martin:2015qta} and Bell inequalities \cite{Martin:2016tbd,Martin:2017zxs}. The interest in entangled states stems from the fact that entanglement is a distinctive feature of quantum mechanics with no classical analogue. It is basically a manifestation of quantum non-locality, famously characterized by Albert Einstein as “spooky action at a distance.” Apart from entanglement, there exist several other signatures of quantumness $-$ such as quantum superposition and the uncertainty principle $-$ which also experience the absence of classical counterparts. Among these, the Bell–Clauser-Horne-Shimony-Holt
(Bell-CHSH) inequality provides an operational framework for quantifying the non-classical correlations present in a quantum system \cite{Dale:2023fnp,Dale:2025nhc,Kanno:2017dci,Martin:2019wta,Choudhury:2016cso}. 

To set the stage for Bell inequalities, one should begin with the Einstein–Podolsky–Rosen (EPR) paradox, which questioned about the completeness of quantum mechanics due to the absence of simultaneous reality for a pair of non-commutating observables \cite{PhysRev.47.777}. Subsequently, several researchers explored local hidden-variable theories in an attempt to account for the probabilistic nature of quantum mechanics through the introduction of underlying, but inaccessible, variables. Then Bell formulated an inequality to distinguish the predictions of local hidden variable theories from that of standard quantum mechanics. The violation of such an inequality implies the presence of non-local quantum correlations that cannot be explained within any local hidden-variable framework \cite{PhysicsPhysiqueFizika.1.195}. Following the experimental verification of Bell inequality violations, the viability of local hidden-variable theories as an alternative description of nature has been largely ruled out. 

Let us delve deeper into the measurement schemes involved in Bell-CHSH experiment. In the context of spatial Bell inequality, measurements are performed on two dichotomic observables, such as $\hat{S}_{\mathbf{k}}$ and $\hat{S}_{\mathbf{-k}}$, defined on the respective sub-systems of a bipartite Hilbert space $\left(\mathcal{H}_{\mathbf{k}} \otimes \mathcal{H}_{\mathbf{-k}} \right)$. These observables are associated with the measurements performed simultaneously at two spatially separated locations $x_{1}$ and $x_{2}$ respectively. The purpose of these measurements is to quantify the correlations that exist between the two spatially separated sub-systems. From the measurement outcomes, one can construct the spin-spin correlators associated with spatially separated observables $E(a, b) = \left\langle \hat{S}_{\mathbf{k, a}} \otimes \hat{S}_{\mathbf{-k, b}} \right\rangle$, which in turn allow for the evaluation of the expectation value of the Bell–CHSH operator. According to CHSH formalism, the spatial Bell operator is defined as follows, 
\begin{eqnarray}\label{Spatial_Bell}
    \hat{\mathcal{B}}_{X} &=& \left(\hat{S}_{\mathbf{k, a}} \otimes \hat{S}_{\mathbf{-k, b}} \right) + \left(\hat{S}_{\mathbf{k, a}} \otimes \hat{S}_{\mathbf{-k, b'}} \right) + \left(\hat{S}_{\mathbf{k, a'}} \otimes \hat{S}_{\mathbf{-k, b}} \right) - \left(\hat{S}_{\mathbf{k, a'}} \otimes \hat{S}_{\mathbf{-k, b'}} \right) \, . 
\end{eqnarray}
Here the subscript $(a, a', b, b')$ denote the possible orientations of the dichotomic, pseudo-spin observables in their respective Hilbert space. Mathematically, the objects within the first bracket can be expressed as $\hat{S}_{\mathbf{k, a}} = \big(\hat{\Vec{S}}_{\mathbf{k}} \, . \, \hat{a} \big)$, where the unit vector $\hat{a}$ specifies the possible measurement direction. 

Under the assumptions of locality and realism, the upper bound for the expectation value of spatial Bell-CHSH operator is found to be $2$. Here, the locality principle asserts that measurements performed on space-like separated events cannot influence one another, while the realism assumption states that every physical systems possess definite outcomes against all possible measurements. However, numerous studies have demonstrated that quantum systems can violate this classical bound under appropriate conditions \cite{Thearle:2018sdl,Praxmeyer:2004jsd,Yarnall:2007ofr}. Such violations indicate that at least one of these underlying assumptions must be relaxed. Mathematically, the violation of the spatial Bell–CHSH inequality can be expressed as 
\begin{eqnarray}
    2 \hspace{0.2 cm} \leq \hspace{0.2 cm} \left\langle \psi \right| \hat{\mathcal{B}}_{X} \left| \psi \right\rangle \hspace{0.2 cm} \leq \hspace{0.2 cm} 2 \sqrt{2} \, . 
\end{eqnarray}
The upper limit $2\sqrt{2}$ is known as the Tsirelson's bound \cite{cirel1980quantum}, which represents the maximum violation allowed by quantum mechanics and cannot be exceeded by any quantum system. 

In contrast to the spatial scenario, there exist alternative formulations of Bell inequalities in which measurements are performed at two distinct instants of time say $t_{1}$ and $t_{2}$, instead of distinct spatial locations say $x_{1}$ and $x_{2}$. These are known as temporal Bell inequalities and are designed to quantify correlations that exist between measurement outcomes performed at different times. The derivation of temporal Bell inequalities requires a slight modification of the underlying assumptions used in the spatial scenario. While the realism condition is retained, the locality assumption is now replaced by the principle of non-invasive measurability. This principle states that it is possible, in principle, to perform a measurement without disturbing the subsequent evolution of the quantum state associated with the system. However, within the Copenhagen interpretation of quantum mechanics, such non-invasive measurements are generally unattainable due to the collapse of the wave function induced by measurement. Consequently, the assumption of non-invasive measurability is typically violated in quantum systems; however, this relaxation alone is not sufficient to ensure the violation of temporal Bell inequalities. 

Building upon the discussion of spatial and temporal Bell inequalities, we now focus on a third class of Bell inequality, widely known as the bipartite temporal Bell inequality \cite{Ando:2020kdz}. In this framework, measurements are performed on two dichotomic observables  $\hat{S}_{\mathbf{k}}$ and $\hat{S}_{\mathbf{-k}}$ defined on a bipartite Hilbert space $ \big(\mathcal{H}_{\mathbf{k}} \otimes \mathcal{H}_{\mathbf{-k}} \big)$, where the individual sub-systems are located at two spatially separated positions $x_{1}$ and $x_{2}$, and measurements are performed at two distinct instants of time $t_{1}$ and $t_{2}$. Such measurement protocols enable the quantification of correlations that exist between the events which are separated in both ways $-$ spatially and temporally. In this setting, consecutive measurements are carried out on two spatially separated subsystems, one after another. \textit{More precisely, the assumption of space-like separation leads to causally disconnected behaviour between the individual sub-systems $\mathcal{H}_{\mathbf{k}}$ and $\mathcal{H}_{\mathbf{-k}}$.} Consequently, the measurement performed on the first sub-system, at time $t_{1}$, cannot influence the quantum state of the second sub-system on which the subsequent measurement is carried out at a later time $t_{2}$. For this reason, bipartite temporal Bell inequalities are formulated under the assumptions of locality and realism, rather than relying on the principle of non-invasive measurability. 

To avoid the appearance of non-Hermitian observables, which would lead to complex eigenvalues for physical observables, it is convenient to adopt the projective measurement formalism of quantum mechanics \cite{Fritz:2010qzm}. Within this framework, the bipartite temporal Bell operator can be defined as follows: 
\begin{eqnarray}\label{Bipartite_Temporal_Bell}
    \hat{\mathcal{B}}_{T} &=& \frac{1}{2} \bigg[ \left\{\hat{S}_{\mathbf{k}}(t_{a}), \hat{S}_{\mathbf{-k}}(t_{b}) \right\} + \left\{\hat{S}_{\mathbf{k}}(t_{a}), \hat{S}_{\mathbf{-k}}(t'_{b}) \right\} + \left\{\hat{S}_{\mathbf{k}}(t'_{a}), \hat{S}_{\mathbf{-k}}(t_{b}) \right\} - \left\{\hat{S}_{\mathbf{k}}(t'_{a}), \hat{S}_{\mathbf{-k}}(t'_{b}) \right\} \bigg] \, . 
\end{eqnarray}
Here $(t_{a}, t_{b}, t'_{a}, t'_{b})$ denote the respective instants of time at which the measurements are performed on the individual sub-systems. The sequential nature of the measurement protocol requires that the later measurements, performed on the second sub-system $\mathcal{H}_{\mathbf{-k}}$, should occur after the earlier ones, performed on the first sub-system $\mathcal{H}_{\mathbf{k}}$, i.e. $(t'_{a}, t'_{b}) > (t_{a}, t_{b})$. The notation within the curly brackets represents the anti-commutator between the corresponding observables, defined as 
\begin{eqnarray}
    \left\{\hat{S}_{\mathbf{k}}(t_{a}), \hat{S}_{\mathbf{-k}}(t_{b}) \right\} &=& \left\{\hat{S}_{\mathbf{k}}(t_{a}) \otimes \hat{I}_{\mathbf{-k}}\, , \, \hat{I}_{\mathbf{k}} \otimes \hat{S}_{\mathbf{-k}}(t_{b}) \right\} \, .  
\end{eqnarray}
Here the operators $\hat{I}_{\mathbf{k}}$ and $\hat{I}_{\mathbf{-k}}$ denote the identity operators acting on the respective Hilbert spaces $\mathcal{H}_{\mathbf{k}}$ and $\mathcal{H}_{\mathbf{-k}}$. 

Under the same assumptions of locality and realism, the classical upper bound of bipartite temporal Bell operator is found to be $2$, identical to the bound appearing in the standard Bell–CHSH inequality. However, quantum mechanics predicts that this classical bound can be violated under suitable conditions, signalling the breakdown of at least one of these underlying assumptions \cite{Li:2011jvq}. Mathematically, the violation of the bipartite temporal Bell inequality can be expressed as follows: 
\begin{equation}
    2 \hspace{0.2 cm} \leq \hspace{0.2 cm} \left\langle \psi \right| \hat{\mathcal{B}}_{T} \left| \psi \right\rangle \hspace{0.2 cm} \leq \hspace{0.2 cm} 2 \sqrt{2} \, . 
\end{equation}
Another alternative formulation of temporal Bell inequalities is provided by the Leggett–Garg inequality, in which measurements are performed sequentially at three different times $t_{1}, t_{2}$ and $t_{3}$. This inequality tests the compatibility of quantum mechanics with the classical assumptions of macro-realism and non-invasive measurability. For detailed discussion, one should go through \cite{Martin:2016nrr,Chatterjee:2024los}. 



We now return to the investigation of quantum measures in the context of inflation. After quantizing the cosmological perturbations, a natural question arises regarding the choice of the initial quantum state of the universe. The Bunch–Davies (BD) vacuum is the conventional choice for the initial state of inflationary perturbations. However, several studies have explored non Bunch–Davies initial conditions \cite{Kundu:2011sg,Ragavendra:2024qpj,Mukherjee:2025dcv,Mondal:2024glo,Bertuzzo:2026gkj}. Among the possible alternatives, coherent states have received considerable attention, as they can lead to a breakdown of statistical homogeneity and isotropy in a cosmological setting \cite{Ragavendra:2024qpj,Mukherjee:2025dcv}. Under inflationary dynamics, an initially chosen coherent state evolves into a squeezed coherent state, in contrast to the squeezed vacuum that arises from the BD initial condition. While the violation of spatial Bell inequalities for squeezed coherent state has already been investigated \cite{Mondal:2024glo}, the corresponding analysis for temporal Bell inequalities remains largely unexplored. 

In the article, the primary objective is to examine the violation of bipartite temporal Bell inequalities for two-mode coherent state, within a cosmological framework and to assess whether it leads to any deviation from the results of Bunch-Davis vacuum.   

The organization of the article is as follows. In \ref{sec:2}, we provide a brief introduction to the theory of inflationary perturbations. In \ref{subsec:2.1}, we discuss the squeezing mechanism arising from inflationary dynamics and present the wave function of the squeezed coherent state obtained from the time evolution of an initially chosen coherent state. In \ref{subsec:3.1}, we outline the basic formalism of bipartite temporal Bell inequalities in terms of projective measurements. Subsequently, in \ref{subsec:3.2}, we present explicit calculations of unequal-time spin–spin correlations evaluated with respect to the initially chosen two-mode coherent state. In \ref{subsec:3.3}, to capture the dynamics of inflationary perturbations, we adopt the de Sitter model of inflation and present all numerical results and plots within this framework. Finally, in \ref{sec:4}, we summarize the main results of our work. 

Throughout the article, we used the following metric signature: $(-, +, +, +)$. To study the background dynamics of cosmological spacetime, we used the spatially flat Friedmann-Lemaitre-Robertson-Walker (FLRW) metric written in conformal time $\eta$ as follows: $ds^{2} = a^{2}(\eta)\left(- d\eta^{2} + \delta_{ij}dx^{i}dx^{j} \right)$. Here overdot $\dot{H}$ denotes the derivative of respective quantity with respect to cosmic time and prime $H'$ denotes the derivative of the respective quantity with respect to conformal time. The symbol $H$ denotes the Hubble parameter of FLRW background.

\section{Inflationary perturbations}\label{sec:2}

To explain the origin of the CMB temperature fluctuations and the formation of large-scale structures in the universe, it is necessary to study the perturbations generated during the inflationary epoch \cite{Brandenberger:1993zc,Mukhanov:1990me,riotto2002inflation,Sriramkumar:2009kg,Martin:2004um,KurkiSuonio2015CosmologicalPT,kurki2005cosmological}. In this article, we are primarily interested in the scalar component of inflationary perturbations, as it leads to the power spectrum of temperature anisotropies observed in the CMB map. On the other hand, vector perturbations decay rapidly with time, while tensor perturbations are responsible for the generation of primordial gravitational waves in the early universe. In this section, we therefore focus on reviewing the dynamics of scalar perturbations. 

The action governing the dynamics of inflation, can be expressed in terms of the action of a real scalar field along with the Einstein-Hilbert term. The entire action can be written as,  
\begin{eqnarray}\label{EH action}
    \mathcal{A} &=& \int d^{4}x \sqrt{-g} \left[\frac{R}{16\pi G} + \frac{1}{2} g^{\mu\nu}\partial_{\mu}\phi \partial_{\nu}\phi - V(\phi) \right] . 
\end{eqnarray}
Here $R$ denotes the Ricci scalar of the perturbed Friedmann-Lemaitre-Robertson-Walker (FLRW) metric, $\phi$ is the canonical scalar field which drives inflation (commonly referred to as the inflaton) and $V(\phi)$ represents the inflaton potential which is primarily responsible for the accelerated expansion of early universe. 

Next we will introduce the perturbations on the geometric and the matter $-$ both degrees of freedom of the above action \ref{EH action}. In terms of perturbed cosmological line element, the fluctuations to the FLRW metric can be expressed as follows,  
\begin{eqnarray}\label{metric_perturb}
    ds^{2} &=& a^{2}(\eta) \left\{-(1-2\psi_{1}) \hspace{0.06cm} d\eta^{2} + 2(\partial_{i}B) \hspace{0.06cm} dx^{i}d\eta + \left[(1-2\psi_{2})\delta_{ij}+2(\partial_{i}\partial_{j}E) \right] \hspace{0.06cm} dx^{i}dx^{j} \right\} , 
\end{eqnarray}
where $\psi_{1}, \psi_{2}, B$ and $E$ are four spacetime-dependent scalar perturbations of FLRW metric. Furthermore, the fluctuations to the inflaton field is also expressed as follows
\begin{eqnarray}\label{matter_perturb}
    \phi(x^{i}, \eta) &=& \overline{\phi}(\eta) + \delta\phi(x^{i}, \eta) \, , 
\end{eqnarray}
where $\overline{\phi}(\eta)$ is the background inflaton field and $\delta\phi(x^{i}, \eta)$ denotes the fluctuation over this smooth background configuration. 
Substituting the perturbative expansion of the metric tensor $g_{\mu\nu}$ and the inflaton field $\phi$, as given by \ref{metric_perturb} and \ref{matter_perturb}, into the action \ref{EH action}, one obtains the following simplified form 
\begin{eqnarray}\label{action_2nd_order}
    ^{(2)}\mathcal{A} &=& \frac{1}{2} \int d^{4}x \left(v'^{2} - \delta^{ij}\partial_{i}v\partial_{j}v + \frac{z''}{z}v^{2} \right) .   
\end{eqnarray}
Interestingly, the dynamics of inflationary perturbations can be expressed in terms of a single degree of freedom $v(x^{i}, \eta)$, known as the Mukhanov-Sasaki variable. It is defined in terms of a gauge invariant combination of the metric perturbation and matter perturbation, 
\begin{eqnarray}
    v(x^{i}, \eta) &=& z \bigg(\psi +  \frac{H \delta\phi}{\dot{\overline{\phi}}} \bigg) \, . 
\end{eqnarray}
The quantities $\psi$ and $z$ are defined as, 
\begin{eqnarray}
    \psi \hspace{0.2 cm} = \hspace{0.2 cm} \psi_{1} \hspace{0.2 cm} = \hspace{0.2 cm} \psi_{2}  \hspace{0.8 cm} \text{and} \hspace{0.8 cm} z \hspace{0.2 cm} = \hspace{0.2 cm} a \sqrt{2\epsilon} \hspace{0.2 cm} = \hspace{0.2 cm}  a \sqrt{2\dot{H}/H^{2}} \, . 
\end{eqnarray}
In general, it is convenient to work in the Fourier space on a constant-time hypersurface, since all the Fourier modes evolve independently with time. Taking the Fourier transform of all the terms present in the second-order action \ref{action_2nd_order}, one obtains the following expression 
\begin{eqnarray}\label{Fourier}
    ^{(2)}\mathcal{A} &=& \frac{1}{2} \int d\eta \hspace{0.07cm} d^{3}k \left[v'_{\mathbf{k}}v'^{*}_{\mathbf{k}} - \left(k^{2}- \frac{z''}{z} \right)v_{\mathbf{k}}v^{*}_{\mathbf{k}} \right]. 
\end{eqnarray}
Taking the variation of the above action as given in \ref{Fourier}, yields the following equation of motion for the gauge invariant Mukhanov-Sasaki variable $v_{\mathbf{k}}$ : 
\begin{eqnarray}\label{EOM}
    v''_{\mathbf{k}} + \bigg(k^{2}-\frac{z''}{z} \bigg)v_{\mathbf{k}} &=& 0 \, . 
\end{eqnarray}
Careful observations will reveal the coincidence between the \ref{EOM} and the equation of motion of a harmonic oscillator with a time-dependent frequency, $\omega(\eta) = \left(k^{2} - z''/z \right)^{1/2}$. Such systems are usually referred to as the parametric oscillator. 

It is evident that the equation of motions \ref{EOM} always remain invariant under the addition of a total derivative term to the action \ref{Fourier}. This freedom introduces a degeneracy in determining the explicit form of the action. Now we can make use of this freedom by adding the total derivative term  $\frac{1}{2} \left(\frac{z'}{z} v_{\mathbf{k}}v^{*}_{\mathbf{k}} \right)'$ with the action \ref{Fourier} \cite{Martin:2015qta}. As a result, the action can be rewritten in the following form: 
\begin{eqnarray}\label{modified_action}
     ^{(2)}\mathcal{A} &=&\int_{\mathbb{R}^{3+}} d\eta \hspace{0.07cm} d^{3}k \left[v'_{\mathbf{k}}v'^{*}_{\mathbf{k}} - \frac{z'}{z}\big(v_{\mathbf{k}}v'^{*}_{\mathbf{k}} + v^{*}_{\mathbf{k}}v'_{\mathbf{k}} \big) + \bigg(\frac{z'^{2}}{z^{2}}-k^{2} \bigg) v_{\mathbf{k}}v^{*}_{\mathbf{k}} \right] \, . 
\end{eqnarray}
Once the explicit form of the action has been specified, our next task is to determine the corresponding Hamiltonian density of the given system. Using the standard definition of the Hamiltonian in terms of Legendre transformation, one obtains the following 
\begin{eqnarray}\label{Hamiltonian}
    H &=& \int d^{3}k \hspace{0.06cm} \left[p_{\mathbf{k}}p^{*}_{\mathbf{k}} + \frac{z'}{z} \big(p_{\mathbf{k}}v^{*}_{\mathbf{k}} + p^{*}_{\mathbf{k}}v_{\mathbf{k}} \big) + k^{2}v_{\mathbf{k}}v^{*}_{\mathbf{k}} \right] \, . 
\end{eqnarray}
In \ref{Hamiltonian}, $p_{\mathbf{k}}$ denotes the conjugate momentum of the Fourier space Mukhanov-Sasaki variable. Once again the expression of the Hamiltonian exactly coincides with that of a parametric oscillator.

\subsection{Squeezed coherent state}\label{subsec:2.1}

So far, we have discussed the classical dynamics of inflationary perturbations. Our next task is to perform the canonical quantization of these perturbations and investigate their quantum dynamics. To analyse the quantum evolution of the corresponding wave function, it is necessary to determine the time-evolution operator $\hat{U}(t_{0}, t)$ associated with the Hamiltonian given in \ref{Hamiltonian}. Upon promoting the classical Mukhanov-Sasaki variable $v_{\mathbf{k}}$ and it's conjugate momentum $p_{\mathbf{k}}$ into quantum mechanical operators $(\hat{v}_{\mathbf{k}}, \hat{p}_{\mathbf{k}})$, it can be shown that the time-evolution operator associated with the cosmological Hamiltonian \ref{Hamiltonian} can be factorised into the Squeezing $\hat{S}_{\mathbf{k}}(r_{k}, \phi_{k})$ and Rotation $\hat{R}_{\mathbf{k}}(\theta_{k})$ operator as given below \cite{Martin:2015qta,Grain:2019vnq},  
\begin{eqnarray}
    \hat{U}_{\mathbf{k}}(r_{k}, \theta_{k}, \phi_{k}) \hspace{0.15 cm} = \hspace{0.15 cm} \hat{S}_{\mathbf{k}}(r_{k},\phi_{k}) \, \hat{R}_{\mathbf{k}}(\theta_{k}) \, . 
\end{eqnarray}
The explicit form of Squeezing $\hat{S}_{\mathbf{k}}(r_{k}, \phi_{k})$ and Rotation $\hat{R}_{\mathbf{k}}(\theta_{k})$ operator, expressed in terms of creation $\hat{a}^{\dagger}_{\mathbf{k}}$ and annihilation $\hat{a}_{\mathbf{k}}$ operator are found to be the following \cite{Martin:2015qta,Grain:2019vnq}, 
\begin{eqnarray}
    \hat{S}_{\mathbf{k}}(r_{k}, \phi_{k}) &=& \exp \left\{r_{k} \left[e^{-2i\phi_{k}}  \hat{a}_{\mathbf{k}}(\eta_{0}) \, \hat{a}_{-\mathbf{k}}(\eta_{0}) - e^{2i\phi_{k}} \hat{a}^{\dagger}_{\mathbf{k}}(\eta_{0}) \, \hat{a}^{\dagger}_{-\mathbf{k}}(\eta_{0}) \right]\right\} \, , \nonumber \\
    \hat{R}_{\mathbf{k}}(\theta_{k}) &=& \exp \left\{-i\theta_{k} \left[\hat{a}^{\dagger}_{\mathbf{k}}(\eta_{0}) \, \hat{a}_{\mathbf{k}}(\eta_{0}) + \hat{a}^{\dagger}_{-\mathbf{k}}(\eta_{0}) \, \hat{a}_{-\mathbf{k}}(\eta_{0}) \right] \right\} \, . 
\end{eqnarray}
Here the quantities $r_{k}, \phi_{k}$ and $\theta_{k}$ are known as the squeezing parameter, squeezing angle and rotation angle respectively. On the other hand, the quantity $\eta_{0}$ represents the conformal time when the Fourier modes are well inside the Hubble radius. 

Having established the structure of the time-evolution operator, the next step is to specify the initial quantum state of the cosmological perturbations and determine its subsequent time evolution. In the standard treatment, the initial state is taken to be the Bunch–Davies vacuum. However, one may also consider alternative states as an initial choice. Following Ref. \cite{Kundu:2011sg,Ragavendra:2024qpj,Mondal:2024glo}, we choose the two-mode coherent state $\ket{\alpha_{\mathbf{k}}, \alpha_{-\mathbf{k}}}$ as a possible alternative to the Bunch–Davies vacuum $\ket{0_{\mathbf{k}}, 0_{-\mathbf{k}}}$. The formal definition of the two-mode coherent state is given below: 
\begin{equation}\label{coherent_def}
    \ket{\alpha_{\mathbf{k}}, \alpha_{-\mathbf{k}}} \, = \, \hat{D}(\alpha_{\mathbf{k}}) \hat{D}(\alpha_{-\mathbf{k}}) \ket{0_{\mathbf{k}}, 0_{-\mathbf{k}}} \hspace{0.3 cm} = \hspace{0.3 cm} \exp(\alpha_{\mathbf{k}} \hat{a}^{\dagger}_{\mathbf{k}} - \alpha^{*}_{\mathbf{k}} \hat{a}_{\mathbf{k}}) \ket{0_{\mathbf{k}}} \, \otimes \, \exp(\alpha_{\mathbf{-k}} \hat{a}^{\dagger}_{\mathbf{-k}} - \alpha^{*}_{\mathbf{-k}} \hat{a}_{\mathbf{-k}}) \ket{0_{\mathbf{-k}}} \, . 
\end{equation}
Upon the action of time-evolution operator $\hat{U}_{\mathbf{k}}(r_{k}, \theta_{k}, \phi_{k})$ on the chosen two-mode coherent state, yields the following expression for the final quantum state, denoted by $\ket{r, \alpha_{\mathbf{k}}, \alpha_{\mathbf{-k}}} $. The expression is given below, 
\begin{eqnarray}
    \ket{r,\alpha_{\mathbf{k}}, \alpha_{\mathbf{-k}}} &=& \hat{U}_{\mathbf{k}}(r_{k}, \theta_{k}, \phi_{k}) \hspace{0.07 cm} \ket{\alpha_{\mathbf{k}},\alpha_{-\mathbf{k}}} \hspace{0.4 cm} = \hspace{0.2 cm} \hat{S}_{\mathbf{k}}(r_{k},\phi_{k}) \hat{R}_{\mathbf{k}}(\theta_{k}) \hspace{0.07 cm} \hat{D}(\alpha_{\mathbf{k}}) \hat{D}(\alpha_{-\mathbf{k}}) \ket{0_{\mathbf{k}}, 0_{-\mathbf{k}}} \, . 
\end{eqnarray}
The resulting state $\ket{r, \alpha_{\mathbf{k}}, \alpha_{\mathbf{-k}}} $ is known as the squeezed coherent state. Its explicit form in terms of number basis $\ket{n_{\mathbf{k}}, n_{\mathbf{-k}}}$ can be found in the Ref. \cite{Mondal:2024glo}. 

To investigate the Bell inequalities, it is necessary to determine the wave function of the squeezed coherent state in the quadrature basis $(\hat{q}_{\mathbf{k}}, \hat{q}_{\mathbf{-k}})$. In this representation, the squeezed coherent state is found to possess a displaced Gaussian wave function. Upon explicit calculation, its analytical form can be expressed as follows \cite{Mondal:2024glo}: 
\begin{eqnarray}\label{wave_func}
    \psi_{r}(q_{\mathbf{k}},q_{-\mathbf{k}}) &=& N_{0} \exp[-A_{0}\big(q_{\mathbf{k}}-C_{0} \big)^{2} - A_{0}\big(q_{-\mathbf{k}}-D_{0} \big)^{2} + B_{0} \hspace{0.06cm} q_{\mathbf{k}}q_{-\mathbf{k}} ] \, . 
\end{eqnarray}
In terms of squeezing parameters $(r_{k}, \theta_{k}, \phi_{k})$ and coherent parameters $(\alpha_{\mathbf{k}}, \alpha_{-\mathbf{k}})$, the coefficients $A_{0}, B_{0}, C_{0}, D_{0}$ and the normalization factor $N_{0}$ are expressed as follows:  
\begin{eqnarray}\label{coff_wave_function}
    A_{0} &=& \frac{1}{2} \bigg(\frac{1+e^{-4i\phi_{k}}\tanh^{2}{r_{k}}}{1-e^{-4i\phi_{k}}\tanh^{2}{r_{k}}} \bigg) \, , \label{coff_1} \\ 
    B_{0} &=& \bigg(\frac{2 \, e^{-2i\phi_{k}}\tanh{r_{k}}}{1-e^{-4i\phi_{k}}\tanh^{2}{r_{k}}} \bigg)\, , \label{coff_2} \\ 
    C_{0} &=& \frac{2}{\big(1 + e^{-4i\phi_{k}}\tanh^{2}{r_{k}} \big)} \left[\left(\frac{\zeta_{\mathbf{k}}}{\sqrt{2}} + e^{-4i\phi_{k}}\tanh^{2}{r_{k}} \hspace{0.07cm} \frac{\zeta^{*}_{\mathbf{k}}}{\sqrt{2}} \right) - 2e^{-2i\phi_{k}}\tanh{r_{k}} \Re\left(\frac{\Upsilon_{\mathbf{k}}}{\sqrt{2}} \right) \right] \, , \label{coff_3} \\ 
    D_{0} &=& \frac{2}{\big(1 + e^{-4i\phi_{k}}\tanh^{2}{r_{k}} \big)} \left[\left(\frac{\Upsilon_{\mathbf{k}}}{\sqrt{2}} + e^{-4i\phi_{k}}\tanh^{2}{r_{k}} \hspace{0.07cm} \frac{\Upsilon^{*}_{\mathbf{k}}}{\sqrt{2}} \right) - 2e^{-2i\phi_{k}}\tanh{r_{k}} \Re\left(\frac{\zeta_{\mathbf{k}}}{\sqrt{2}} \right) \right] \, , \label{coff_4} \\ 
    N_{0} &=& \frac{\sech{r_{k}}}{\sqrt{2 \pi (1-e^{-4i\phi_{k}}\tanh^{2}{r_{k}})}} \exp\left\{-4i\Im\big(A_{0} \big) \left[\Re\left(\frac{\zeta_{\mathbf{k}}}{\sqrt{2}} \right)^{2} + \Re\left(\frac{\Upsilon_{\mathbf{k}}}{\sqrt{2}} \right)^{2} \right] + 4i\Im\big(B_{0} \big) \,  \Re\left(\frac{\zeta_{\mathbf{k}}}{\sqrt{2}} \right) \Re\left(\frac{\Upsilon_{\mathbf{k}}}{\sqrt{2}} \right) \right\} \nonumber \\ 
    && \cross \, \exp\left\{A_{0} \left(C_{0}^{2} + D_{0}^{2} \right) - 2i \, \left[\Re\left(\frac{\zeta_{\mathbf{k}}}{\sqrt{2}} \right) \Im\left(\frac{\zeta_{\mathbf{k}}}{\sqrt{2}} \right) + \Re\left(\frac{\Upsilon_{\mathbf{k}}}{\sqrt{2}} \right) \Im\left(\frac{\Upsilon_{\mathbf{k}}}{\sqrt{2}} \right) \right] - \frac{1}{2} \left(g_{01} \, x^{2}_{0} + g_{02} \, y^{2}_{0} \right) \right\} \, \label{Normalization_sc} .  
\end{eqnarray}
The parameters $\zeta_{\mathbf{k}}$ and $\Upsilon_{\mathbf{k}}$ are expressed as
\begin{eqnarray}
    \zeta_{\mathbf{k}} &=& \big(e^{-i\theta_{k}}\cosh{r_{k}} \hspace{0.1cm} \alpha_{\mathbf{k}} - \hspace{0.1cm} e^{i(\theta_{k}+2\phi_{k})}\sinh{r_{k}} \hspace{0.1cm} \alpha^{*}_{-\mathbf{k}} \big) \, , \label{zeta_def} \\ 
    \Upsilon_{\mathbf{k}} &=& \big(e^{-i\theta_{k}}\cosh{r_{k}} \hspace{0.1cm} \alpha_{-\mathbf{k}} - \hspace{0.1cm} e^{i(\theta_{k}+2\phi_{k})}\sinh{r_{k}} \hspace{0.1cm} \alpha^{*}_{\mathbf{k}} \big) \, . \label{upsilon_def} 
\end{eqnarray}
The auxiliary quantities $(g_{01}, \, g_{02}, \, x_{0}, \, y_{0})$ are defined as follows, 
\begin{eqnarray}
    g_{01} &=& \Re\left(\frac{2A_{0} - B_{0}}{2} \right) \hspace{0.7 cm} = \hspace{0.4 cm} \frac{\sech^{2}{r_{k}}}{2} \bigg(\frac{1 + \tanh^{2}{r_{k}} - 2\cos{2\phi_{k}} \, \tanh{r_{k}}}{1 + \tanh^{4}{r_{k}} - 2\cos{4\phi_{k}} \, \tanh^{2}{r_{k}}} \bigg) \label{def-g_1} \, , \\ 
    g_{02} &=& \Re\left(\frac{2A_{0} + B_{0}}{2} \right) \hspace{0.7 cm} = \hspace{0.4 cm} \frac{\sech^{2}{r_{k}}}{2} \bigg(\frac{1 + \tanh^{2}{r_{k}} + 2\cos{2\phi_{k}} \, \tanh{r_{k}}}{1 + \tanh^{4}{r_{k}} - 2\cos{4\phi_{k}} \, \tanh^{2}{r_{k}}} \bigg) \label{def-g_2} \, , \\ 
    x_{0} \hspace{-0.1 cm} &=& \hspace{-0.1 cm} \frac{\Re\left(A_{0}C_{0} + A_{0}D_{0} \right)}{g_{01}} \hspace{0.4 cm} = \hspace{0.4 cm} \sqrt{2} \, \Re\left(\zeta_{\mathbf{k}} + \Upsilon_{\mathbf{k}} \right) \, , \label{def-x_0} \\ 
    y_{0} \hspace{-0.1 cm} &=& \hspace{-0.1 cm} \frac{\Re\left(A_{0}C_{0} - A_{0}D_{0} \right)}{g_{02}} \hspace{0.4 cm} = \hspace{0.4 cm} \sqrt{2} \, \Re\left(\zeta_{\mathbf{k}} - \Upsilon_{\mathbf{k}} \right) \, . \label{def-y_0} 
\end{eqnarray}

\section{Bipartite temporal Bell Inequality}\label{sec:3}

\subsection{Basic formalism}\label{subsec:3.1}

To investigate the violations of bipartite temporal Bell inequality, one needs to define the two point correlation function of pseudo-spin operators $\hat{S}_{i}$ evaluated at different instants of time, say $t_{1}$ and $t_{2}$. In the temporal scenario, where quantum operators evolve with time, it is convenient to adopt the Heisenberg picture. Within this formalism, the time evolution of the operators is given by: 
\begin{eqnarray}\label{spin_1st}
    \hat{S}_{a}(t_{1}) &=& \hat{U}^{\dagger}(t_{1}) \, \hat{S}_{a}(0) \, \hat{U}(t_{1}) \, , 
\end{eqnarray}
where $\hat{U}(t_{1})$ denotes the time-evolution operator of the corresponding system under consideration and $\hat{S}_{a}(0)$ is the pseudo-spin operator belonging to the first subspace $\mathcal{H}_{a}$ at time $t=0$. Furthermore, the time-evolution of the corresponding operators belonging to the second subspace $\mathcal{H}_{b}$ can be also defined in an analogous manner, 
\begin{eqnarray}\label{spin_2nd}
    \hat{S}_{b}(t_{2}) &=& \hat{U}^{\dagger}(t_{2}) \, \hat{S}_{b}(0) \, \hat{U}(t_{2}) \, . 
\end{eqnarray}

Our next task is to define the two-point spin–spin correlation function $\overline{E}(t_{1}, t_{2})$, evaluated at two different times, in such a way that it yields a real quantity. A naive definition of this correlation function can be written as 
\begin{eqnarray}\label{spin_correlation_old}
    \overline{E}(t_{1}, t_{2}) &=& \left\langle \psi \right| \hat{S}_{a}(t_{1}) \hat{S}_{b}(t_{2}) \left| \psi \right\rangle \, . 
\end{eqnarray}
However, the pseudo-spin operators as defined in the \ref{spin_1st} and \ref{spin_2nd} do not commute with each other. Consequently, their product $\hat{S}_{a}(t_{1}) \hat{S}_{b}(t_{2})$ is found to be a non-Hermitian operator, and therefore the corresponding correlation function defined in the \ref{spin_correlation_old} yields a complex quantity. From an observational perspective, one should always expect a real-valued quantity for the experimental verification of the inequality. For this reason, it is more appropriate to define the correlation functions in terms of the quantum expectation value of projective measurements \cite{Ando:2020kdz, Fritz:2010qzm}. 

According to the measurement postulate of quantum mechanics, the wave function collapse associated with the measurement of a quantum system, can be described in terms of projection operators. In the context of Bell inequalities, one typically considers dichotomic observables that admit only two possible outcomes $(+ 1)$ and $(- 1)$. Accordingly, the projection operators $\hat{\mathcal{P}}^{(a)}_{1}$ and $\hat{\mathcal{P}}^{(a)}_{-1}$, which are associated with the $(+1)$ and $(-1)$ eigen-space of the respective pseudo-spin operator $\hat{S}_{a}(t_{1})$, are defined as follows: 
\begin{eqnarray}
    \hat{\mathcal{P}}^{(a)}_{1} &=& \frac{1}{2} \, \big[\hat{I}_{a} + \hat{S}_{a}(t_{1}) \big] \hspace{0.7 cm} , \hspace{0.7 cm} \hat{\mathcal{P}}^{(a)}_{-1} \hspace{0.2 cm} = \hspace{0.2 cm} \frac{1}{2} \, \big[\hat{I}_{a} - \hat{S}_{a}(t_{1}) \big] \, . 
\end{eqnarray}
Here $\hat{I}_{a}$ denotes the identity operator belonging to the first Hilbert space $\mathcal{H}_{a}$. Similarly, one can also define another set of projection operators $\hat{\mathcal{P}}^{(b)}_{1}$ and $\hat{\mathcal{P}}^{(b)}_{-1}$ associated with the pseudo-spin operator $\hat{S}_{b}(t_{2})$ belonging to the second Hilbert space $\mathcal{H}_{b}$. 

For notational convenience, let us assume that $m$ and $n$ $-$ denote the possible outcomes of the first and second pseudo-spin measurements, associated with the respective operators $\hat{S}_{a}(t_{1})$ and $\hat{S}_{b}(t_{2})$. The corresponding projection operators $\hat{\mathcal{P}}^{(a)}_{m}$ and $\hat{\mathcal{P}}^{(b)}_{n}$ which project the quantum states onto the eigen-space of the respective measurement outcomes $m$ and $n$, are defined as follows: 
\begin{eqnarray}
    \hat{\mathcal{P}}^{(a)}_{m} &=& \frac{1}{2} \, \big[\hat{I}_{a} + m \, \hat{S}_{a}(t_{1}) \big] \hspace{0.7 cm} , \hspace{0.7 cm} \hat{\mathcal{P}}^{(b)}_{n} \hspace{0.2 cm} = \hspace{0.2 cm} \frac{1}{2} \, \big[\hat{I}_{b} + n \, \hat{S}_{b}(t_{2}) \big] \, . 
\end{eqnarray}
From these definitions, one can readily verify the standard properties of projection operators, such as the following:  
\begin{equation}
    \left(\hat{\mathcal{P}}^{(a)}_{m} \right)^{2} \, = \, \hat{\mathcal{P}}^{(a)}_{m} \hspace{0.7 cm} , \hspace{0.7 cm} \hat{\mathcal{P}}^{(a)}_{m} \, = \, \left(\hat{\mathcal{P}}^{(a)}_{m} \right)^{\dagger} \, . 
\end{equation}

In the context of temporal Bell inequalities, observers has to perform two consecutive measurements on a dichotomic observable at two different instants of time, for the evaluation of unequal-time spin-spin correlation function $E(t_{1}, t_{2})$. Let us assume that Alice performs a measurement on $\hat{S}_{a}$ at time $t_{1}$ and later Bob performs another measurement on $\hat{S}_{b}$ at time $t_{2}$. Again we assume that the joint probability for Alice and Bob to get the respective measurement outcomes $- \, m$ and $n \, -$ is denoted by $\mathcal{F}(m, n)$. \textit{This joint probability can be expressed as a product of the individual probabilities associated with the measurement outcomes of Alice and Bob.} Following Alice's measurement, the initial wave function $\ket{\psi}$ collapses into the state $\ket{\psi'} = \hat{\mathcal{P}}^{(a)}_{m} \ket{\psi}$. Subsequently, after Bob's measurement, the intermediate state $\ket{\psi'}$ collapses into the final state $\ket{\psi''} = \hat{\mathcal{P}}^{(b)}_{n} \ket{\psi'}$. According to Born's probability rule, the joint probability is given by: 
\begin{eqnarray}
    \mathcal{F}(m, n) &=& \left\langle \psi' | \psi'' \right\rangle \nonumber \\ 
    &=& \left\langle \psi \right| \hat{\mathcal{P}}^{(a) \dagger}_{m} \, \hat{\mathcal{P}}^{(b)}_{n} \hat{\mathcal{P}}^{(a)}_{m} \left| \psi \right\rangle \nonumber \\ 
    &=& \frac{1}{4} + \frac{m}{4} \left\langle \psi \right| \hat{S}_{a}(t_{1}) \left| \psi \right\rangle + \frac{n}{4} \left\langle \psi \right| \hat{S}_{b}(t_{2}) \left| \psi \right\rangle + \frac{mn}{8} \left\langle \psi \right| \big\{\hat{S}_{a}(t_{1}), \hat{S}_{b}(t_{2}) \big\} \left| \psi \right\rangle + \frac{n}{8} \left\langle \psi \right| \hat{S}_{a}(t_{1}) \hat{S}_{b}(t_{2}) \hat{S}_{a}(t_{1}) \left| \psi \right\rangle \, . \nonumber \\ 
\end{eqnarray}
Here $\{\, ,  \}$ denotes the anti-commutator between the pseudo-spin operators. Our next task is to evaluate the unequal-time spin–spin correlation function using the sequential measurement procedure described above. This correlation is defined as an average over all possible outcomes of the consecutive measurements performed at two different times. According to this definition, the unequal-time correlation function can be written as 
\begin{eqnarray}\label{spin_correlation_new_1}
    E(t_{1}, t_{2}) &=& \sum_{m} \sum_{n} mn \, \mathcal{F}(m, n) \nonumber \\ 
    &=& \mathcal{F}(+1, +1) - \mathcal{F}(+1, -1) - \mathcal{F}(-1, +1) + \mathcal{F}(-1, -1) \nonumber \\ 
    &=& \frac{1}{2} \left\langle \psi \right| \big\{\hat{S}_{a}(t_{1}), \hat{S}_{b}(t_{2}) \big\} \left| \psi \right\rangle \, . 
\end{eqnarray}
The use of projective measurements ensure that the resulting correlation function is real-valued, thereby eliminating the issue of complex correlations encountered earlier. The remaining three unequal-time spin-spin correlators appearing in the bipartite temporal Bell operator \ref{Bipartite_Temporal_Bell} can be defined in an analogous manner, as given below: 
\begin{eqnarray}
    E(t_{1}, t'_{2}) &=& \frac{1}{2} \left\langle \psi \right| \big\{\hat{S}_{a}(t_{1}), \hat{S}_{b}(t'_{2}) \big\} \left| \psi \right\rangle \, , \label{spin_correlations_new_2} \\ 
    E(t'_{1}, t_{2}) &=& \frac{1}{2} \left\langle \psi \right| \big\{\hat{S}_{a}(t'_{1}), \hat{S}_{b}(t_{2}) \big\} \left| \psi \right\rangle \, , \label{spin_correlations_new_3} \\ 
    E(t'_{1}, t'_{2}) &=& \frac{1}{2} \left\langle \psi \right| \big\{\hat{S}_{a}(t'_{1}), \hat{S}_{b}(t'_{2}) \big\} \left| \psi \right\rangle \, . \label{spin_correlations_new_4} 
\end{eqnarray}
The quantum expectation value of bipartite temporal Bell operator $\hat{\mathcal{B}}_{T}$ is defined in the following manner, 
\begin{eqnarray}\label{Temporal_Bell}
    \left\langle \psi \right| \hat{\mathcal{B}}_{T} \left| \psi \right\rangle &=& E(t_{1}, t_{2}) \, + \, E(t_{1}, t'_{2}) \, + \, E(t'_{1}, t_{2}) \, - \, E(t'_{1}, t'_{2}) \, . 
\end{eqnarray}
Substituting the unequal-time correlations as given by \ref{spin_correlation_new_1} - \ref{spin_correlations_new_4} into the definition of temporal Bell operator \ref{Temporal_Bell}, one can examine the possible violations of bipartite temporal Bell inequality for a given quantum state.

\subsection{Unequal-time spin-spin correlation}\label{subsec:3.2}

In this section, we perform the explicit computation of unequal-time spin-spin correlation function $E(t_{i}, t_{j})$ evaluated with respect to two-mode coherent state $\ket{\alpha_{\mathbf{k}}, \alpha_{\mathbf{-k}}} $. In the quadrature basis $(q_{\mathbf{k}}, q_{\mathbf{-k}})$, its wave function is given in \ref{wave_func}. Before proceeding further, let us first introduce the formal definition of pseudo-spin operators $\hat{S}_{z}$ for a continuous variable quantum system. According to Gour-Khanna-Mann-Revzen (GKMR) prescription, the spin-$z$ operator $\hat{S}_{z}(\mathbf{k})$ is defined as follows (in the Schrödinger picture) \cite{Gour:2003wsm,Revzen:2004mzw}: 
\begin{eqnarray}\label{GKMR}
    \hat{S}_{z}(\mathbf{k}) &=& -\int^{\infty}_{-\infty} dq_{\mathbf{k}} \ket{q_{\mathbf{k}}}\bra{-q_{\mathbf{k}}} \, .  
\end{eqnarray}
The definition of another spin-$z$ operator $\hat{S}_{z}(-\mathbf{k})$ belonging to the subspace $\mathcal{H}_{\mathbf{-k}}$ can be found by interchanging the quadrature variable $q_{\mathbf{k}}$ with $q_{-\mathbf{k}}$ in the \ref{GKMR}. It is worth noting that the above definition of $\hat{S}_{z}$ represents the operators $\hat{S}_{a}(0)$ and $\hat{S}_{b}(0)$ involved in the \ref{spin_1st} and \ref{spin_2nd} respectively. More precisely, $\hat{S}_{z}(\mathbf{k}) = \hat{S}_{a}(0)$ and $\hat{S}_{z}(-\mathbf{k}) = \hat{S}_{b}(0)$.  

Let us begin the computation of first unequal-time spin-spin correlator \ref{spin_correlation_new_1} as follows, 
\begin{eqnarray}\label{spin_correlation_1}
    E(t_{1}, t_{2}) \hspace{-0.20 cm} &=& \hspace{-0.22 cm} \Re\left[\left\langle \alpha_{\mathbf{k}}, \alpha_{-\mathbf{k}} \right| \hat{S}_{z}(\mathbf{k}, t_{1}) \, \hat{S}_{z}(-\mathbf{k}, t_{2}) \left| \alpha_{\mathbf{k}}, \alpha_{-\mathbf{k}} \right\rangle \right] \nonumber \\ 
    \hspace{-0.20 cm} &=& \hspace{-0.22 cm} \Re\left[\left\langle \alpha_{\mathbf{k}}, \alpha_{-\mathbf{k}} \right| \hat{U}^{\dagger}(t_{1}) \, \hat{S}_{z}(\mathbf{k}, t_{0}) \, \hat{U}(t_{1}) \, \hat{U}^{\dagger}(t_{2}) \, \hat{S}_{z}(-\mathbf{k}, t_{0}) \, \hat{U}(t_{2}) \left| \alpha_{\mathbf{k}}, \alpha_{-\mathbf{k}} \right\rangle \right] \nonumber \\ 
    \hspace{-0.20 cm} &=& \hspace{-0.22 cm} \Re\left[\left\langle r_{t_{1}}, \alpha_{\mathbf{k}}, \alpha_{\mathbf{-k}} \right| \hat{S}_{z}(\mathbf{k}, t_{0}) \, \hat{U}(t_{1}) \, \hat{U}^{\dagger}(t_{2}) \, \hat{S}_{z}(-\mathbf{k}, t_{0}) \left| r_{t_{2}}, \alpha_{\mathbf{k}}, \alpha_{\mathbf{-k}} \right\rangle \right] \nonumber \\ 
    \hspace{-0.20 cm} &=& \hspace{-0.22 cm} \int dq_{\mathbf{k_{1}}} dq_{-\mathbf{k_{2}}} \int d\widetilde{q}_{\mathbf{k_{1}}} d\widetilde{q}_{-\mathbf{k_{2}}} \, \Re\bigg[\left\langle r_{t_{1}}, \alpha_{\mathbf{k}}, \alpha_{\mathbf{-k}} | q_{\mathbf{k_{1}}}, q_{-\mathbf{k_{2}}} \right\rangle \, \left\langle q_{\mathbf{k_{1}}}, q_{-\mathbf{k_{2}}} \right| \hat{S}_{z}(\mathbf{k}, t_{0}) \, \hat{U}(t_{1}) \, \hat{U}^{\dagger}(t_{2}) \, \hat{S}_{z}(-\mathbf{k}, t_{0}) \left| \widetilde{q}_{\mathbf{k_{1}}}, \widetilde{q}_{-\mathbf{k_{2}}} \right\rangle \bigg. \nonumber \\ 
    && \hspace{4.6 cm} \bigg. \cross \, \left\langle \widetilde{q}_{\mathbf{k_{1}}}, \widetilde{q}_{-\mathbf{k_{2}}} | r_{t_{2}}, \alpha_{\mathbf{k}}, \alpha_{\mathbf{-k}} \right\rangle \, \bigg] \nonumber \\ 
    \hspace{-0.20 cm} &=& \hspace{-0.22 cm} \int dq_{\mathbf{k_{1}}} dq_{-\mathbf{k_{2}}} \int d\widetilde{q}_{\mathbf{k_{1}}} d\widetilde{q}_{-\mathbf{k_{2}}} \, \Re\bigg[\psi^{*}_{r, \alpha} \left(q_{\mathbf{k_{1}}}, q_{-\mathbf{k_{2}}}, t_{1} \right) \, \left\langle q_{\mathbf{k_{1}}}, q_{-\mathbf{k_{2}}} \right| \hat{S}_{z}(\mathbf{k}, t_{0}) \, \hat{U}(t_{1}) \, \hat{U}^{\dagger}(t_{2}) \, \hat{S}_{z}(-\mathbf{k}, t_{0}) \left| \widetilde{q}_{\mathbf{k_{1}}}, \widetilde{q}_{-\mathbf{k_{2}}} \right\rangle \bigg. \nonumber \\ 
    && \hspace{4.6 cm} \bigg. \cross \,  \psi_{r, \alpha} \left(\widetilde{q}_{\mathbf{k_{1}}}, \widetilde{q}_{-\mathbf{k_{2}}}, t_{2} \right) \, \bigg] \, . 
\end{eqnarray}
In the fourth line of the \ref{spin_correlation_1}, we plugged in the completeness relation for the eigen states of quadrature operators $\hat{q}_{\mathbf{k}}$ as follows, 
\begin{eqnarray}\label{complete_quadrature}
    \int dq_{\mathbf{k}} \, dq_{\mathbf{-k}} \hspace{0.1 cm} \ket{q_{\mathbf{k}}, q_{\mathbf{-k}}} \bra{q_{\mathbf{k}}, q_{\mathbf{-k}}} &=& \hat{I}_{\mathbf{k}} \otimes \hat{I}_{\mathbf{-k}} \, . 
\end{eqnarray}
In \ref{complete_quadrature}, $\hat{I}_{\mathbf{k}}$ and $\hat{I}_{\mathbf{-k}}$ denote the identity operators of the respective Hilbert space $\mathcal{H}_{\mathbf{k}}$ and $\mathcal{H}_{\mathbf{-k}}$ respectively. 

The definitions for the wave function of squeezed coherent state which we used in the fifth line of the \ref{spin_correlation_1}, are given below: 
\begin{eqnarray}\label{intermediate}
    \psi_{r, \alpha} \left(\widetilde{q}_{\mathbf{k_{1}}}, \widetilde{q}_{-\mathbf{k_{2}}}, t_{2} \right) &=& \left\langle \widetilde{q}_{\mathbf{k_{1}}}, \widetilde{q}_{-\mathbf{k_{2}}} \, | \, r_{t_{2}}, \alpha_{\mathbf{k}}, \alpha_{\mathbf{-k}} \right\rangle \, , \\ 
    \psi^{*}_{r, \alpha} \left(q_{\mathbf{k_{1}}}, q_{-\mathbf{k_{2}}}, t_{1} \right) &=& \left\langle r_{t_{1}}, \alpha_{\mathbf{k}}, \alpha_{\mathbf{-k}} \, | \, q_{\mathbf{k_{1}}}, q_{-\mathbf{k_{2}}} \right\rangle \, . 
\end{eqnarray}
Let us denote the matrix element of the particular combination $\hat{S}_{z}(\mathbf{k}, t_{0}) \, \hat{U}(t_{1}) \, \hat{U}^{\dagger}(t_{2}) \, \hat{S}_{z}(-\mathbf{k}, t_{0})$ as arrived in the \ref{spin_correlation_1}, by $J_{0}$. Substituting the GKMR definition of $\hat{S}_{z}$ in the above mentioned matrix element, one obtains the following expression 
\begin{eqnarray}\label{intermediate_1}
    J_{0} &=& \left\langle q_{\mathbf{k_{1}}}, q_{-\mathbf{k_{2}}} \right| \hat{S}_{z}(\mathbf{k}, t_{0}) \, \hat{U}(t_{1}) \, \hat{U}^{\dagger}(t_{2}) \, \hat{S}_{z}(-\mathbf{k}, t_{0}) \left| \widetilde{q}_{\mathbf{k_{1}}}, \widetilde{q}_{-\mathbf{k_{2}}} \right\rangle \nonumber \\ 
    &=& \int\int dq_{\mathbf{k}} dq_{-\mathbf{k}} \left\langle q_{\mathbf{k_{1}}}, q_{-\mathbf{k_{2}}} | q_{\mathbf{k}} \right\rangle \left\langle -q_{\mathbf{k}} \right| \hat{U}(t_{1}) \, \hat{U}^{\dagger}(t_{2}) \left| q_{-\mathbf{k}} \right\rangle \left\langle -q_{-\mathbf{k}} | \widetilde{q}_{\mathbf{k_{1}}}, \widetilde{q}_{-\mathbf{k_{2}}} \right\rangle \nonumber \\ 
    &=& \int dq_{\mathbf{k}} dq_{-\mathbf{k}} \int dq_{-\mathbf{k_{3}}} dq_{\mathbf{k_{4}}} \left\langle q_{\mathbf{k_{1}}}, q_{-\mathbf{k_{2}}} | q_{\mathbf{k}}, q_{-\mathbf{k_{3}}} \right\rangle \left\langle -q_{\mathbf{k}}, q_{-\mathbf{k_{3}}} \right| \hat{U}(t_{1}) \, \hat{U}^{\dagger}(t_{2}) \left| q_{\mathbf{k_{4}}}, q_{-\mathbf{k}} \right\rangle \left\langle q_{\mathbf{k_{4}}}, -q_{-\mathbf{k}} | \widetilde{q}_{\mathbf{k_{1}}}, \widetilde{q}_{-\mathbf{k_{2}}} \right\rangle \nonumber \\ 
    &=& \left\langle -q_{\mathbf{k_{1}}}, q_{-\mathbf{k_{2}}} \right| \hat{U}(t_{1}) \, \hat{U}^{\dagger}(t_{2}) \left| \widetilde{q}_{\mathbf{k_{1}}}, -\widetilde{q}_{-\mathbf{k_{2}}} \right\rangle \, . 
\end{eqnarray}
In the above expression, we used the completeness relation and orthogonality condition for the eigen-states of quadrature operator $\hat{q}_{\mathbf{k}}$. Explicit form of the orthogonality condition is given below, 
\begin{eqnarray}
    \left\langle q_{\mathbf{k_{1}}}, q_{-\mathbf{k_{2}}} | q_{\mathbf{k}}, q_{-\mathbf{k_{3}}} \right\rangle &=& \delta\left(q_{\mathbf{k_{1}}} - q_{\mathbf{k}} \right) \, \delta\left(q_{-\mathbf{k_{2}}} - q_{-\mathbf{k_{3}}} \right) \, . 
\end{eqnarray}
Inserting the completeness relation of coherent state $\ket{\beta_{\mathbf{k}}}$ 
\begin{equation}
    \iint \frac{d^{2}\beta_{\mathbf{k}} \, d^{2}\beta_{\mathbf{-k}}}{\pi^{2}} \hspace{0.1 cm} \ket{\beta_{\mathbf{k}}, \, \beta_{\mathbf{-k}}} \bra{\beta_{\mathbf{k}}, \, \beta_{\mathbf{-k}}} \hspace{0.2 cm} = \hspace{0.2 cm} \hat{I}_{\mathbf{k}} \otimes \hat{I}_{\mathbf{-k}} \, , 
\end{equation}
in the last line of \ref{intermediate_1}, one obtains the following expression 
\begin{eqnarray}\label{intermediate_2}
    J_{0} &=& \int\int \frac{d^{2}\beta_{\mathbf{k}} d^{2}\beta_{-\mathbf{k}}}{\pi^{2}} \left\langle -q_{\mathbf{k_{1}}}, q_{-\mathbf{k_{2}}} \right| \hat{U}(t_{1}) \left| \beta_{\mathbf{k}}, \beta_{-\mathbf{k}} \right\rangle \left\langle \beta_{\mathbf{k}}, \beta_{-\mathbf{k}} \right| \hat{U}^{\dagger}(t_{2}) \left| \widetilde{q}_{\mathbf{k_{1}}}, -\widetilde{q}_{-\mathbf{k_{2}}} \right\rangle \nonumber \\ 
    &=& \frac{1}{\pi^{2}} \int\int d^{2}\beta_{\mathbf{k}} d^{2}\beta_{-\mathbf{k}} \hspace{0.15 cm} \psi_{r, \beta}\left(-q_{\mathbf{k_{1}}}, q_{-\mathbf{k_{2}}}, t_{1} \right) \, \psi^{*}_{r, \beta}\left(\widetilde{q}_{\mathbf{k_{1}}}, -\widetilde{q}_{-\mathbf{k_{2}}}, t_{2} \right) \, . 
\end{eqnarray}
Substituting the final expression of $J_{0}$ as given by the \ref{intermediate_2}, into the unequal-time spin correlation $E(t_{1}, t_{2})$ as given by \ref{spin_correlation_1}, one should get the following expression 
\begin{eqnarray}\label{spin_correlation_3}
    E(t_{1}, t_{2}) &=& \int dq_{\mathbf{k_{1}}} dq_{-\mathbf{k_{2}}} \int d\widetilde{q}_{\mathbf{k_{1}}} d\widetilde{q}_{-\mathbf{k_{2}}} \int \frac{d^{2}\beta_{\mathbf{k}} d^{2}\beta_{-\mathbf{k}}}{\pi^{2}} \, \Re\left[\psi^{*}_{r, \alpha} \left(q_{\mathbf{k_{1}}}, q_{-\mathbf{k_{2}}}, t_{1} \right) \, \psi_{r, \alpha} \left(\widetilde{q}_{\mathbf{k_{1}}}, \widetilde{q}_{-\mathbf{k_{2}}}, t_{2} \right) \right. \nonumber \\ 
    && \hspace{4.6 cm} \left. \cross \hspace{0.1 cm} \psi_{r, \beta}\left(-q_{\mathbf{k_{1}}}, q_{-\mathbf{k_{2}}}, t_{1} \right) \, \psi^{*}_{r, \beta}\left(\widetilde{q}_{\mathbf{k_{1}}}, -\widetilde{q}_{-\mathbf{k_{2}}}, t_{2} \right) \, \right] \nonumber \\  
    &=& \Re\left[\int \frac{d^{2}\beta_{\mathbf{k}} d^{2}\beta_{-\mathbf{k}}}{\pi^{2}} \int dq_{\mathbf{k_{1}}} dq_{-\mathbf{k_{2}}} \, \psi^{*}_{r, \alpha} \left(q_{\mathbf{k_{1}}}, q_{-\mathbf{k_{2}}}, t_{1} \right) \psi_{r, \beta}\left(-q_{\mathbf{k_{1}}}, q_{-\mathbf{k_{2}}}, t_{1} \right) \right . \nonumber \\
    && \left . \cross \, \int d\widetilde{q}_{\mathbf{k_{1}}} d\widetilde{q}_{-\mathbf{k_{2}}} \, \psi_{r, \alpha} \left(\widetilde{q}_{\mathbf{k_{1}}}, \widetilde{q}_{-\mathbf{k_{2}}}, t_{2} \right) \psi^{*}_{r, \beta} \left(\widetilde{q}_{\mathbf{k_{1}}}, -\widetilde{q}_{-\mathbf{k_{2}}}, t_{2} \right) \right] \, . 
\end{eqnarray}
Performing all these six integrals as given in the \ref{spin_correlation_3}, one should arrive at the following result:  
\begin{eqnarray}\label{spin_correlation_final}
    E(t_{1}, t_{2}) &=& \Re\left[\frac{\xi_{1} \xi^{*}_{2} \widetilde{N}^{*}_{3} \widetilde{N}_{4}}{\sqrt{\gamma_{1} \gamma_{2} \gamma_{3} \gamma_{4}}} \, \mathcal{J}_{1} \mathcal{J}_{2} \mathcal{J}_{3} \hspace{0.1 cm} \exp\left(\mathbf{\lambda_{k}}^{T} \widetilde{T} \mathbf{\lambda_{k}} + \Delta_{\mathbf{k}} \right) \, \frac{(2\pi)^{2}}{\sqrt{\Det{(-2N - 2K)}}} \, \exp\left\{-\frac{1}{2} Q^{T} \left(2N + 2K \right)^{-1} Q \right\} \right] \, . \nonumber \\ 
\end{eqnarray}
For a detailed derivation of the above expression of unequal-time spin-spin correlation $E(t_{1}, t_{2})$ as given in \ref{spin_correlation_final}, one should go through the \ref{Detailed_derivation} of the Appendix. 

To obtain a compact expression for the unequal-time spin–spin correlation function \ref{spin_correlation_final}, we have introduced several auxiliary quantities, namely $(\xi_{1}, \xi_{2}, \gamma_{i}, \mathcal{J}_{i}, \Delta_{\mathbf{k}})$ which have not yet been explicitly defined. These quantities can be expressed in terms of the fundamental squeezing parameters $(r_{k}, \theta_{k}, \phi_{k})$ and coherent state parameters $(\alpha_{\mathbf{k}}, \alpha_{-\mathbf{k}})$ respectively. We now define each of these quantities explicitly one by one. 

The factors $(\xi_{1}, \xi_{2}, \widetilde{N}_{3}, \widetilde{N}_{4})$ are arising from the normalization factor $N_{j}$ of the wave function $\psi_{r}(q_{\mathbf{k}}, q_{\mathbf{-k}}, t_{j})$ as defined in the \ref{Normalization_sc}. To obtain their explicit forms, we decompose the normalization factor $N_{j}$ in the following manner: 
\begin{eqnarray}\label{Decomposition_normalization}
    N_{j} &=& \xi_{j} \, \exp\left(N_{jj} \right) \hspace{0.3 cm} = \hspace{0.3 cm} \xi_{j} \, \exp\left(\overline{N}_{jj} + \widetilde{N}_{jj} \right) \, , 
\end{eqnarray}
where $\xi_{j}, \overline{N}_{jj}$ and $\widetilde{N}_{jj}$ are defined as follows, 
\begin{eqnarray}
    \xi_{j} &=& \frac{\sech{r_{j, k}}}{\sqrt{2\pi \left(1-e^{-4i\phi_{j, k}}\tanh^{2}{r_{j, k}} \right)}} \, , \\ 
    \overline{N}_{jj} &=& \exp\left[A_{j} \left(C_{j}^{2} + D_{j}^{2} \right) - \frac{1}{2} \left(g_{j, 1}x^{2}_{j, 0} + g_{j, 2}y^{2}_{j, 0} \right) \right] \, , \\ 
    \widetilde{N}_{jj} &=& \exp\left\{-4i\Im\big(A_{j} \big) \left[\Re\left(\frac{\zeta_{j, \mathbf{k}}}{\sqrt{2}} \right)^{2} + \Re\left(\frac{\Upsilon_{j, \mathbf{k}}}{\sqrt{2}} \right)^{2} \right] + 4i\Im\big(B_{j} \big) \,  \Re\left(\frac{\zeta_{j, \mathbf{k}}}{\sqrt{2}} \right) \Re\left(\frac{\Upsilon_{j, \mathbf{k}}}{\sqrt{2}} \right) \right\} \nonumber \\ 
    && \cross \, \exp\left\{ - 2i \, \left[\Re\left(\frac{\zeta_{j, \mathbf{k}}}{\sqrt{2}} \right)\Im\left(\frac{\zeta_{j, \mathbf{k}}}{\sqrt{2}} \right) + \Re\left(\frac{\Upsilon_{j, \mathbf{k}}}{\sqrt{2}} \right) \Im\left(\frac{\Upsilon_{j, \mathbf{k}}}{\sqrt{2}} \right) \right] \right\} \, . \label{modified_normalization}
\end{eqnarray}
In \ref{spin_correlation_3}, one can easily find the appearance of four different wave functions of squeezed coherent state, which are given below: 
\begin{eqnarray}
    \psi^{*}_{r, \alpha} \left(q_{\mathbf{k_{1}}}, q_{-\mathbf{k_{2}}}, t_{1} \right) \hspace{0.3 cm} , \hspace{0.3 cm} \psi_{r, \beta}\left(-q_{\mathbf{k_{1}}}, q_{-\mathbf{k_{2}}}, t_{1} \right) \hspace{0.3 cm} , \hspace{0.3 cm} \psi_{r, \alpha} \left(\widetilde{q}_{\mathbf{k_{1}}}, \widetilde{q}_{-\mathbf{k_{2}}}, t_{2} \right) \hspace{0.3 cm} \text{and} \hspace{0.3 cm} \psi^{*}_{r, \beta} \left(\widetilde{q}_{\mathbf{k_{1}}}, -\widetilde{q}_{-\mathbf{k_{2}}}, t_{2} \right) \, . 
\end{eqnarray}
Due to the existence of these four different wave functions in \ref{spin_correlation_3}, the normalization factor $N_{j}$ and wave function coefficients $(A_{j}, B_{j}, C_{j}, D_{j})$ are usually expressed with a subscript $- \, j$ as presented in the \ref{Decomposition_normalization} - \ref{modified_normalization}. Consequently, the subscript $j$ can take four possible values $j = (1, 2, 3, 4)$ according to the convention specified later in the \ref{Table: 3}. 

The quantities $(A_{j}, B_{j}, C_{j}, D_{j}, N_{j})$ represent the coefficients of the wave function as defined in the \ref{coff_1} - \ref{Normalization_sc}.  The parameters $(g_{j, 1} \, , \, g_{j, 2} \, , \, x_{j, 0} \, , \, y_{j, 0})$ which appear in the definition of $\overline{N}_{jj}$ are defined as follows:  
\begin{eqnarray}
    g_{j, 1} &=& \Re\left(\frac{2A_{j} - B_{j}}{2} \right) \hspace{0.9 cm} , \hspace{0.6 cm} g_{j, 2} \hspace{0.2 cm} = \hspace{0.2 cm} \Re\left(\frac{2A_{j} + B_{j}}{2} \right) \, , \label{def_g_1} \\ 
    x_{j, 0} &=& \frac{\Re\left(A_{j}C_{j} + A_{j}D_{j} \right)}{g_{j, 1}} \hspace{0.6 cm} , \hspace{0.6 cm} y_{j, 0} \hspace{0.2 cm} = \hspace{0.2 cm} \frac{\Re\left(A_{j}C_{j} - A_{j}D_{j} \right)}{g_{j, 2}} \, . \label{def_x_0}
\end{eqnarray}
It is worth noting that the auxiliary quantities $\widetilde{N}_{3}$ and $\widetilde{N}_{4}$ appearing in the \ref{spin_correlation_final}, are defined in the following manner, 
\begin{eqnarray}
    \widetilde{N}_{3} &=& \xi_{3} \, \exp\left(\overline{N}_{33} \right) \hspace{0.75 cm} , \hspace{0.75 cm} \widetilde{N}_{4} \hspace{0.1 cm} = \hspace{0.1 cm} \xi_{4} \, \exp\left(\overline{N}_{44} \right) \, . 
\end{eqnarray}

The second set of factors $(\mathcal{J}_{1}, \mathcal{J}_{2}, \mathcal{J}_{3})$ represent the Jacobians associated with three different coordinate transformations, performed during the computation of the complex $(\beta_{\mathbf{k}}, \beta_{\mathbf{-k}}) \, - $ integral which appear in the spin-spin correlation of \ref{spin_correlation_3}. Their explicit definitions are given below, 
\begin{eqnarray}
    \mathcal{J}_{1} &=& \left|\frac{\partial \left(\beta_{\mathbf{k, R}}, \beta_{\mathbf{k, I}}, \beta_{\mathbf{-k, R}}, \beta_{\mathbf{-k, I}} \right)}{\partial \left(\beta_{\mathbf{k}}, \beta^{*}_{\mathbf{k}}, \beta_{\mathbf{-k}}, \beta^{*}_{\mathbf{-k}} \right)} \right| \hspace{0.2 cm} = \hspace{0.2 cm} \frac{1}{(2i)^{2}} \, , \\ 
    \mathcal{J}_{2} &=& \left|\frac{\partial \left(\beta_{\mathbf{k}}, \beta^{*}_{\mathbf{k}}, \beta_{\mathbf{-k}}, \beta^{*}_{\mathbf{-k}} \right)}{\partial \left(\delta_{\mathbf{k}}, \delta^{*}_{\mathbf{k}}, \Psi_{\mathbf{k}}, \Psi^{*}_{\mathbf{k}} \right)} \right| \hspace{1.25 cm} = \hspace{0.2 cm} \frac{1}{4} \hspace{0.16 cm} , \\ 
    \mathcal{J}_{3} &=& \left|\frac{\partial \left(\delta_{\mathbf{k}}, \delta^{*}_{\mathbf{k}}, \Psi_{\mathbf{k}}, \Psi^{*}_{\mathbf{k}} \right)}{\partial \left(\delta_{\mathbf{k, R}}, \delta_{\mathbf{k, I}}, \Psi_{\mathbf{k, R}}, \Psi_{\mathbf{k, I}} \right)} \right| \hspace{0.6 cm} = \hspace{0.2 cm} (2i)^{2} \, . 
\end{eqnarray}
The explicit definition of the coordinate transformations $(\beta_{\mathbf{k}}, \beta_{\mathbf{-k}}) \to (\delta_{\mathbf{k}}, \psi_{\mathbf{k}})$ are provided later in the \ref{coordinate_6}, \ref{coordinate_3} - \ref{coordinate_5} of the \ref{Detailed_derivation} of the Appendix. The information about the coordinate transformations and their corresponding Jacobians, are provided in the \ref{Table: 5}.

\begin{table}[h]
    \centering
    
\begin{tabular}{| c | c | c |}
    \hline 
    \rule{0 pt}{11 pt} Jacobians & Old coordinate system & New coordinate system \\ 
    \hline
    \rule{0 pt}{11 pt} $\mathcal{J}_{1}$ & $\beta_{\mathbf{k, R}}, \, \beta_{\mathbf{k, I}}, \, \beta_{\mathbf{-k, R}}, \, \beta_{\mathbf{-k, I}}$ & $\beta_{\mathbf{k}}, \, \beta^{*}_{\mathbf{k}}, \, \beta_{\mathbf{-k}}, \, \beta^{*}_{\mathbf{-k}}$ \\ 
    \hline 
    \rule{0 pt}{11 pt} $\mathcal{J}_{2}$ & $\beta_{\mathbf{k}}, \, \beta^{*}_{\mathbf{k}}, \, \beta_{\mathbf{-k}}, \, \beta^{*}_{\mathbf{-k}}$ & $\delta_{\mathbf{k}}, \, \delta^{*}_{\mathbf{k}}, \, \Psi_{\mathbf{k}}, \, \Psi^{*}_{\mathbf{k}}$ \\ 
    \hline 
    \rule{0 pt}{11 pt} $\mathcal{J}_{3}$ & $\delta_{\mathbf{k}}, \, \delta^{*}_{\mathbf{k}}, \, \Psi_{\mathbf{k}}, \, \Psi^{*}_{\mathbf{k}}$ & $\delta_{\mathbf{k, R}}, \, \delta_{\mathbf{k, I}}, \, \Psi_{\mathbf{k, R}}, \, \Psi_{\mathbf{k, I}}$ \\ 
    \hline 
\end{tabular}

\caption{Jacobians and the associated coordinate transformations, performed during the computation of complex $(\beta_{\mathbf{k}}, \beta_{\mathbf{-k}}) \, -$ integral.}
\label{Table: 5}
\end{table}

The third factor $\Delta_{\mathbf{k}}$ is defined as follows, 
\begin{equation}
    \Delta_{\mathbf{k}} \hspace{0.3 cm} = \hspace{0.3 cm} \exp\left(\frac{\gamma_{1} \, \Sigma^{2}_{4} + \gamma_{2} \, \Sigma^{2}_{3} - \Sigma_{5}}{4\gamma_{1}\gamma_{2}} + \frac{\gamma_{3} \, \Sigma^{2}_{7} + \gamma_{4} \, \Sigma^{2}_{6} - \Sigma_{8}}{4\gamma_{3}\gamma_{4}} \right) \, . 
\end{equation}
The unknown quantities $\Sigma_{i}$ utilized in the definition of $\Delta_{\mathbf{k}}$, are defined below
\begin{eqnarray}
    \Sigma_{3} &=& A^{*}_{3} \left(C^{*}_{3} + D^{*}_{3} \right) \hspace{1.8 cm} , \hspace{0.6 cm} \Sigma_{6} \hspace{0.2 cm} = \hspace{0.2 cm} A_{4} \left(C_{4} + D_{4} \right) \, , \\ 
    \Sigma_{4} &=& A^{*}_{3} \left(C^{*}_{3} - D^{*}_{3} \right) \hspace{1.8 cm} , \hspace{0.6 cm} \Sigma_{7} \hspace{0.2 cm} = \hspace{0.2 cm} A_{4} \left(C_{4} - D_{4} \right) \, , \\ 
    \Sigma_{5} &=& 4\gamma_{1}\gamma_{2} A^{*}_{3} \left(C^{2*}_{3} + D^{2*}_{3} \right) \hspace{0.6 cm} , \hspace{0.6 cm} \Sigma_{8} \hspace{0.2 cm} = \hspace{0.2 cm} 4\gamma_{3}\gamma_{4} A_{4} \left(C^{2}_{4} + D^{2}_{4} \right) \, . 
\end{eqnarray}
The other unknowns $\gamma_{i}$ utilized in the definition of $\Delta_{\mathbf{k}}$, are defined as follows
\begin{eqnarray}
    \gamma_{1} &=& \frac{2A_{0} + B_{0}}{2} \hspace{0.9 cm} = \hspace{0.7 cm} \frac{2 (A_{1} + A^{*}_{3}) + (B_{1} - B^{*}_{3})}{4} \, , \\ 
    \gamma_{2} &=& \frac{2A_{0} - B_{0}}{2} \hspace{0.9 cm} = \hspace{0.7 cm} \frac{2 (A_{1} + A^{*}_{3}) - (B_{1} - B^{*}_{3})}{4} \, , \\ 
    \gamma_{3} &=& \frac{2A_{00} - B_{00}}{2} \hspace{0.7 cm} = \hspace{0.7 cm} \frac{2(A_{4} + A^{*}_{2}) - (B_{4} - B^{*}_{2})}{4} \, , \\ 
    \gamma_{4} &=& \frac{2A_{00} + B_{00}}{2} \hspace{0.7 cm} = \hspace{0.7 cm} \frac{2(A_{4} + A^{*}_{2}) + (B_{4} - B^{*}_{2})}{4} \, . 
\end{eqnarray}

Furthermore, the quantities $(N, K ,\widetilde{T}) \, - $ represent three different $(4 \cross 4)$ matrices arising from the gaussian integrals over the complex coherent state parameters $(\beta_{\mathbf{k}}, \beta_{\mathbf{-k}})$ as given in the \ref{spin_correlation_3}. The remaining unknown $Q \, - $ represents a $(4 \cross 1)$ matrix arising from the same gaussian integral. The explicit expressions of all these matrix elements, written in terms of the fundamental squeezing parameters $(r_{k}, \theta_{k}, \phi_{k})$ and coherent state parameters $(\alpha_{\mathbf{k}}, \alpha_{\mathbf{-k}})$, are provided in the \ref{Matrix_elements} of the Appendix.     

The only remaining unknown $\lambda_{\mathbf{k}} \, - $ again represents a $(4 \cross 1)$ matrix, constructed from the coherent state parameters $(\alpha_{\mathbf{k}}, \alpha_{\mathbf{-k}})$ as follows, 
\begin{eqnarray}
    \mathbf{\lambda_{k}} &=& 
    \begin{bmatrix}
        \hspace{0.1 cm} \Re \, (\alpha_{\mathbf{k}} + \alpha_{\mathbf{-k}}) \hspace{0.1 cm} \\ 
        \hspace{0.1 cm} \Im \, (\alpha_{\mathbf{k}} + \alpha_{\mathbf{-k}}) \hspace{0.1 cm} \\ 
        \hspace{0.1 cm} \Re \, (\alpha_{\mathbf{k}} - \alpha_{\mathbf{-k}}) \hspace{0.1 cm} \\ 
        \hspace{0.1 cm} \Im \, (\alpha_{\mathbf{k}} - \alpha_{\mathbf{-k}}) \hspace{0.1 cm} 
    \end{bmatrix} \, . 
\end{eqnarray}
Once the expression for the first unequal-time spin-spin correlation $E(t_{1}, t_{2})$ has been obtained, our next task is to determine the explicit expressions for the remaining three spin-spin correlations $E(t_{1},t'_{2}), E(t'_{1}, t_{2})$ and $E(t'_{1}, t'_{2})$. To derive the explicit expression for the second spin-spin correlation $E(t_{1}, t'_{2})$ from that of the first one $E(t_{1}, t_{2})$, one has to replace all the squeezing parameters $[r_{k}(t_{2}) \, , \, \theta_{k}(t_{2}) \, , \, \phi_{k}(t_{2})]$ evaluated at a time $t_{2}$, with a second set of squeezing parameters $[r_{k}(t'_{2}) \, , \, \theta_{k}(t'_{2}) \, , \, \phi_{k}(t'_{2})]$ evaluated at time $t'_{2} \, - $ in the expression of $E(t_{1}, t_{2})$. Following the same procedure, one can also determine the expressions for the remaining two correlators $ - \, E(t'_{1}, t_{2})$ and $E(t'_{1}, t'_{2})$ respectively. 

It is important to note that the analytical expressions for all the four unequal-time spin–spin correlation functions $E(t_{i}, t_{j})$, which enter into the expectation value of the bipartite temporal Bell operator in \ref{Temporal_Bell}, retain the same functional form as given in \ref{spin_correlation_final}. The only difference among these correlators arises solely from the time arguments at which the squeezing parameters $(r_{k}, \theta_{k}, \phi_{k})$ are being evaluated. Explicit dependency of all the correlation functions $E(t_{i}, t_{j})$ on different time arguments, is summarized in the \ref{Table: 4}.  

Substituting the expressions of all these four spin-spin correlators in the quantum expectation value of bipartite temporal Bell operator $\left\langle \alpha_{\mathbf{k}}, \alpha_{\mathbf{-k}} \right| \hat{\mathcal{B}}_{T} \left| \alpha_{\mathbf{k}}, \alpha_{\mathbf{-k}} \right\rangle$ as given in \ref{Temporal_Bell}, yields the final result in terms of squeezing parameters $(r_{k}, \theta_{k}, \phi_{k})$ and coherent state parameters $(\alpha_{\mathbf{k}}, \alpha_{\mathbf{-k}})$, evaluated at four different times $t_{1}, t_{2}, t'_{1}$ and $t'_{2}$ respectively.

\begin{table}[h]
    \centering
    
\begin{tabular}{| c | c | c | c |}
    \hline 
    \rule{0 pt}{11 pt} Correlation function & \makecell{Parameters dependent \\ on 1st time argument} & \makecell{Parameters dependent \\ on 2nd time argument} & Coherent state parameters \\ 
    \hline  
    \rule{0 pt}{11 pt} $E(t_{1}, t_{2})$ & $r_{k}(t_{1}), \, \theta_{k}(t_{1}), \, \phi_{k}(t_{1})$ & $r_{k}(t_{2}), \, \theta_{k}(t_{2}), \, \phi_{k}(t_{2})$ & $\alpha_{\mathbf{k}}, \alpha_{\mathbf{-k}}$ \\ 
    \hline 
    \rule{0 pt}{11 pt} $E(t_{1}, t'_{2})$ & $r_{k}(t_{1}), \, \theta_{k}(t_{1}), \, \phi_{k}(t_{1})$ & $r_{k}(t'_{2}), \, \theta_{k}(t'_{2}), \, \phi_{k}(t'_{2})$ & $\alpha_{\mathbf{k}}, \alpha_{\mathbf{-k}}$ \\ 
    \hline 
    \rule{0 pt}{11 pt} $E(t'_{1}, t_{2})$ & $r_{k}(t'_{1}), \, \theta_{k}(t'_{1}), \, \phi_{k}(t'_{1})$ & $r_{k}(t_{2}), \, \theta_{k}(t_{2}), \, \phi_{k}(t_{2})$ & $\alpha_{\mathbf{k}}, \alpha_{\mathbf{-k}}$ \\ 
    \hline  
    \rule{0 pt}{11 pt} $E(t'_{1}, t'_{2})$ & $r_{k}(t'_{1}), \, \theta_{k}(t'_{1}), \, \phi_{k}(t'_{1})$ & $r_{k}(t'_{2}), \, \theta_{k}(t'_{2}), \, \phi_{k}(t'_{2})$ & $\alpha_{\mathbf{k}}, \alpha_{\mathbf{-k}}$ \\ 
    \hline 
\end{tabular}

    \caption{Dependency of unequal-time spin-spin correlators $E(t_{i}, t_{j})$ on the squeezing parameters, evaluated at the first time argument $t_{i}$ and second time argument $t_{j}$.}
    \label{Table: 4}
\end{table}

\subsection{De sitter inflation}\label{subsec:3.3}

To know the exact dynamics of the squeezing parameters $(r_{k}, \theta_{k}, \phi_{k})$, one needs to impose a particular model of inflation in the above mentioned calculation. The de Sitter expansion is one of the well known models of inflation in which the Hubble parameter of the background universe is approximated to be a constant i.e. $H = \dot{a}/a = \text{const.}$ Assuming de Sitter expansion, the solution to the dynamical equations of motion obeyed by the squeezing parameters, are found to be the following \cite{Martin:2015qta}: 
\begin{eqnarray}\label{De_Sitter}
    r_{k}(\eta) &=& - \sinh^{-1}{\left(\frac{1}{2k\eta} \right)} \,, \label{sque_parameter} \\ 
    \phi_{k}(\eta) &=& \frac{\pi}{4} - \frac{1}{2} \tan^{-1}{\left(\frac{1}{2k\eta} \right)} \,, \label{squeezing} \\ 
    \theta_{k}(\eta) &=& k\eta + \tan^{-1}{\left(\frac{1}{2k\eta} \right)} \,\label{rotation} . 
\end{eqnarray}
Here $\eta$ is the conformal time of the cosmological spacetime. Due to the analytical simplicity of the \ref{sque_parameter}-\ref{rotation}, one can express the rotation angle $\theta_{k}$ and squeezing angle $\phi_{k}$ in terms of the squeezing parameter $r_{k}$. As a consequence, total number of independent variables has been reduced from three to one. 

Our next task is to re-express the expectation value of temporal Bell operator evaluated with respect to two-mode coherent state $\left\langle \alpha_{\mathbf{k}}, \alpha_{\mathbf{-k}} \right| \hat{\mathcal{B}}_{T} \left| \alpha_{\mathbf{k}}, \alpha_{\mathbf{-k}} \right\rangle$, in terms of squeezing parameter $r_{k}(t_{j})$ alone. In the context of temporal Bell inequality, the unequal-time spin-spin correlators $E(t_{i}, t_{j})$ are evaluated at four different pair of times $- \, (t_{1}, t_{2})$, $(t_{1}, t'_{2})$, $(t'_{1}, t_{2})$ and $(t'_{1}, t'_{2})$ respectively. That's why the expectation value of temporal Bell operator depends on four distinct squeezing parameters $r_{k}(t_{1}), \, r_{k}(t_{2}), \, r_{k}(t'_{1})$ and $r_{k}(t'_{2})$, along with the coherent state parameters $\alpha_{\mathbf{k}}$ and $\alpha_{\mathbf{-k}}$ respectively.  

\begin{figure}[htbp]
    \centering
    \includegraphics[width=0.49\linewidth]{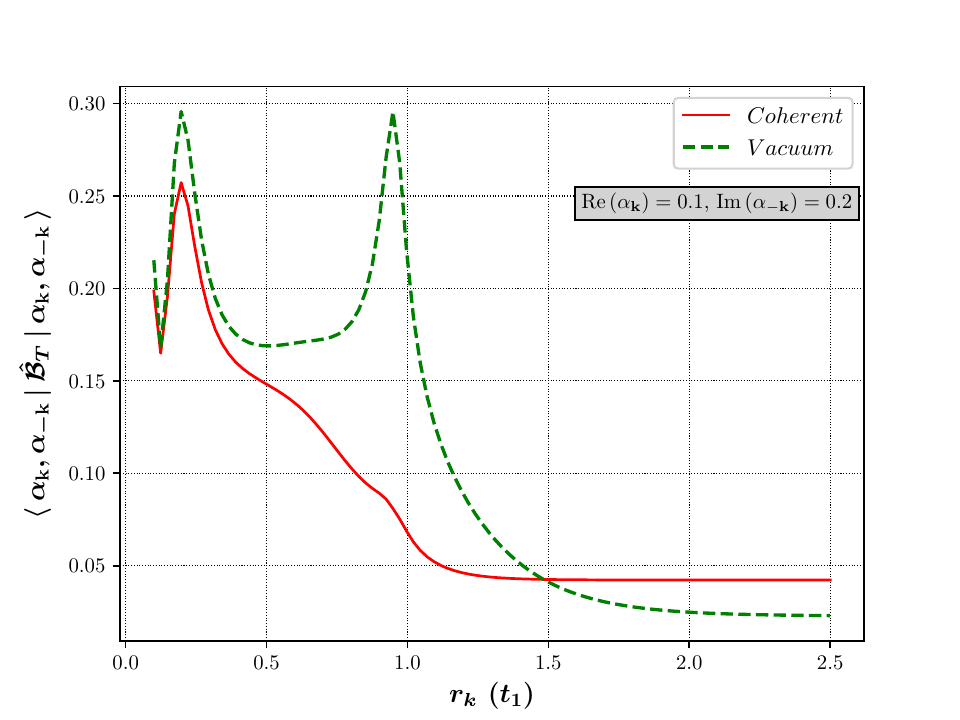}
    \includegraphics[width=0.49\linewidth]{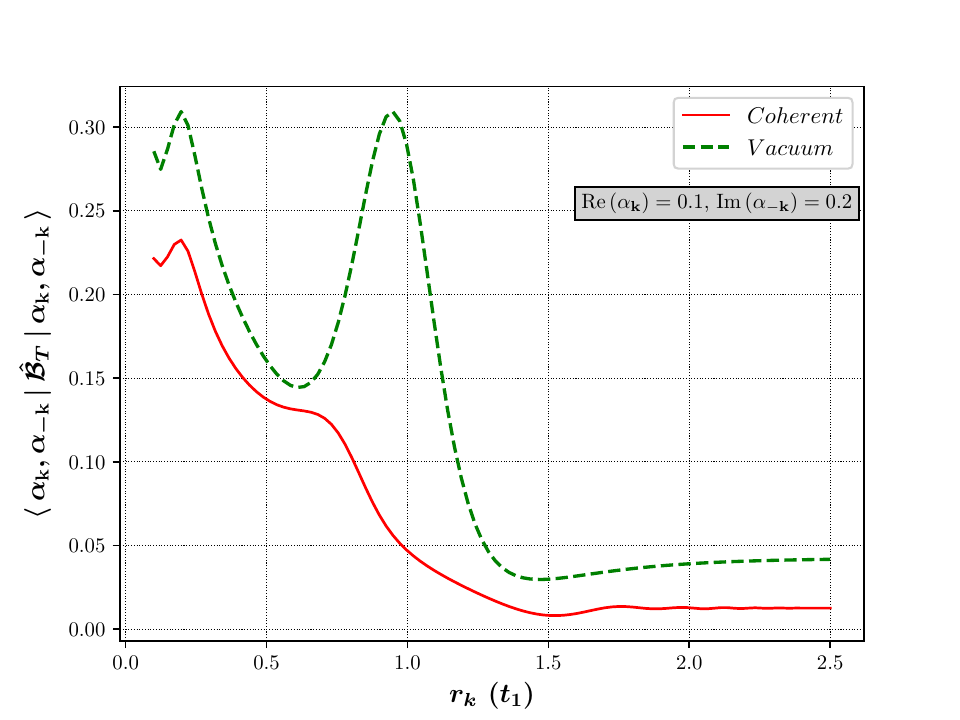} \\
    \includegraphics[width=0.49\textwidth]{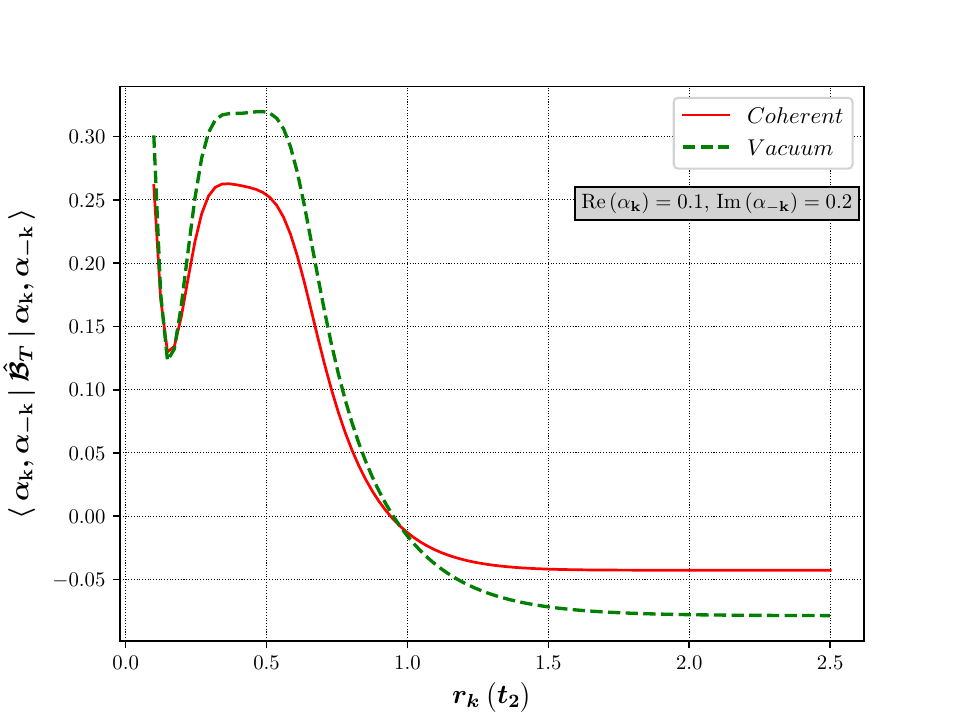}
    \includegraphics[width=0.49\textwidth]{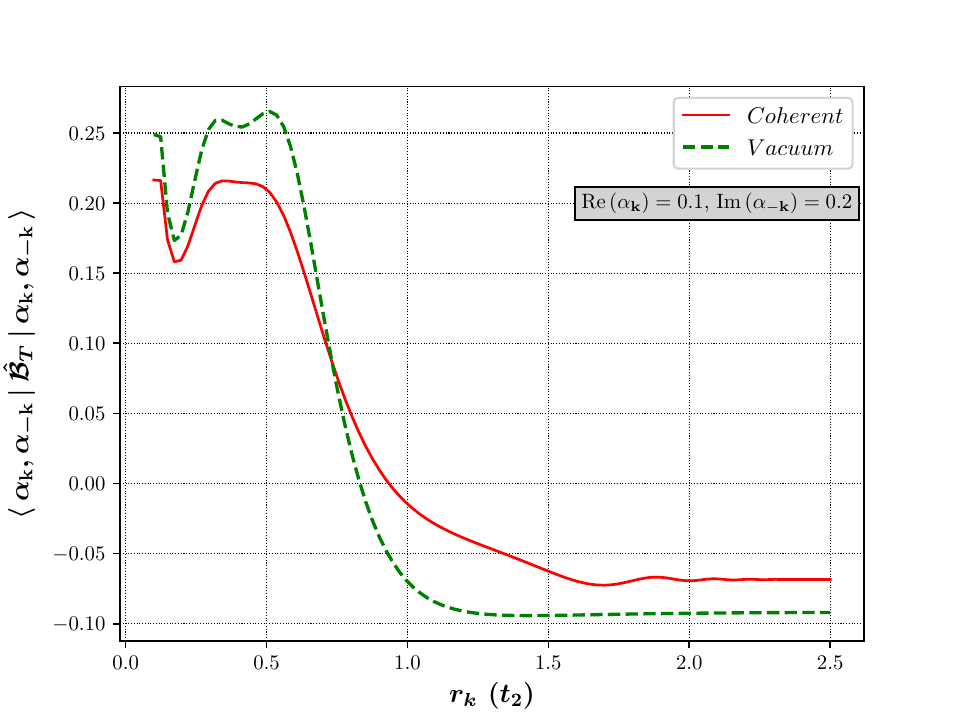} \\ 
    \includegraphics[width=0.49\textwidth]{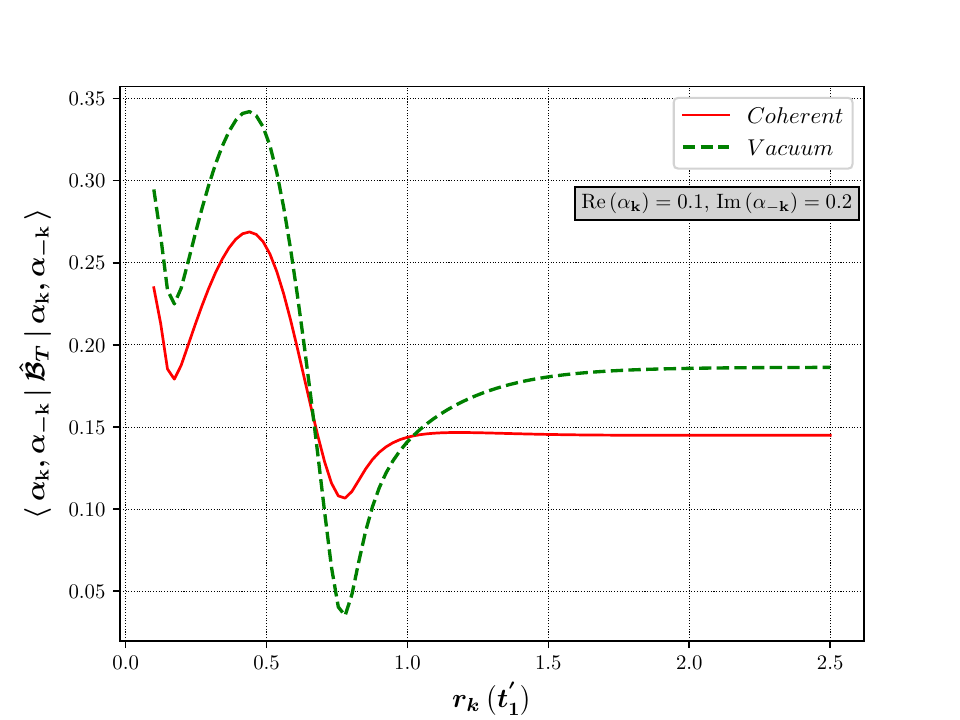}
    \includegraphics[width=0.49\textwidth]{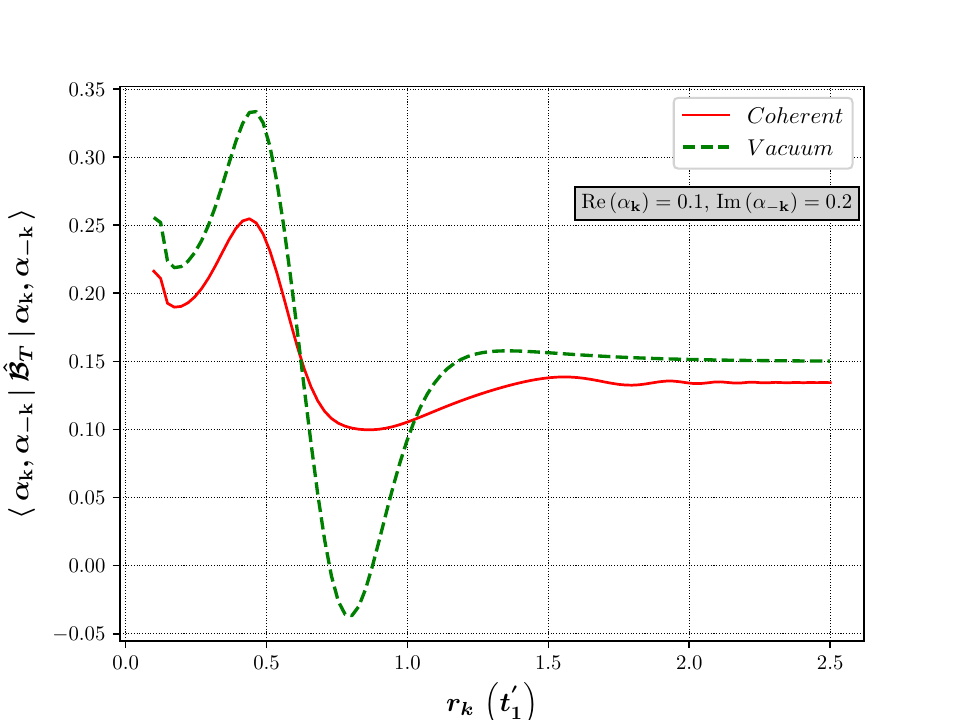} \\ 
    \includegraphics[width=0.49\textwidth]{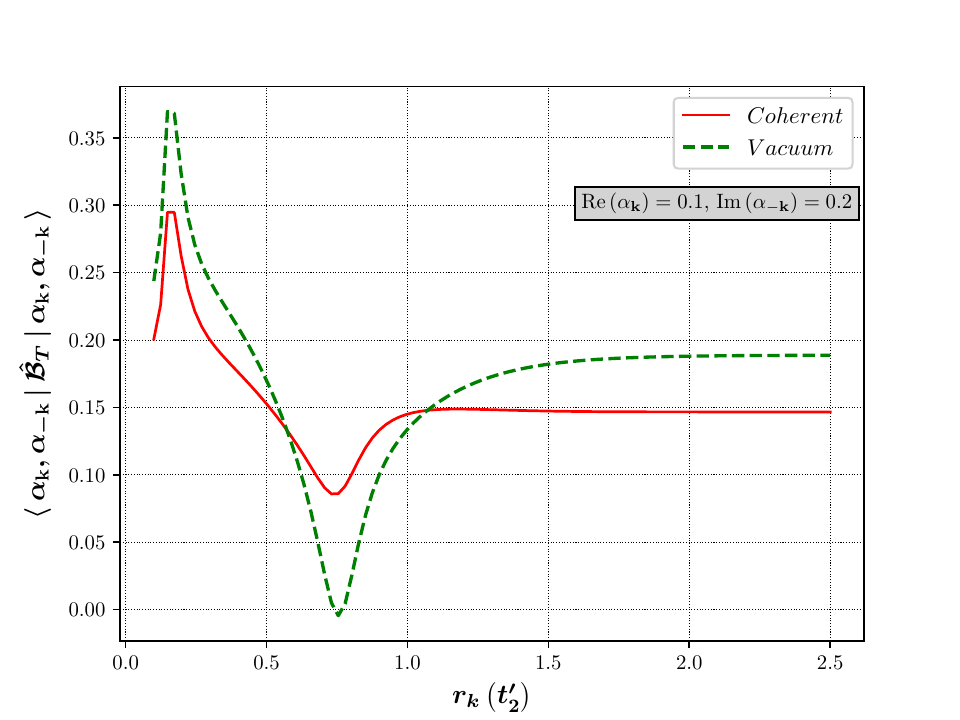} 
    \includegraphics[width=0.49\textwidth]{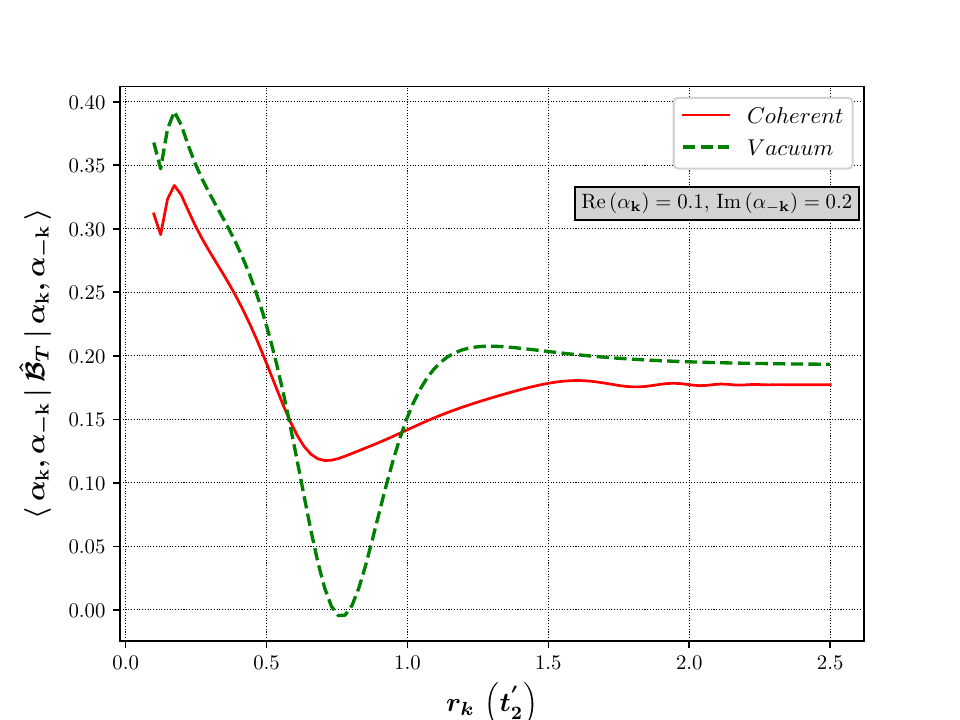}
    \captionsetup{width=0.96\linewidth}
    \caption{Bipartite temporal Bell operator as a function of squeezing parameters $r_{k}(t_{j})$ evaluated at different times $t_{j}$ while the coherent state parameters $\alpha_{\mathbf{k}}$ and $\alpha_{\mathbf{-k}}$ kept fixed at, $\alpha_{\mathbf{k}} = 0.1 + i \, 0.1$ and $\alpha_{\mathbf{-k}} = 0.2 + i \, 0.2$. }
    \label{fig:1}
\end{figure}

In the first row of \ref{fig:1}, we plot the expectation value of the temporal Bell operator $\left\langle \alpha_{\mathbf{k}}, \alpha_{\mathbf{-k}} \right| \hat{B} \left| \alpha_{\mathbf{k}}, \alpha_{\mathbf{-k}} \right\rangle$ as a function of the squeezing parameter $r_{k}(t_{1})$, while the other squeezing parameters are fixed at $r_{k}(t_{2}) = 0.20$, $r_{k}(t'_{1}) = 0.60$ and $r_{k}(t'_{2}) = 0.95$. The coherent state parameters are chosen to be $\alpha_{\mathbf{k}} = (0.1 + i \, 0.1)$ and $\alpha_{-\mathbf{k}} = (0.2 + i \, 0.2)$. 

To explicitly examine the role of the time-dependent imaginary phase factor $\widetilde{N}_{jj}$ appearing in the normalization constant of the wave function $\psi_{r,\alpha}(q_{1}, q_{2}, t_{1})$ as given in \ref{modified_normalization}, we present two different plots in the first row of \ref{fig:1}. In the left panel, the imaginary phase factor $\widetilde{N}_{jj}$ is completely omitted from the normalization constant of the wave function (by hand). Whereas in the right panel, this phase factor is retained in the wave function to isolate and analyse its individual contribution to the expectation value of the temporal Bell operator. A comparison between these two plots shows that the imaginary phase factor $\widetilde{N}_{jj}$ does not affect the overall behaviour of the temporal Bell operator, except for a slight shift in its magnitude. This observation suggests that the influence of the imaginary phase is relatively mild at the level of temporal Bell operator. Nevertheless, the unequal-time spin-spin correlation in \ref{spin_correlation_final} contains two additional terms that arise solely due to the imaginary phase factor of the wave function. These terms are presented below: 
\begin{eqnarray}
    \exp\left(\lambda^{T}_{\mathbf{k}} \, \widetilde{T} \, \lambda_{\mathbf{k}} \right) \hspace{0.6 cm} \text{and} \hspace{0.6 cm} \frac{1}{\sqrt{\Det{\left(-2N - 2K \right)}}} \exp\left[-\frac{1}{2} Q^{T} (2K)^{-1} Q \right] \, .     
\end{eqnarray}

In contrast to the temporal scenario, the imaginary phase factor $\widetilde{N}_{jj}$ does not have any contribution to the expectation value of spatial Bell-CHSH operator \cite{Mondal:2024glo}. In the spatial scenario, the spin–spin correlators are evaluated at the same time, with the measurements performed simultaneously at different spatial locations. Consequently, the imaginary phase factors $\widetilde{N}_{jj}$, associated with different spin components of the correlation function, exactly cancel and render the spatial correlations independent of the imaginary phase.  

Furthermore, the red curve in \ref{fig:1} represents the quantum expectation value of the temporal Bell operator evaluated with respect to initially chosen coherent state, whereas the green curve represents the same for Bunch-Davis vacuum. This simultaneous representation provides a possible way to distinguish the coherent state from the vacuum one through the investigation of the temporal Bell inequality. Although the difference between the results of coherent state and Bunch-Davis vacuum becomes very small for large values of the squeezing parameter $r_{k}(t_{1})$, the two curves do not exactly coincide. This suggests a possible avenue for distinguishing the initial state of primordial perturbations, generated during inflation. 

\begin{figure}[ht]
    \centering
    \includegraphics[width=0.49\linewidth]{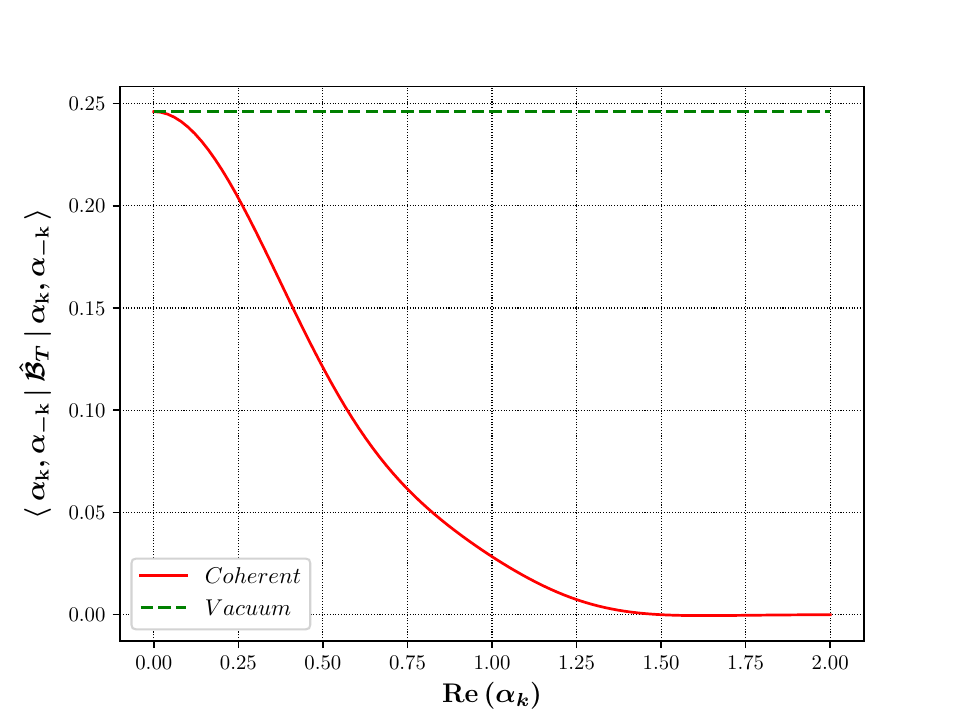} 
    \includegraphics[width=0.49\textwidth]{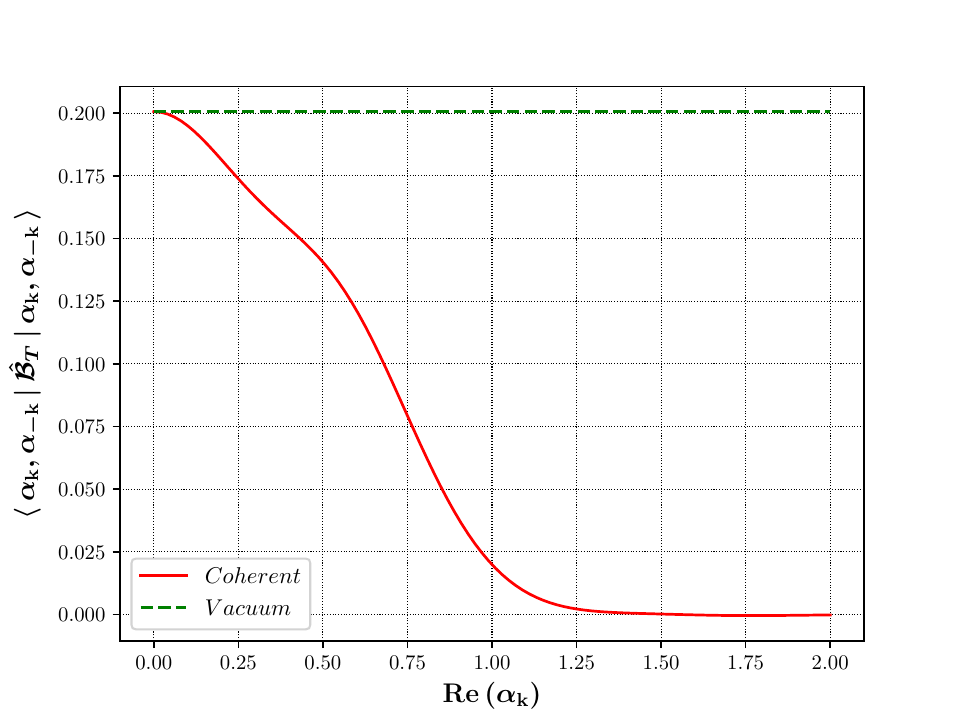} \\ 
    \includegraphics[width=0.49\textwidth]{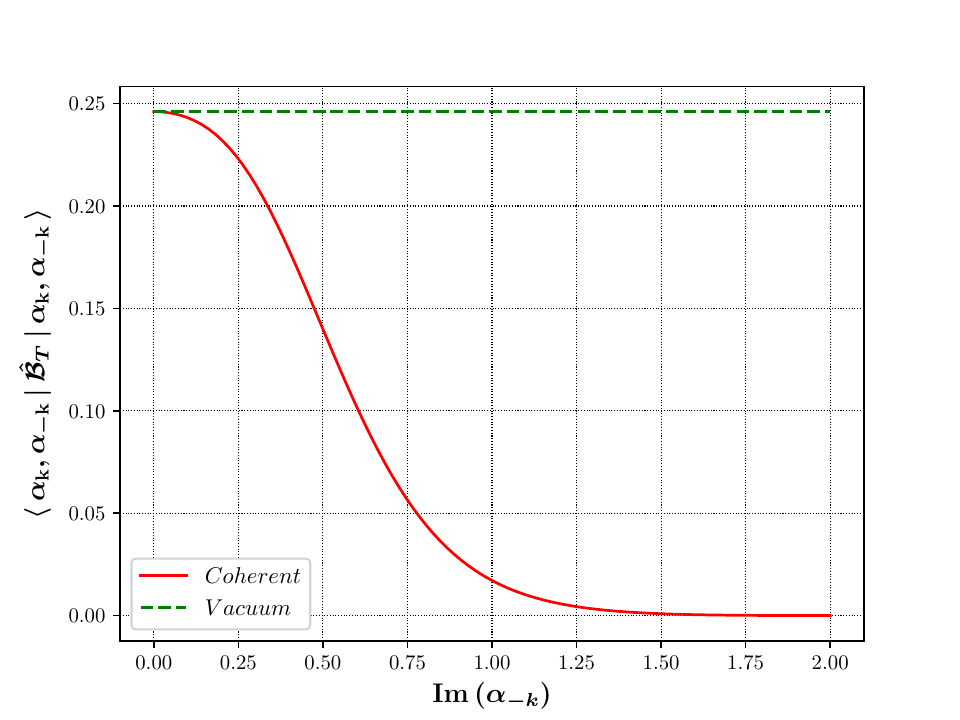} 
    \includegraphics[width=0.49\textwidth]{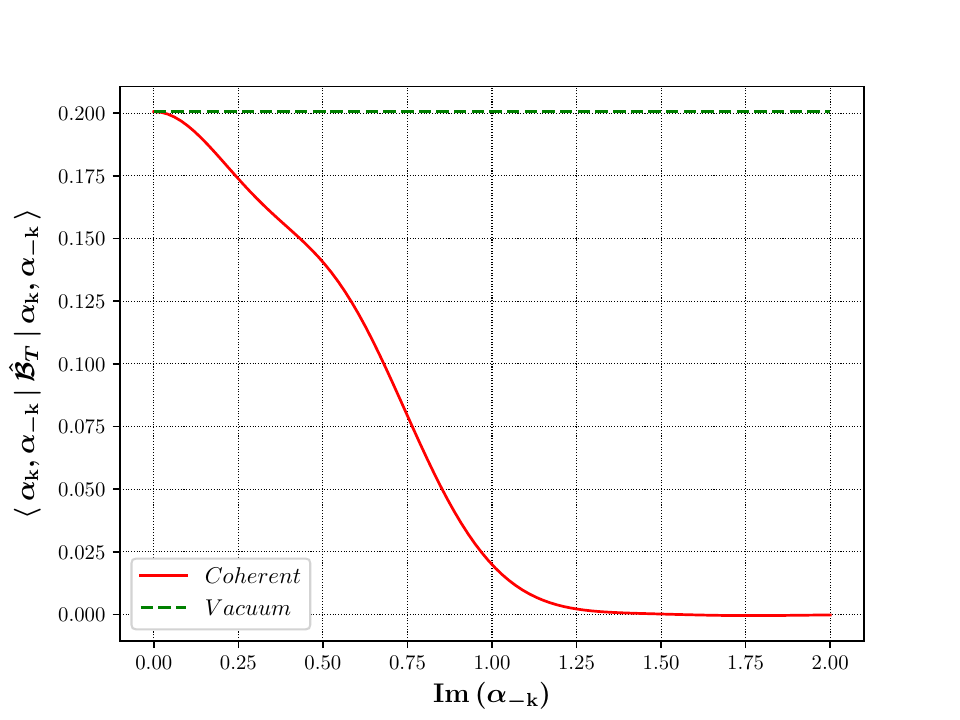} 
    \captionsetup{width=0.96\linewidth} 
    \caption{Bipartite temporal Bell operator as a function of the real part of $\alpha_{\mathbf{k}}$ and imaginary part of $\alpha_{-\mathbf{k}}$ while the squeezing parameters are fixed at, $r_{k}(t_{1}) = 0.26$, $r_{k}(t_{2}) = 0.54$, $r_{k}(t'_{1}) = 0.64$ and $r_{k}(t'_{2}) = 0.83$.}
    \label{fig:2}
\end{figure}

Similarly, in the second, third, and fourth rows of \ref{fig:1}, we plot the expectation value of temporal Bell operator as a function of the remaining squeezing parameters $r_{k}(t_{2})$, $r_{k}(t'_{1})$ and $r_{k}(t'_{2})$ respectively. The respective values at which the squeezing parameters and coherent state parameters kept fixed, are provided accordingly in \ref{Table:1}. The difference between the plots appeared in the left and right panel of \ref{fig:1} and \ref{fig:2}, in terms of the $\zeta_{\mathbf{k}}$ and $\Upsilon_{\mathbf{k}}$ dependent purely imaginary phase of the normalization factor $\widetilde{N}_{jj}$, is provided in the \ref{Table: 2}. 

Similarly, in the first and second row of the \ref{fig:2}, we plot the expectation value of temporal Bell operator as a function of the real part of $\alpha_{\mathbf{k}}$ and imaginary part of $\alpha_{\mathbf{-k}}$ respectively. The values at which the squeezing parameters $r_{k}(t_{1}), r_{k}(t_{2}), r_{k}(t_{1}'), r_{k}(t_{2}')$ and the remaining parts of coherent state parameters $\alpha_{\mathbf{k}}, \alpha_{\mathbf{-k}}$ kept fixed, are provided in the \ref{Table:1}. However, the left panel plots of \ref{fig:2} correspond to those results, obtained by omitting the purely imaginary phase of the normalization factor i.e. $\widetilde{N}_{jj} = 0$. Whereas the right panel plots of \ref{fig:2} correspond to those results, obtained by retaining the purely imaginary phase of the normalization factor i.e. $\widetilde{N}_{jj} \neq 0$.

\begin{table}[h]
    \centering
    
    \begin{tabular}{| c | c | c | c | c | c | c | c |}
        \hline
        \rule{0 pt}{11 pt} Corresponding Figure & Row No. & $r_{k}(t_{1})$ & $r_{k}(t_{2})$ & $r_{k}(t'_{1})$ & $r_{k}(t'_{2})$ & $\alpha_{\mathbf{k}}$ & $\alpha_{\mathbf{-k}}$ \\ 
        \hline 
        
        \multirow{4}{*}{\ref{fig:1}}
        \rule{0 pt}{10 pt} & 1 & $-$ & 0.20 & 0.60 & 0.95 & $(0.1 + i \, 0.1)$ & $(0.2 + i \, 0.2)$ \\ 
        \cline{2-8} 
        \rule{0 pt}{10 pt} & 2 & 0.30 & $-$ & 0.56 & 0.74 & $(0.1 + i \, 0.1)$ & $(0.2 + i \, 0.2)$ \\ 
        \cline{2-8} 
        \rule{0 pt}{10 pt} & 3 & 0.25 & 0.48 & $-$ & 0.76 & $(0.1 + i \, 0.1)$ & $(0.2 + i \, 0.2)$ \\ 
        \cline{2-8} 
        \rule{0 pt}{10 pt} & 4 & 0.16 & 0.42 & 0.75 & $-$ & $(0.1 + i \, 0.1)$ & $(0.2 + i \, 0.2)$ \\ 
        \hline 

        \multirow{2}{*}{\ref{fig:2}}
        \rule{0 pt}{10 pt} & 1 & 0.26 & 0.54 & 0.64 & 0.83 & $(``-" \, + i \, 0.0)$ & $(0.0 + i \, 0.0)$ \\ 
        \cline{2-8} 
        \rule{0 pt}{10 pt} & 2 & 0.26 & 0.54 & 0.64 & 0.83 & $(0.0 + i \, 0.0)$ & $(0.0 \, + i \, ``-")$ \\ 
        \hline 
    \end{tabular}
    
    \caption{Table containing the respective fixed values of the squeezing parameters and coherent state parameters, used during the plotting of bipartite temporal Bell operator $\left\langle \alpha_{\mathbf{k}}, \alpha_{\mathbf{-k}} \right| \hat{\mathcal{B}}_{T} \left| \alpha_{\mathbf{k}}, \alpha_{\mathbf{-k}} \right\rangle$ as shown in \ref{fig:1} and \ref{fig:2}.}
    \label{Table:1}
\end{table}

\begin{table}[]
    \centering
    
    \begin{tabular}{| c | c | c |}
        \hline
        \rule{0 pt}{12 pt} Column side / Column No. & Imaginary phase of Normalization & Presence of that imaginary phase \\  
        \hline 
        \rule{0 pt}{12 pt} Left Panel / 1st column & $\widetilde{N}_{11} \, , \, \widetilde{N}_{22} \, , \, \widetilde{N}_{33} \, , \, \widetilde{N}_{44}$ & Absent $ \hspace{0.2 cm} \Rightarrow \hspace{0.2 cm} \widetilde{N}_{jj} = 0$ \\ 
        \hline 
        \rule{0 pt}{12 pt} Right Panel / 2nd column & $\widetilde{N}_{11} \, , \, \widetilde{N}_{22} \, , \, \widetilde{N}_{33} \, , \, \widetilde{N}_{44}$ & Present $ \hspace{0.2 cm} \Rightarrow \hspace{0.2 cm} \widetilde{N}_{jj} \neq 0$ \\ 
        \hline 
    \end{tabular}

    \caption{Table containing information about the imaginary phase of the normalization factor $\widetilde{N}_{jj}$, used during the plotting of \ref{fig:1} and \ref{fig:2}.}
    \label{Table: 2}
\end{table}

\section{Conclusion}\label{sec:4}

In recent years, significant efforts have been devoted to investigating the quantum nature of primordial perturbations imprinted in the temperature anisotropies of the cosmic microwave background (CMB). In particular, several studies have explored the possible existence of quantum correlations in cosmological perturbations by examining various quantum information–theoretic measures, such as entanglement entropy, quantum discord, and violations of Bell inequalities.

In the literature, the standard assumption is that the initial state of cosmological perturbations is the Bunch–Davies vacuum. However, several alternative choices for the initial state and for Bunch-Davis initial conditions $-$ have been already proposed \cite{Chen:2024ckx,Wood-Saanaoui:2026bma,Yin:2023jlv,Gessey-Jones:2021yky}. In some earlier works \cite{Kundu:2011sg,Ragavendra:2024qpj,Mondal:2024glo}, the two-mode coherent state has been considered as a viable alternative to Bunch–Davies vacuum. Under the dynamical evolution during the inflationary epoch, an initially chosen coherent state evolves into a squeezed coherent state, which differs slightly from that of the squeezed vacuum state. 

In this article, our primary objective is to investigate the role of the bipartite temporal Bell inequality in the context of cosmological perturbations when the Bunch–Davies vacuum is replaced by a two-mode coherent state as the initial condition. Following the GKMR prescription, we construct the relevant components of the pseudo-spin operators $\hat{S}_x$, $\hat{S}_y$ and $\hat{S}_z$, one of which is required to define the temporal Bell operator. We then evaluate the two-point correlation functions of these pseudo-spin operators, measured at different instants of time, with respect to the resulting squeezed coherent state.

After deriving the explicit expression for the temporal Bell operator, we find that the bipartite temporal Bell inequality is not violated for squeezed coherent state. This result indicates that, despite the intrinsically quantum origin of primordial perturbations and the presence of squeezing, the temporal correlations do not exhibit non-classical features strong enough to violate a Bell-type bound. Furthermore, for sufficiently large values of the squeezing parameter $r(t_j)$ — corresponding to the late stages of inflation — the expectation value of the temporal Bell operator approaches nearly zero. This asymptotic behaviour suggests that the temporal correlations become progressively suppressed as inflation proceeds, reinforcing the notion that the large-scale cosmological perturbations acquire an effectively classical character by the end of inflation. 

From the figures, we observe that at the end of inflation the expectation value of the temporal Bell operator saturates to a constant value that differs from the corresponding result obtained for the squeezed vacuum state. This difference persists even though the bipartite temporal Bell inequality is not violated in either case.

Therefore, our analysis demonstrates that the investigation of the bipartite temporal Bell inequality can still provide a way to distinguish between different initial states of primordial perturbations, despite the absence of an actual Bell inequality violation. In other words, the non-violation of a temporal Bell inequality does not preclude the possibility of discriminating between distinct initial quantum states in cosmology. This finding highlights that the quantitative behaviour of the Bell operator itself, rather than merely the presence or absence of its violation, carries physically relevant information about the underlying initial conditions of primordial fluctuations. 

In the wave function of the squeezed coherent state $\psi_{r, \alpha}(q_{1}, q_{2}, t_{j})$, a purely imaginary phase factor $\widetilde{N}_{jj}$ appears in the exponential part of the normalization factor \textcolor{red}{\ref{modified_normalization}}. In the context of equal-time spin-spin correlations, this imaginary phase does not contribute to the expectation value of physical observables, as it cancels out in the relevant expressions. However, in the present work we consider unequal-time spin-spin correlations, where measurements are performed at different instants of time. In this situation, the time-dependent imaginary phase factors do not cancel exactly and consequently contribute to the unequal-time correlators of the pseudo-spin operators. 

As a result, the expectation value of temporal Bell operator acquires an explicit dependence on the imaginary phase of the wave function. Such a phase dependence is entirely absent in the context of the spatial Bell inequality, highlighting a qualitative distinction between temporal and spatial correlations in cosmological settings. 

Although the entire formalism developed in this work is based on the Fourier modes of the Mukhanov–Sasaki variable $\hat{v}_{\mathbf{k}}$, establishing a direct connection with observational data requires reformulating the derivation of bipartite temporal Bell inequality in terms of real-space perturbations $\hat{v}(\mathbf{x})$. Such an analysis would allow the theoretical predictions to be more directly related with observable quantities. In addition, it would be interesting to investigate the effects of decoherence and the quantum-to-classical transition on the possible violation of bipartite temporal Bell inequality in a cosmological setting. Exploring these aspects constitutes a natural extension of the present work and represents promising directions for future research.

\section{Acknowledgements}

AM gratefully acknowledges the valuable discussions with Subir Ghosh and Sumanta Chakraborty during the preparation of this manuscript.  

\vspace{0.4 cm}

\appendix
\noindent\textbf{\Large{Appendix}} 

\section{Detailed derivation of unequal-time spin-spin correlation}\label{Detailed_derivation} 

The primary objective of this section is to present a detailed derivation of the six integrals that appear in the unequal-time spin–spin correlation function \ref{spin_correlation_3}. For convenience, we first separate four of these integrals into two groups and denote them by $I_{3}$ and $I_{4}$ respectively as follows, 
\begin{eqnarray}
    I_{3} &=& \int dq_{\mathbf{k_{1}}} dq_{-\mathbf{k_{2}}} \, \psi^{*}_{r, \alpha} \left(q_{\mathbf{k_{1}}}, q_{-\mathbf{k_{2}}}, t_{1} \right) \psi_{r, \beta}\left(-q_{\mathbf{k_{1}}}, q_{-\mathbf{k_{2}}}, t_{1} \right) \, , \label{Integral_def_q} \\ 
    I_{4} &=& \int d\widetilde{q}_{\mathbf{k_{1}}} d\widetilde{q}_{-\mathbf{k_{2}}} \, \psi_{r, \alpha} \left(\widetilde{q}_{\mathbf{k_{1}}}, \widetilde{q}_{-\mathbf{k_{2}}}, t_{2} \right) \psi^{*}_{r, \beta} \left(\widetilde{q}_{\mathbf{k_{1}}}, -\widetilde{q}_{-\mathbf{k_{2}}}, t_{2} \right) \, \label{Integral_def_tilde_q} . 
\end{eqnarray}
The wave functions of squeezed coherent state $\psi_{r,\alpha}(q_{\mathbf{k}}, q_{\mathbf{-k}}, t_{j})$ appeared in the \ref{Integral_def_q} and \ref{Integral_def_tilde_q}, with four different sets of arguments, are characterized by the distinct coefficients $A_{i}, B_{i}, C_{i}, D_{i}$ and $N_{i}$ respectively. Their explicit expressions are presented below,  
\begin{eqnarray}
    \psi_{r, \beta}\left(-q_{\mathbf{k_{1}}}, q_{-\mathbf{k_{2}}}, t_{1} \right) &=& N_{1} \, \exp[-A_{1}\big(q_{\mathbf{k}}+C_{1} \big)^{2} - A_{1}\big(q_{-\mathbf{k}}-D_{1} \big)^{2} - B_{1} \hspace{0.06cm} q_{\mathbf{k}}q_{-\mathbf{k}} ] \, , \label{wave_func_integral_1} \\ 
    \psi^{*}_{r, \beta} \left(\widetilde{q}_{\mathbf{k_{1}}}, -\widetilde{q}_{-\mathbf{k_{2}}}, t_{2} \right) &=& N^{*}_{2} \, \exp[-A^{*}_{2}\big(\widetilde{q}_{\mathbf{k}}-C^{*}_{2} \big)^{2} - A^{*}_{2}\big(\widetilde{q}_{-\mathbf{k}}+D^{*}_{2} \big)^{2} - B^{*}_{2} \hspace{0.06cm} \widetilde{q}_{\mathbf{k}}\widetilde{q}_{-\mathbf{k}} ] \, , \label{wave_func_integral_2} \\ 
    \psi^{*}_{r, \alpha} \left(q_{\mathbf{k_{1}}}, q_{-\mathbf{k_{2}}}, t_{1} \right) &=& N^{*}_{3} \, \exp[-A^{*}_{3}\big(q_{\mathbf{k}}-C^{*}_{3} \big)^{2} - A^{*}_{3}\big(q_{-\mathbf{k}}-D^{*}_{3} \big)^{2} + B^{*}_{3} \hspace{0.06cm} q_{\mathbf{k}}q_{-\mathbf{k}} ] \, , \label{wave_func_integral_3} \\ 
    \psi_{r, \alpha} \left(\widetilde{q}_{\mathbf{k_{1}}}, \widetilde{q}_{-\mathbf{k_{2}}}, t_{2} \right) &=& N_{4} \, \exp[-A_{4}\big(\widetilde{q}_{\mathbf{k}}-C_{4} \big)^{2} - A_{4}\big(\widetilde{q}_{-\mathbf{k}}-D_{4} \big)^{2} + B_{4} \hspace{0.06cm} \widetilde{q}_{\mathbf{k}}\widetilde{q}_{-\mathbf{k}} ] \, . \label{wave_func_integral_4} 
\end{eqnarray}
Owing to the presence of different time arguments $(t_{1}, t_{2})$ and different coherent-state parameters $(\alpha_{\mathbf{k}}, \alpha_{\mathbf{-k}}, \beta_{\mathbf{k}}, \beta_{\mathbf{-k}})$, four distinct sets of coefficients $(A_{i}, B_{i}, C_{i}, D_{i}, N_{i})$ are introduced for the above-mentioned wave functions. The fundamental dependence of these coefficients on the squeezing parameters and the coherent-state parameters are provided in \ref{coff_1} - \ref{Normalization_sc}. 

However, there is a slight change in the definition of the auxiliary variables $\zeta_{j, \mathbf{k}}$ and $\Upsilon_{j, \mathbf{k}}$ as appeared in the \ref{zeta_def} and \ref{upsilon_def}. First two set of auxiliary quantities $(\zeta_{1, \mathbf{k}}, \Upsilon_{1, \mathbf{k}})$ and $(\zeta_{2, \mathbf{k}}, \Upsilon_{2, \mathbf{k}})$, appeared in the coefficients $(C_{1}, D_{1})$ and $(C_{2}, D_{2})$ of the corresponding wave functions $\psi_{r, \beta}\left(-q_{\mathbf{k_{1}}}, q_{-\mathbf{k_{2}}}, t_{1} \right)$ and $\psi^{*}_{r, \beta} \left(\widetilde{q}_{\mathbf{k_{1}}}, -\widetilde{q}_{-\mathbf{k_{2}}}, t_{2} \right)$, are defined below: 
\begin{eqnarray}
    \zeta_{j, \mathbf{k}} &=& \left[e^{-i\theta_{j}}\cosh{r_{j}} \hspace{0.1cm} \beta_{\mathbf{k}} - \hspace{0.1cm} e^{i(\theta_{j}+2\phi_{j})}\sinh{r_{j}} \hspace{0.1cm} \beta^{*}_{-\mathbf{k}} \right] \, , \label{beta_t1} \\ 
    \Upsilon_{j, \mathbf{k}} &=& \left[e^{-i\theta_{j}}\cosh{r_{j}} \hspace{0.1cm} \beta_{-\mathbf{k}} - \hspace{0.1cm} e^{i(\theta_{j}+2\phi_{j})}\sinh{r_{j}} \hspace{0.1cm} \beta^{*}_{\mathbf{k}} \right] \, . \label{beta_t2}
\end{eqnarray}
In \ref{beta_t1} - \ref{beta_t2}, the subscript $j \, -$ associated with all the squeezing and auxiliary parameters, can take only two possible values $j = (1, 2)$. 

Likewise, the remaining two set of auxiliary quantities $(\zeta_{3, \mathbf{k}}, \Upsilon_{3, \mathbf{k}})$ and $(\zeta_{4, \mathbf{k}}, \Upsilon_{4, \mathbf{k}})$ which appeared in the coefficients $(C_{3}, D_{3})$ and $(C_{4}, D_{4})$ of the corresponding wave functions $\psi^{*}_{r, \alpha} \left(q_{\mathbf{k_{1}}}, q_{-\mathbf{k_{2}}}, t_{1} \right)$ and $\psi_{r, \alpha} \left(\widetilde{q}_{\mathbf{k_{1}}}, \widetilde{q}_{-\mathbf{k_{2}}}, t_{2} \right)$, are defined below: 
\begin{eqnarray}
    \zeta_{j, \mathbf{k}} &=& \left[e^{-i\theta_{j}}\cosh{r_{j}} \hspace{0.1cm} \alpha_{\mathbf{k}} - \hspace{0.1cm} e^{i(\theta_{j}+2\phi_{j})}\sinh{r_{j}} \hspace{0.1cm} \alpha^{*}_{-\mathbf{k}} \right] \, , \label{alpha_t1} \\ 
    \Upsilon_{j, \mathbf{k}} &=& \left[e^{-i\theta_{j}}\cosh{r_{j}} \hspace{0.1cm} \alpha_{-\mathbf{k}} - \hspace{0.1cm} e^{i(\theta_{j}+2\phi_{j})}\sinh{r_{j}} \hspace{0.1cm} \alpha^{*}_{\mathbf{k}} \right] \, . \label{alpha_t2} 
\end{eqnarray}
In \ref{alpha_t1} - \ref{alpha_t2}, the subscript $j \, -$ associated with all the squeezing and auxiliary parameters, can take only two possible values $j = (3, 4)$. 

Furthermore, the dependency of all the wave function coefficients $(A_{i}, B_{i}, C_{i}, D_{i}, N_{i})$ on the squeezing parameters and coherent state parameters, are provided in the \ref{Table: 3}.

\begin{table}[h]
    \centering
    
    \begin{tabular}{| c | c | c | c | c |}
        \hline 
        \rule{0 pt}{12 pt} Wave function & \makecell{Corresponding \\ coefficients} & Squeezing parameters & \makecell{1st Auxiliary \\ parameter} & \makecell{2nd Auxiliary \\ parameter} \\ 
        \hline 
        \rule{0 pt}{12 pt} $\psi_{r, \beta}\left(-q_{\mathbf{k_{1}}}, q_{-\mathbf{k_{2}}}, t_{1} \right)$ & $N_{1}, A_{1}, B_{1}, C_{1}, D_{1}$ & $r(t_{1}), \, \theta(t_{1}), \, \phi(t_{1})$ & $\zeta_{1} = \zeta_{1}(\beta_{\mathbf{k}}, \beta_{\mathbf{-k}})$ & $\Upsilon_{1} = \Upsilon_{1}(\beta_{\mathbf{k}}, \beta_{\mathbf{-k}})$ \\ 
        \hline 
        \rule{0 pt}{12 pt} $\psi^{*}_{r, \beta} \left(\widetilde{q}_{\mathbf{k_{1}}}, -\widetilde{q}_{-\mathbf{k_{2}}}, t_{2} \right)$ & $N_{2}, A_{2}, B_{2}, C_{2}, D_{2}$ & $r(t_{2}), \, \theta(t_{2}), \, \phi(t_{2})$ & $\zeta_{2} = \zeta_{2}(\beta_{\mathbf{k}}, \beta_{\mathbf{-k}})$ & $\Upsilon_{2} = \Upsilon_{2}(\beta_{\mathbf{k}}, \beta_{\mathbf{-k}})$ \\ 
        \hline 
        \rule{0 pt}{12 pt} $\psi^{*}_{r, \alpha} \left(q_{\mathbf{k_{1}}}, q_{-\mathbf{k_{2}}}, t_{1} \right)$ & $N_{3}, A_{3}, B_{3}, C_{3}, D_{3}$ & $r(t_{1}), \, \theta(t_{1}), \, \phi(t_{1})$ & $\zeta_{3} = \zeta_{3}(\alpha_{\mathbf{k}}, \alpha_{\mathbf{-k}})$ & $\Upsilon_{3} = \Upsilon_{3}(\alpha_{\mathbf{k}}, \alpha_{\mathbf{-k}})$ \\ 
        \hline 
        \rule{0 pt}{12 pt} $\psi_{r, \alpha} \left(\widetilde{q}_{\mathbf{k_{1}}}, \widetilde{q}_{-\mathbf{k_{2}}}, t_{2} \right)$ & $N_{4}, A_{4}, B_{4}, C_{4}, D_{4}$ & $r(t_{2}), \, \theta(t_{2}), \, \phi(t_{2})$ & $\zeta_{4} = \zeta_{4}(\alpha_{\mathbf{k}}, \alpha_{\mathbf{-k}})$ & $\Upsilon_{4} = \Upsilon_{4}(\alpha_{\mathbf{k}}, \alpha_{\mathbf{-k}})$ \\ 
        \hline 
    \end{tabular}

    \caption{Dependency of all wave function coefficients on the squeezing parameters and coherent state parameters.}
    \label{Table: 3}
\end{table}




\hspace{-0.5 cm} $\boldsymbol{\bullet}$ \textbf{First integral:} To evaluate the first integral $I_{3}$ as defined by \ref{Integral_def_q}, first we have to perform a coordinate transformation from the old variables $(q_{\mathbf{k_{1}}}, q_{-\mathbf{k_{2}}})$ to the new variables $(x_{\mathbf{k}}, y_{\mathbf{k}})$. The new variables are defined as follows, 
\begin{equation}
    x_{\mathbf{k}} \hspace{0.1 cm} = \hspace{0.1 cm} \left(q_{\mathbf{k_{1}}} + q_{-\mathbf{k_{2}}} \right) \hspace{0.6 cm} , \hspace{0.6 cm} y_{\mathbf{k}} \hspace{0.1 cm} = \hspace{0.1 cm} \left(q_{\mathbf{k_{1}}} - q_{-\mathbf{k_{2}}} \right) \, . 
\end{equation}
After carrying out all the necessary algebraic manipulations, the integral $I_{3}$ reduces to the following expression:  
\begin{eqnarray}\label{Integral_q}
    I_{3} &=& \iint dq_{\mathbf{k_{1}}} \, dq_{-\mathbf{k_{2}}} \, \psi^{*}_{r, \alpha} \left(q_{\mathbf{k_{1}}}, q_{-\mathbf{k_{2}}}, t_{1} \right) \, \psi_{r, \beta}\left(-q_{\mathbf{k_{1}}}, q_{-\mathbf{k_{2}}}, t_{1} \right) \nonumber \\
    &=& \iint dx_{\mathbf{k}} \, dy_{\mathbf{k}} \, \left|\frac{\partial \, (q_{\mathbf{k_{1}}}, q_{-\mathbf{k_{2}}})}{\partial \, (x_{\mathbf{k}}, y_{\mathbf{k}})} \right| \, \psi^{*}_{r, \alpha} \left(x_{\mathbf{k}}, y_{\mathbf{k}}, t_{1} \right) \, \psi_{r, \beta}\left(x_{\mathbf{k}}, y_{\mathbf{k}}, t_{1} \right) \nonumber \\ 
    &=& \frac{1}{2} N_{1} N^{*}_{3} \, \exp\left(\kappa_{1} \right) \int^{\infty}_{-\infty} dx_{\mathbf{k}} \, \exp\left[-\gamma_{1} \left(x_{\mathbf{k}} + \frac{\epsilon_{1}}{\gamma_{1}} \right)^{2} \right] \, \int^{\infty}_{-\infty} dy_{\mathbf{k}} \, \exp\left[-\gamma_{2} \left(y_{\mathbf{k}} + \frac{\epsilon_{2}}{\gamma_{2}} \right)^{2} \right] \nonumber \\ 
    &=& \frac{\pi}{2} \, \frac{N_{1} N^{*}_{3}}{\sqrt{\gamma_{1} \gamma_{2}}} \, \exp\left(\kappa_{1} \right) \, . 
\end{eqnarray}
The auxiliary quantities $(\gamma_{1}, \gamma_{2}, \epsilon_{1}, \epsilon_{2}, \kappa_{1})$ which appear in the third and fourth line of the \ref{Integral_q}, are defined as follows: 
\begin{eqnarray}
    \gamma_{1} &=& \frac{2A_{0} + B_{0}}{2} \hspace{1.8 cm} = \hspace{1.0 cm} \frac{2 \, (A_{1} + A^{*}_{3}) + (B_{1} - B^{*}_{3})}{4} \, , \label{unknown_11} \\ 
    \gamma_{2} &=& \frac{2A_{0} - B_{0}}{2} \hspace{1.8 cm} = \hspace{1.0 cm} \frac{2 \, (A_{1} + A^{*}_{3}) - (B_{1} - B^{*}_{3})}{4} \, , \label{unknown_12} \\ 
    \epsilon_{1} &=& C_{0} - D_{0} \hspace{2.08 cm} = \hspace{1.0 cm} \frac{(A_{1}C_{1} - A^{*}_{3}C^{*}_{3}) - (A_{1}D_{1} + A^{*}_{3}D^{*}_{3})}{2} \, , \label{unknown_21} \\  
    \epsilon_{2} &=& C_{0} + D_{0} \hspace{2.08 cm} = \hspace{1.0 cm} \frac{(A_{1}C_{1} - A^{*}_{3}C^{*}_{3}) + (A_{1}D_{1} + A^{*}_{3}D^{*}_{3})}{2} \, , \label{unknown_22} \\ 
    \kappa_{1} &=& \left(\frac{\epsilon^{2}_{1}}{\gamma_{1}} + \frac{\epsilon^{2}_{2}}{\gamma_{2}} \right) - \kappa_{0} \hspace{0.8 cm} = \hspace{0.8 cm} \left(\frac{\epsilon^{2}_{1}}{\gamma_{1}} + \frac{\epsilon^{2}_{2}}{\gamma_{2}} \right) - \left[A_{1} \left(C^{2}_{1} + D^{2}_{1} \right) + A^{*}_{3} \left(C^{*2}_{3} + D^{*2}_{3} \right) \right] \, . \label{unknown_33} 
\end{eqnarray}

\hspace{-0.5 cm} $\boldsymbol{\bullet}$ \textbf{Second integral:} For the evaluation of the second integral $I_{4}$ as defined in \ref{Integral_def_tilde_q}, one has to perform another coordinate transformation from the old variables $(\widetilde{q}_{\mathbf{k_{1}}}, \widetilde{q}_{-\mathbf{k_{2}}})$ to the new variables $(\widetilde{x}_{\mathbf{k}}, \widetilde{y}_{\mathbf{k}})$. The new variables are defined as follows, 
\begin{equation}
    \widetilde{x}_{\mathbf{k}} \hspace{0.1 cm} = \hspace{0.1 cm} \left(\widetilde{q}_{\mathbf{k_{1}}} + \widetilde{q}_{-\mathbf{k_{2}}} \right) \hspace{0.6 cm} , \hspace{0.6 cm} \widetilde{y}_{\mathbf{k}} \hspace{0.1 cm} = \hspace{0.1 cm} \left(\widetilde{q}_{\mathbf{k_{1}}} - \widetilde{q}_{-\mathbf{k_{2}}} \right) \, . 
\end{equation}
After performing all the necessary algebraic manipulations, the integral $I_{4}$ reduces to the following expression: 
\begin{eqnarray}\label{Integral_q_tilde}
    I_{4} &=& \iint d\widetilde{q}_{\mathbf{k_{1}}} \, d\widetilde{q}_{-\mathbf{k_{2}}} \, \psi_{r, \alpha} \left(\widetilde{q}_{\mathbf{k_{1}}}, \widetilde{q}_{-\mathbf{k_{2}}}, t_{2} \right) \, \psi^{*}_{r, \beta} \left(\widetilde{q}_{\mathbf{k_{1}}}, -\widetilde{q}_{-\mathbf{k_{2}}}, t_{2} \right) \nonumber \\
    &=& \iint d\widetilde{x}_{\mathbf{k}} \, d\widetilde{y}_{\mathbf{k}} \, \left|\frac{\partial \, (\widetilde{q}_{\mathbf{k_{1}}}, \widetilde{q}_{-\mathbf{k_{2}}})}{\partial \, (\widetilde{x}_{\mathbf{k}}, \widetilde{y}_{\mathbf{k}})} \right| \, \psi_{r, \alpha} \left(\widetilde{x}_{\mathbf{k}}, \widetilde{y}_{\mathbf{k}}, t_{2} \right) \, \psi^{*}_{r, \beta}\left(\widetilde{x}_{\mathbf{k}}, \widetilde{y}_{\mathbf{k}}, t_{2} \right) \nonumber \\ 
    &=& \frac{1}{2} N^{*}_{2} N_{4} \, \exp\left(\kappa_{11} \right) \, \int^{\infty}_{-\infty} d\widetilde{x}_{\mathbf{k}} \, \exp\left[-\gamma_{3} \left(\widetilde{x}_{\mathbf{k}} - \frac{\epsilon_{3}}{\gamma_{3}} \right)^{2} \right] \, \int^{\infty}_{-\infty} d\widetilde{y}_{\mathbf{k}} \, \exp\left[-\gamma_{4} \left(\widetilde{y}_{\mathbf{k}} + \frac{\epsilon_{4}}{\gamma_{4}} \right)^{2} \right] \nonumber \\ 
    &=& \frac{\pi}{2} \, \frac{N^{*}_{2} N_{4}}{\sqrt{\gamma_{3} \gamma_{4}}} \, \exp\left(\kappa_{11} \right) \, . 
\end{eqnarray}
The auxiliary quantities $(\gamma_{3}, \gamma_{4}, \epsilon_{3}, \epsilon_{4}, \kappa_{11})$ appearing in the third and fourth line of the \ref{Integral_q_tilde}, are defined as follows: 
\begin{eqnarray}
    \gamma_{3} &=& \frac{2A_{00} - B_{00}}{2} \hspace{1.5 cm} = \hspace{1.0 cm} \frac{2 (A_{4} + A^{*}_{2}) - (B_{4} - B^{*}_{2})}{4} \, , \label{unknown_41} \\ 
    \gamma_{4} &=& \frac{2A_{00} + B_{00}}{2} \hspace{1.5 cm} = \hspace{1.0 cm} \frac{2 (A_{4} + A^{*}_{2}) + (B_{4} - B^{*}_{2})}{4} \, , \label{unknown_42} \\ 
    \epsilon_{3} &=& C_{00} + D_{00} \hspace{1.8 cm} = \hspace{1.0 cm} \frac{(A_{4}C_{4} + A^{*}_{2}C^{*}_{2}) + (A_{4}D_{4} - A^{*}_{2}D^{*}_{2})}{2} \, , \label{unknown_51} \\ 
    \epsilon_{4} &=& C_{00} - D_{00} \hspace{1.8 cm} = \hspace{1.0 cm} \frac{(A_{4}C_{4} + A^{*}_{2}C^{*}_{2}) - (A_{4}D_{4} - A^{*}_{2}D^{*}_{2})}{2} \, , \label{unknown_52} \\ 
    \kappa_{11} &=& \left(\frac{\epsilon^{2}_{3}}{\gamma_{3}} + \frac{\epsilon^{2}_{4}}{\gamma_{4}} \right) - \kappa_{00} \hspace{0.7 cm} = \hspace{0.7 cm} \left(\frac{\epsilon^{2}_{3}}{\gamma_{3}} + \frac{\epsilon^{2}_{4}}{\gamma_{4}} \right) - \left[A_{4} \left(C^{2}_{4} + D^{2}_{4} \right) + A^{*}_{2} \left(C^{*2}_{2} + D^{*2}_{2} \right) \right] \hspace{0.1 cm} . \label{unknown_66} 
\end{eqnarray}

\hspace{-0.5 cm} $\boldsymbol{\bullet}$ \textbf{Final $(\beta_{\mathbf{k}}, \beta_{\mathbf{-k}})$ integral:} To evaluate the complex $(\beta_{\mathbf{k}}, \beta_{\mathbf{-k}})$ integrals appearing in \ref{spin_correlation_3}, it is necessary to explicitly isolate all the terms in the integrand that depend on the parameters $\beta_{\mathbf{k}}$ and $\beta_{\mathbf{-k}}$. The integrand is simply given by the product of the previously derived integrals $I_{3}$ and $I_{4}$, as defined in \ref{Integral_q} and \ref{Integral_q_tilde} respectively. Among all the terms appearing in these expressions, only four quantities $(N_{1}, N^{*}_{2}, \kappa_{1}, \kappa_{11})$ exhibit explicit dependence on the parameters $\beta_{\mathbf{k}}$ and $\beta_{\mathbf{-k}}$. All the remaining terms of $I_{3}$ and $I_{4}$ are independent of the coherent state parameters $\beta_{\mathbf{k}}$ and $\beta_{\mathbf{-k}}$. Consequently, they can be taken outside the corresponding $(\beta_{\mathbf{k}}, \beta_{\mathbf{-k}})$ integrals. 

Before proceeding further, let us decompose the complex parameters $\beta_{\mathbf{k}}$ and $\beta_{\mathbf{-k}}$ into their real and imaginary parts respectively. The explicit decomposition is given below, 
\begin{equation}
    \beta_{\mathbf{k}} \hspace{0.2 cm} = \hspace{0.2 cm} \beta_{\mathbf{k,R}} + i \, \beta_{\mathbf{k, I}} \hspace{0.78 cm} , \hspace{0.78 cm} \beta_{\mathbf{-k}} \hspace{0.2 cm} = \hspace{0.2 cm} \beta_{\mathbf{-k,R}} + i \, \beta_{\mathbf{-k, I}} \, . \label{coordinate_6} 
\end{equation}
In addition, we further decompose the normalization factor $N_{j}$ of the wave function $\psi_{r, \alpha}(q_{\mathbf{k}}, q_{\mathbf{-k}}, t_{j})$ defined in the \ref{Normalization_sc}, in the following manner: 
\begin{eqnarray}
    N_{j} &=& \xi_{j} \, \exp\left(N_{jj} \right) \hspace{0.3 cm} = \hspace{0.3 cm} \xi_{j} \, \exp\left(\overline{N}_{jj} + \widetilde{N}_{jj} \right) \, , 
\end{eqnarray}
where $\xi_{j}, \overline{N}_{jj}$ and $\widetilde{N}_{jj}$ are defined as follows, 
\begin{eqnarray}
    \xi_{j} &=& \frac{\sech{r_{j, k}}}{\sqrt{2\pi (1-e^{-4i\phi_{j, k}}\tanh^{2}{r_{j, k}})}} \, , \\ 
    \overline{N}_{jj} &=& \exp\left[A_{j} \left(C_{j}^{2} + D_{j}^{2} \right) - \frac{1}{2} \left(g_{j, 1} \, x^{2}_{j, 0} + g_{j, 2} \, y^{2}_{j, 0} \right) \right] \, , \\ 
    \widetilde{N}_{jj} &=& \exp\left\{-4i\Im\big(A_{j} \big) \left[\Re\left(\frac{\zeta_{j, \mathbf{k}}}{\sqrt{2}} \right)^{2} + \Re\left(\frac{\Upsilon_{j, \mathbf{k}}}{\sqrt{2}} \right)^{2} \right] + 4i\Im\big(B_{j} \big) \,  \Re\left(\frac{\zeta_{j, \mathbf{k}}}{\sqrt{2}} \right) \Re\left(\frac{\Upsilon_{j, \mathbf{k}}}{\sqrt{2}} \right) \right\} \nonumber \\ 
    && \cross \, \exp\left\{ - 2i \, \left[\Re\left(\frac{\zeta_{j, \mathbf{k}}}{\sqrt{2}} \right)\Im\left(\frac{\zeta_{j, \mathbf{k}}}{\sqrt{2}} \right) + \Re\left(\frac{\Upsilon_{j, \mathbf{k}}}{\sqrt{2}} \right) \Im\left(\frac{\Upsilon_{j, \mathbf{k}}}{\sqrt{2}} \right) \right] \right\} \, . \label{modified_normalization_1}
\end{eqnarray}
Here the subscript $j$ can take four possible values $j = (1, 2, 3, 4)$, by following the convention introduced earlier in the \ref{Table: 3}. Definition of the auxiliary quantities $(\zeta_{j, \mathbf{k}}, \Upsilon_{j, \mathbf{k}})$ in terms of the coherent state parameters $(\alpha_{\mathbf{k}}, \alpha_{\mathbf{-k}}, \beta_{\mathbf{k}}, \beta_{\mathbf{-k}})$, are provided in \ref{beta_t1} - \ref{alpha_t2}. Furthermore, definition of the remaining auxiliary quantities $(g_{j, 1} \, , \, g_{j, 2} \, , \, x_{j, 0} \, , \, y_{j, 0})$ which appear in the expression of $\overline{N}_{jj}$, are provided in \ref{def_g_1} and \ref{def_x_0} respectively.  
    
Substituting the results of the respective integrals $I_{3}$ and $I_{4}$ as given by the \ref{Integral_q} and \ref{Integral_q_tilde}, into the expression of unequal-time spin-spin correlation \ref{spin_correlation_3}, one arrives at the following: 
\begin{eqnarray}\label{intermediate_3}
    E(t_{1}, t_{2}) &=& \Re\left[\iint \frac{d^{2}\beta_{\mathbf{k}} \hspace{0.1 cm} d^{2}\beta_{-\mathbf{k}}}{\pi^{2}} \hspace{0.1 cm} I_{3} \, I_{4} \right] \nonumber \\ 
    &=& \Re\left[\frac{\xi_{1} \, \xi^{*}_{2} \, N^{*}_{3} N_{4}}{\sqrt{\gamma_{1} \gamma_{2} \gamma_{3} \gamma_{4}}} \iint d^{2}\beta_{\mathbf{k}} \hspace{0.1 cm} d^{2}\beta_{-\mathbf{k}} \hspace{0.1 cm} \exp\big(\kappa_{1} + \kappa_{11} \big) \, \exp\big(N_{11} + N^{*}_{22} \big) \right] \nonumber \\ 
    &=& \Re\left[\frac{\xi_{1} \, \xi^{*}_{2} \, N^{*}_{3} N_{4}}{\sqrt{\gamma_{1} \gamma_{2} \gamma_{3} \gamma_{4}}} \iint d\beta_{\mathbf{k, R}} \, d\beta_{\mathbf{k, I}} \, d\beta_{\mathbf{-k, R}} \, d\beta_{\mathbf{-k, I}} \hspace{0.1 cm} \exp\bigg(\kappa_{1} + \kappa_{11} + N_{11} + N^{*}_{22} \bigg) \right] \nonumber \\ 
    &=& \Re\left[\frac{\xi_{1} \, \xi^{*}_{2} \, N^{*}_{3} N_{4}}{\sqrt{\gamma_{1} \gamma_{2} \gamma_{3} \gamma_{4}}} \left|\frac{\partial \left(\beta_{\mathbf{k, R}}, \beta_{\mathbf{k, I}}, \beta_{\mathbf{-k, R}}, \beta_{\mathbf{-k, I}} \right)}{\partial \left(\beta_{\mathbf{k}}, \beta^{*}_{\mathbf{k}}, \beta_{\mathbf{-k}}, \beta^{*}_{\mathbf{-k}} \right)} \right| \iint d\beta_{\mathbf{k}} \, d\beta^{*}_{\mathbf{k}} \, d\beta_{\mathbf{-k}} \, d\beta^{*}_{\mathbf{-k}} \hspace{0.1 cm} \exp\bigg(\kappa_{1} + \kappa_{11} + N_{11} + N^{*}_{22} \bigg) \right] . \nonumber \\  
\end{eqnarray}
To simplify the subsequent calculations, we perform a coordinate transformation from the original variables $\left(\beta_{\mathbf{k}}, \beta^{*}_{\mathbf{k}}, \right. \\ \left. \beta_{\mathbf{-k}}, \beta^{*}_{\mathbf{-k}} \right)$ to the new variables $\left(\delta_{\mathbf{k}}, \delta^{*}_{\mathbf{k}}, \Psi_{\mathbf{k}}, \Psi^{*}_{\mathbf{k}} \right)$. The new variables are defined as follows, 
\begin{eqnarray}
    \delta_{\mathbf{k}} &=& \beta_{\mathbf{k}} + \beta_{\mathbf{-k}} \hspace{0.45 cm} , \hspace{0.45 cm} \delta^{*}_{\mathbf{k}} \hspace{0.2 cm} = \hspace{0.2 cm} \beta^{*}_{\mathbf{k}} + \beta^{*}_{\mathbf{-k}} \, , \label{coordinate_3} \\ 
    \Psi_{\mathbf{k}} &=& \beta_{\mathbf{k}} - \beta_{\mathbf{-k}} \hspace{0.45 cm} , \hspace{0.45 cm} \Psi^{*}_{\mathbf{k}} \hspace{0.2 cm} = \hspace{0.2 cm} \beta^{*}_{\mathbf{k}} - \beta^{*}_{\mathbf{-k}} \, . \label{coordinate_4}
\end{eqnarray}
For later convenience, we decompose the complex parameters $\delta_{\mathbf{k}}$ and $\Psi_{\mathbf{k}}$ into their real and imaginary parts. The explicit decomposition is given below, 
\begin{equation}
    \delta_{\mathbf{k}} \hspace{0.2 cm} = \hspace{0.2 cm} \delta_{\mathbf{k, R}} + i \, \delta_{\mathbf{k, I}} \hspace{0.75 cm} , \hspace{0.75 cm} \Psi_{\mathbf{k}} \hspace{0.2 cm} = \hspace{0.2 cm} \Psi_{\mathbf{k, R}} + i \, \Psi_{\mathbf{k, I}} \, . \label{coordinate_5} 
\end{equation}
For convenience, we denote the Jacobians associated with the above-mentioned coordinate transformations as follows, 
\begin{eqnarray}
    \mathcal{J}_{1} &=& \left|\frac{\partial \left(\beta_{\mathbf{k, R}}, \beta_{\mathbf{k, I}}, \beta_{\mathbf{-k, R}}, \beta_{\mathbf{-k, I}} \right)}{\partial \left(\beta_{\mathbf{k}}, \beta^{*}_{\mathbf{k}}, \beta_{\mathbf{-k}}, \beta^{*}_{\mathbf{-k}} \right)} \right| \hspace{0.2 cm} = \hspace{0.2 cm} \frac{1}{(2i)^{2}} \, , \\ 
    \mathcal{J}_{2} &=& \left|\frac{\partial \left(\beta_{\mathbf{k}}, \beta^{*}_{\mathbf{k}}, \beta_{\mathbf{-k}}, \beta^{*}_{\mathbf{-k}} \right)}{\partial \left(\delta_{\mathbf{k}}, \delta^{*}_{\mathbf{k}}, \Psi_{\mathbf{k}}, \Psi^{*}_{\mathbf{k}} \right)} \right| \hspace{1.25 cm} = \hspace{0.2 cm} \frac{1}{4} \hspace{0.16 cm} , \\ 
    \mathcal{J}_{3} &=& \left|\frac{\partial \left(\delta_{\mathbf{k}}, \delta^{*}_{\mathbf{k}}, \Psi_{\mathbf{k}}, \Psi^{*}_{\mathbf{k}} \right)}{\partial \left(\delta_{\mathbf{k, R}}, \delta_{\mathbf{k, I}}, \Psi_{\mathbf{k, R}}, \Psi_{\mathbf{k, I}} \right)} \right| \hspace{0.6 cm} = \hspace{0.2 cm} (2i)^{2} \, . 
\end{eqnarray}
The information about the Jacobians and the corresponding coordinate transformations, are provided in \ref{Table: 5}. 
    
Performing the above-mentioned coordinate transformations defined by \ref{coordinate_3} - \ref{coordinate_5} and plugging them into the unequal-time spin-spin correlation \ref{intermediate_3}, one obtains the following expression: 
\begin{eqnarray}
    E(t_{1}, t_{2}) &=& \Re\left[\frac{\xi_{1} \, \xi^{*}_{2} \, N^{*}_{3} N_{4}}{\sqrt{\gamma_{1} \gamma_{2} \gamma_{3} \gamma_{4}}} \, \mathcal{J}_{1} \left|\frac{\partial \left(\beta_{\mathbf{k}}, \beta^{*}_{\mathbf{k}}, \beta_{\mathbf{-k}}, \beta^{*}_{\mathbf{-k}} \right)}{\partial \left(\delta_{\mathbf{k}}, \delta^{*}_{\mathbf{k}}, \Psi_{\mathbf{k}}, \Psi^{*}_{\mathbf{k}} \right)} \right| \hspace{0.1 cm} \iint d\delta_{\mathbf{k}} d\delta^{*}_{\mathbf{k}} \, d\Psi_{\mathbf{k}} d\Psi^{*}_{\mathbf{k}} \hspace{0.1 cm} \exp\bigg(\kappa_{1} + \kappa_{11} + N_{11} + N^{*}_{22} \bigg) \right] \nonumber 
\end{eqnarray}
\begin{eqnarray}
    E(t_{1}, t_{2}) &=& \Re\left[\frac{\xi_{1} \, \xi^{*}_{2} \, N^{*}_{3} N_{4}}{\sqrt{\gamma_{1} \gamma_{2} \gamma_{3} \gamma_{4}}} \, \mathcal{J}_{1} \mathcal{J}_{2} \left|\frac{\partial \left(\delta_{\mathbf{k}}, \delta^{*}_{\mathbf{k}}, \Psi_{\mathbf{k}}, \Psi^{*}_{\mathbf{k}} \right)}{\partial \left(\delta_{\mathbf{k, R}}, \delta_{\mathbf{k, I}}, \Psi_{\mathbf{k, R}}, \Psi_{\mathbf{k, I}} \right)} \right| \hspace{0.1 cm} \iint d\delta_{\mathbf{k, R}} d\delta_{\mathbf{k, I}} \, d\Psi_{\mathbf{k, R}} d\Psi_{\mathbf{k, I}} \hspace{0.1 cm} \right . \nonumber \\ 
    && \bigg. \hspace{7.5 cm} \cross \hspace{0.1 cm} \exp\bigg(\kappa_{1} + \kappa_{11} + N_{11} + N^{*}_{22} \bigg) \bigg] \nonumber \\ 
    &=& \Re\left[\frac{\xi_{1} \, \xi^{*}_{2} \, \widetilde{N}^{*}_{3} \widetilde{N}_{4}}{\sqrt{\gamma_{1} \gamma_{2} \gamma_{3} \gamma_{4}}} \, \mathcal{J}_{1} \mathcal{J}_{2} \mathcal{J}_{3} \hspace{0.1 cm} \exp\left(\mathbf{\lambda_{k}}^{T} \widetilde{T} \mathbf{\lambda_{k}} + \Delta_{\mathbf{k}} \right) \iint d^{4}\mathbf{\sigma_{k}} \, \exp\bigg(\mathbf{\sigma_{k}}^{T} N \mathbf{\sigma_{k}} + Q^{T} \mathbf{\sigma_{k}} + \mathbf{\sigma_{k}}^{T} K \mathbf{\sigma_{k}} \bigg) \right] \nonumber \\ 
    &=& \Re\left[\frac{\xi_{1} \, \xi^{*}_{2} \, \widetilde{N}^{*}_{3} \widetilde{N}_{4}}{\sqrt{\gamma_{1} \gamma_{2} \gamma_{3} \gamma_{4}}} \, \mathcal{J}_{1} \mathcal{J}_{2} \mathcal{J}_{3} \hspace{0.1 cm} \exp\left(\mathbf{\lambda_{k}}^{T} \widetilde{T} \mathbf{\lambda_{k}} + \Delta_{\mathbf{k}} \right) \, \frac{(2\pi)^{2}}{\sqrt{\Det{(-2N - 2K)}}} \, \exp\left\{-\frac{1}{2} Q^{T} \big(2N + 2K \big)^{-1} Q \right\} \right] \, . \nonumber \\ \label{intermediate_4} 
\end{eqnarray}
In the third line of \ref{intermediate_4}, we introduce a four-dimensional vector $\mathbf{\sigma_{k}}$ over which the integrations are carried out, together with three $(4 \cross 4)$ matrices $N, K$ and $\widetilde{T}$, and a single $(4 \cross 1)$ matrix $Q$. In the same line, we also introduced another four-dimensional vector $\lambda_{\mathbf{k}}$ which arises from the exponential part of the normalization factors $N^{*}_{3}$ and $N_{4}$. The matrices $N$ and $K \, - $ both of them are constructed from the coefficients that are associated with the terms quadratic in $\mathbf{\sigma_{k}}$, while the matrix $Q$ is constructed from the coefficients which are associated with the terms linear in $\mathbf{\sigma_{k}}$. The matrix $\widetilde{T}$ is constructed from the coefficients that are associated with terms quadratic in $\lambda_{\mathbf{k}}$. The explicit expressions for all the elements of the matrix $N, K$ and $Q$ are provided in the \ref{Matrix_elements} of the Appendix.   

We first define the vector $\mathbf{\sigma_{k}}$ in terms of the integration variables $\delta_{\mathbf{k}}$ and $\Psi_{\mathbf{k}}$ as follows, 
\begin{eqnarray}\label{sigma_def}
    \mathbf{\sigma_{k}} &=& 
    \begin{pmatrix}
        \delta_{\mathbf{k, R}} \\ 
        \delta_{\mathbf{k, I}} \\ 
        \Psi_{\mathbf{k, R}} \\ 
        \Psi_{\mathbf{k, I}}
    \end{pmatrix} \, . 
    \hspace{0.5 cm} = \hspace{0.5 cm} 
    \begin{bmatrix}
        \hspace{0.1 cm} \Re \, (\beta_{\mathbf{k}} + \beta_{\mathbf{-k}}) \hspace{0.1 cm} \\ 
        \hspace{0.1 cm} \Im \, (\beta_{\mathbf{k}} + \beta_{\mathbf{-k}}) \hspace{0.1 cm} \\
        \hspace{0.1 cm} \Re \, (\beta_{\mathbf{k}} - \beta_{\mathbf{-k}}) \hspace{0.1 cm} \\ 
        \hspace{0.1 cm} \Im \, (\beta_{\mathbf{k}} - \beta_{\mathbf{-k}}) \hspace{0.1 cm} 
    \end{bmatrix} \, . 
\end{eqnarray}
The definition of another $4D$ vector $\mathbf{\lambda_{k}}$ in terms of the coherent state parameters $\alpha_{\mathbf{k}}, \alpha_{-\mathbf{k}}$ are given below, 
\begin{eqnarray}\label{lambda_def}
    \mathbf{\lambda_{k}} &=& 
    \begin{pmatrix}
        \chi_{\mathbf{k, R}} \\ 
        \chi_{\mathbf{k, I}} \\ 
        \Xi_{\mathbf{k, R}} \\ 
        \Xi_{\mathbf{k, I}} 
    \end{pmatrix} 
    \hspace{0.75 cm} = \hspace{0.75 cm} 
    \begin{bmatrix}
        \hspace{0.1 cm} \Re \, (\alpha_{\mathbf{k}} + \alpha_{\mathbf{-k}}) \hspace{0.1 cm} \\ 
        \hspace{0.1 cm} \Im \, (\alpha_{\mathbf{k}} + \alpha_{\mathbf{-k}}) \hspace{0.1 cm} \\ 
        \hspace{0.1 cm} \Re \, (\alpha_{\mathbf{k}} - \alpha_{\mathbf{-k}}) \hspace{0.1 cm} \\ 
        \hspace{0.1 cm} \Im \, (\alpha_{\mathbf{k}} - \alpha_{\mathbf{-k}}) \hspace{0.1 cm} 
    \end{bmatrix} \, . 
\end{eqnarray}
The complex variables $\chi_{\mathbf{k}}$ and $\Xi_{\mathbf{k}}$ are defined as follows, 
\begin{eqnarray}
    \chi_{\mathbf{k}} &=& \alpha_{\mathbf{k}} + \alpha_{\mathbf{-k}} \hspace{0.8 cm} , \hspace{0.8 cm} \chi^{*}_{\mathbf{k}} \hspace{0.2 cm} = \hspace{0.2 cm} \alpha^{*}_{\mathbf{k}} + \alpha^{*}_{\mathbf{-k}} \, , \label{coordinate_33} \\ 
    \Xi_{\mathbf{k}} &=& \alpha_{\mathbf{k}} - \alpha_{\mathbf{-k}} \hspace{0.8 cm} , \hspace{0.8 cm} \Xi^{*}_{\mathbf{k}} \hspace{0.2 cm} = \hspace{0.2 cm} \alpha^{*}_{\mathbf{k}} - \alpha^{*}_{\mathbf{-k}} \, . \label{coordinate_44}
\end{eqnarray}
In terms of real and imaginary part, the complex variables can be expressed as: 
\begin{eqnarray}
    \chi_{\mathbf{k}} &=& \chi_{\mathbf{k, R}} + i \, \chi_{\mathbf{k, I}} \hspace{0.8 cm} , \hspace{0.8 cm} \Xi_{\mathbf{k}} \hspace{0.3 cm} = \hspace{0.3 cm} \Xi_{\mathbf{k, R}} + i \, \Xi_{\mathbf{k, I}} \, . 
\end{eqnarray}
It is worth noting that the auxiliary quantities $\widetilde{N}_{3}$ and $\widetilde{N}_{4}$ appearing in the third line of \ref{intermediate_4}, are defined in the following manner, 
\begin{eqnarray}
    \widetilde{N}_{3} &=& \xi_{3} \, \exp\left(\overline{N}_{33} \right) \hspace{0.75 cm} , \hspace{0.75 cm} \widetilde{N}_{4} \hspace{0.1 cm} = \hspace{0.1 cm} \xi_{4} \, \exp\left(\overline{N}_{44} \right) \, . 
\end{eqnarray}

Furthermore, the scalar quantity $\Delta_{\mathbf{k}}$ is defined as follows, 
\begin{equation}
    \Delta_{\mathbf{k}} \hspace{0.3 cm} = \hspace{0.3 cm} \exp\left(\frac{\gamma_{1} \, \Sigma^{2}_{4} + \gamma_{2} \, \Sigma^{2}_{3} - \Sigma_{5}}{4\gamma_{1}\gamma_{2}} + \frac{\gamma_{3} \, \Sigma^{2}_{7} + \gamma_{4} \, \Sigma^{2}_{6} - \Sigma_{8}}{4\gamma_{3}\gamma_{4}} \right) \, . 
\end{equation}
The coefficients $\Sigma_{i}$ appearing in the definition of $\Delta_{\mathbf{k}}$, are specified below 
\begin{eqnarray}
    \Sigma_{3} &=& A^{*}_{3} \left(C^{*}_{3} + D^{*}_{3} \right) \hspace{1.8 cm} , \hspace{0.6 cm} \Sigma_{6} \hspace{0.2 cm} = \hspace{0.2 cm} A_{4} \left(C_{4} + D_{4} \right) \, , \\ 
    \Sigma_{4} &=& A^{*}_{3} \left(C^{*}_{3} - D^{*}_{3} \right) \hspace{1.8 cm} , \hspace{0.6 cm} \Sigma_{7} \hspace{0.2 cm} = \hspace{0.2 cm} A_{4} \left(C_{4} - D_{4} \right) \, , \\ 
    \Sigma_{5} &=& 4\gamma_{1}\gamma_{2} A^{*}_{3} \left(C^{*2}_{3} + D^{*2}_{3} \right) \hspace{0.6 cm} , \hspace{0.6 cm} \Sigma_{8} \hspace{0.2 cm} = \hspace{0.2 cm} 4\gamma_{3}\gamma_{4} A_{4} \left(C^{2}_{4} + D^{2}_{4} \right) \, . 
\end{eqnarray}
The coefficients $(\gamma_{1}, \gamma_{2}, \gamma_{3}, \gamma_{4})$ appearing in the definition of $\Delta_{\mathbf{k}}$, are already provided in the \ref{unknown_11} - \ref{unknown_12} and \ref{unknown_41} - \ref{unknown_42} respectively.

    

\section{Explicit expressions of $M, N$ and $K, \widetilde{T} \, -$ matrix}\label{Matrix_elements} 

Before proceeding further, we introduce a set of parameters $\left(p_{j}, q_{j}, m_{j}, n_{j} \right)$ which allow the matrix elements to be expressed in a more compact form. In terms of squeezing parameters $(r_{j}, \theta_{j}, \phi_{j})$ and the new quantity $w_{j}$, defined as,
\begin{eqnarray}
    w_{j} &=& \frac{1}{2} \, e^{-2i\phi_{j}} \tanh{r_{j}} \, , 
\end{eqnarray}
these new set of parameters are defined as follows, 
\begin{eqnarray}
    p_{1} &=& e^{-i\theta_{1}} \cosh{r_{1}} \hspace{1.2 cm} , \hspace{1.0 cm} q_{1} \hspace{0.2 cm} = \hspace{0.2 cm} e^{i(\theta_{1} + 2\phi_{1})} \sinh{r_{1}} \hspace{0.1 cm} , \label{squeezing_t_1} \\ 
    p_{2} &=& e^{-i\theta_{2}} \cosh{r_{2}} \hspace{1.2 cm} , \hspace{1.0 cm} q_{2} \hspace{0.2 cm} = \hspace{0.2 cm} e^{i(\theta_{2} + 2\phi_{2})} \sinh{r_{2}} \hspace{0.1 cm} , \label{squeezing_t_2} \\ 
    m_{1} &=& p_{1} + 2w_{1} \, q^{*}_{1} \hspace{1.4 cm} , \hspace{1.0 cm} n_{1} \hspace{0.2 cm} = \hspace{0.2 cm} q_{1} + 2w_{1} \, p^{*}_{1} \hspace{0.1 cm} , \label{squeezing_t_11} \\ 
    m_{2} &=& p_{2} + 2w_{2} \, q^{*}_{2} \hspace{1.4 cm} , \hspace{1.0 cm} n_{2} \hspace{0.2 cm} = \hspace{0.2 cm} q_{2} + 2w_{2} \, p^{*}_{2} \hspace{0.1 cm} . \label{squeezing_t_22} 
\end{eqnarray}
For convenience, we introduce another set of parameters $(\omega_{1}, \, \omega_{2}, \, \omega_{3}, \, \omega_{4})$ which are defined as given below: 
\begin{eqnarray}
    \omega_{1} &=& 2 \, \left(\frac{1-2w_{1}}{1 + 4w^{2}_{1}} \right) \hspace{0.6 cm} , \hspace{0.6 cm} \omega_{2} \hspace{0.2 cm} = \hspace{0.2 cm} 2 \, \left(\frac{1 - 2w_{2}}{1 + 4w^{2}_{2}} \right) \, , \\ 
    \omega_{3} &=& 2 \, \left(\frac{1+2w_{1}}{1 + 4w^{2}_{1}} \right) \hspace{0.6 cm} , \hspace{0.6 cm} \omega_{4} \hspace{0.2 cm} = \hspace{0.2 cm} 2 \, \left(\frac{1 + 2w_{2}}{1 + 4w^{2}_{2}} \right) \, . 
\end{eqnarray}
Note that the subscript $j$ appearing in the squeezing parameters $(r_{j}, \theta_{j}, \phi_{j})$, as given in \ref{squeezing_t_1} - \ref{squeezing_t_22}, denotes the time $t_{j}$ at which these parameters are being evaluated. Explicitly, this can be written as follows, 
\begin{eqnarray}
    (r_{1}, \theta_{1}, \phi_{1}) &:=& \left[ \, r(t_{1}), \theta(t_{1}), \phi(t_{1}) \, \right] \, , \\ 
    (r_{2}, \theta_{2}, \phi_{2}) &:=& \left[ \, r(t_{2}), \theta(t_{2}), \phi(t_{2}) \, \right] \, . 
\end{eqnarray}

To make the expressions more compact, we introduce some new quantities $\left(g_{1}, g_{2}, \widetilde{g}_{1}, \widetilde{g}_{2} \right)$ which will appear in the elements of $N$-matrix. The definitions are the following, 
\begin{eqnarray}
    g_{1} &=& \Re\left(\frac{2A_{1} - B_{1}}{2} \right) \hspace{0.6 cm} , \hspace{0.6 cm} g_{2} \hspace{0.2 cm} = \hspace{0.2 cm} \Re\left(\frac{2A_{1} + B_{1}}{2} \right) \, , \\ 
    \widetilde{g}_{1} &=& \Re\left(\frac{2A_{2} - B_{2}}{2} \right) \hspace{0.6 cm} , \hspace{0.6 cm} \widetilde{g}_{2} \hspace{0.2 cm} = \hspace{0.2 cm} \Re\left(\frac{2A_{2} + B_{2}}{2} \right) \, . 
\end{eqnarray}

In the section, the primary goal is to provide explicit expressions for all the elements of the $N$-matrix. To simplify the expressions, we introduce a new $(4 \cross 4)$ matrix $M$, in terms of which all the elements of the $N$-matrix can be conveniently expressed. The elements of the $M$-matrix are constructed from the integrand of the complex-variable integral over the parameters $(\beta_{\mathbf{k}}, \beta_{\mathbf{-k}})$. This construction is achieved by re-organizing the integrand in terms of the complex variables $\delta_{\mathbf{k}}, \delta^{*}_{\mathbf{k}}, \psi_{\mathbf{k}}$ and $\psi^{*}_{\mathbf{k}}$, as defined in \ref{coordinate_3} and \ref{coordinate_4}. Mathematically, this procedure can be expressed as follows: 
\begin{eqnarray}\label{M_construction}
    \exp\left(\kappa_{1} + \kappa_{11} + \overline{N}_{11} + \overline{N}^{*}_{22} \right) &=& \exp\bigg(M_{11} \hspace{0.1 cm} \delta^{2}_{\mathbf{k}} \, + \, M_{22} \hspace{0.1 cm} \delta^{*2}_{\mathbf{k}} \, + \, M_{12} \hspace{0.1 cm} \delta_{\mathbf{k}}\delta^{*}_{\mathbf{k}} \, + \, M_{21} \hspace{0.1 cm} \delta^{*}_{\mathbf{k}}\delta_{\mathbf{k}} \, + \, R_{1} \hspace{0.05 cm} \delta_{\mathbf{k}} \, + \, R_{2} \hspace{0.05 cm} \delta^{*}_{\mathbf{k}} \bigg. \nonumber \\ 
    && \bigg. + \, M_{33} \hspace{0.1 cm} \psi^{2}_{\mathbf{k}} \, + \, M_{44} \hspace{0.1 cm} \psi^{*2}_{\mathbf{k}} \, + \, M_{34} \hspace{0.1 cm} \psi_{\mathbf{k}}\psi^{*}_{\mathbf{k}} \, + \, M_{43} \hspace{0.1 cm} \psi^{*}_{\mathbf{k}}\psi_{\mathbf{k}} \, + \, R_{3} \hspace{0.05 cm} \psi_{\mathbf{k}} \, + \, R_{4} \hspace{0.05 cm} \psi^{*}_{\mathbf{k}} \bigg) \nonumber \\ 
    &=& \exp\bigg(\Lambda^{T}_{\mathbf{k}} \, M \, \Lambda_{\mathbf{k}} \, + \, R^{T} \Lambda_{\mathbf{k}} \bigg) \, . 
\end{eqnarray}
In the last line of \ref{M_construction}, the argument of the exponential is written in a more compact matrix notation, involving the matrix $M$ and the four dimensional vectors $\Lambda_{\mathbf{k}}$ and $R \, -$ respectively. The vector $\Lambda_{\mathbf{k}}$ is defined as follows, 
\begin{eqnarray}
    \Lambda^{T}_{\mathbf{k}} &=& 
    \begin{pmatrix}
        \delta_{\mathbf{k}} & \delta^{*}_{\mathbf{k}} & \psi_{\mathbf{k}} & \psi^{*}_{\mathbf{k}} 
    \end{pmatrix} \, . 
\end{eqnarray}

The explicit expressions for the elements of the $M$-matrix, associated with the terms which are quadratic in $\delta_{\mathbf{k}}$ and $\delta^{*}_{\mathbf{k}}$, are given below: 
\begin{eqnarray}
    M_{11} &=& \left\{(8 \, g_{1}g_{2})^{-1} \left[g_{2}A_{1} \left(4g_{1}-A_{1} \right) \left(m_{1}\omega_{1} \right)^{2} - g_{2}A^{*2}_{1} \left(n^{*}_{1}\omega^{*}_{1} \right)^{2} + 2g_{2} \, m_{1}n^{*}_{1} \abs{A_{1}\omega_{1}}^{2} \right] \right . \nonumber \\ 
    && + \, (8 \, \widetilde{g}_{1}\widetilde{g}_{2})^{-1} \left[\widetilde{g}_{2}A^{*}_{2} \left(4\widetilde{g}_{1}-A^{*}_{2} \right) \left(n^{*}_{2}\omega^{*}_{2} \right)^{2} - \widetilde{g}_{2}A^{2}_{2} \left(m_{2}\omega_{2} \right)^{2} + 2\widetilde{g}_{2} \, m_{2}n^{*}_{2} \abs{A_{2}\omega_{2}}^{2} \right] \nonumber \\ 
    && \left . + \, (4 \, \gamma_{1}\gamma_{2})^{-1} \, \gamma_{1}A_{1} \left(A_{1}-2\gamma_{2} \right) \left(m_{1}\omega_{1} \right)^{2} \hspace{0.1 cm} + \hspace{0.1 cm} (4 \, \gamma_{3}\gamma_{4})^{-1} \, \gamma_{3}A^{*}_{2} \left(A^{*}_{2}-2\gamma_{4} \right) \left(n^{*}_{2}\omega^{*}_{2} \right)^{2} \right\} \, , \\ 
    M_{22} &=& \left\{(8 \, g_{1}g_{2})^{-1} \left[g_{2}A_{1} \left(4g_{1}-A_{1} \right) \left(n_{1}\omega_{1} \right)^{2} - g_{2}A^{*2}_{1} \left(m^{*}_{1}\omega^{*}_{1} \right)^{2} + 2g_{2} \, m^{*}_{1}n_{1} \abs{A_{1}\omega_{1}}^{2} \right] \right . \nonumber \\ 
    && + \, (8 \, \widetilde{g}_{1}\widetilde{g}_{2})^{-1} \left[\widetilde{g}_{2}A^{*}_{2} \left(4\widetilde{g}_{1}-A^{*}_{2} \right) \left(m^{*}_{2}\omega^{*}_{2} \right)^{2} - \widetilde{g}_{2}A^{2}_{2} \left(n_{2}\omega_{2} \right)^{2} + 2\widetilde{g}_{2} \, m^{*}_{2}n_{2} \abs{A_{2}\omega_{2}}^{2} \right] \nonumber \\ 
    && \left . + \, (4 \, \gamma_{1}\gamma_{2})^{-1} \, \gamma_{1}A_{1} \left(A_{1}-2\gamma_{2} \right) \left(n_{1}\omega_{1} \right)^{2} \hspace{0.1 cm} + \hspace{0.1 cm} (4 \, \gamma_{3}\gamma_{4})^{-1} \, \gamma_{3}A^{*}_{2} \left(A^{*}_{2}-2\gamma_{4} \right) \left(m^{*}_{2}\omega^{*}_{2} \right)^{2} \right\} \, , \\ 
    M_{12} &=& \left\{(8 \, g_{1}g_{2})^{-1} \left[g_{2} \, m^{*}_{1}n^{*}_{1} \left(A^{*}_{1}\omega^{*}_{1} \right)^{2} - g_{2}A_{1} \left(4g_{1}-A_{1} \right) \, m_{1}n_{1} \omega^{2}_{1} - g_{2}\abs{A_{1}\omega_{1}}^{2} \left(\abs{m_{1}}^{2} + \abs{n_{1}}^{2} \right) \right] \right . \nonumber \\ 
    && + \, (8 \, \widetilde{g}_{1}\widetilde{g}_{2})^{-1} \left[\widetilde{g}_{2} \, m_{2}n_{2} \left(A_{2}\omega_{2} \right)^{2} - \widetilde{g}_{2}A^{*}_{2} \left(4\widetilde{g}_{1}-A^{*}_{2} \right) m^{*}_{2}n^{*}_{2}\omega^{*2}_{2}  - \widetilde{g}_{2} \abs{A_{2}\omega_{2}}^{2} \left(\abs{m_{2}}^{2} + \abs{n_{2}}^{2} \right) \right] \nonumber \\ 
    && \bigg . - \, (4 \, \gamma_{1}\gamma_{2})^{-1} \, \gamma_{1}A_{1} \left(A_{1}-2\gamma_{2} \right) m_{1}n_{1}\omega^{2}_{1} \hspace{0.1 cm} - \hspace{0.1 cm} (4 \, \gamma_{3}\gamma_{4})^{-1} \, \gamma_{3}A^{*}_{2} \left(A^{*}_{2}-2\gamma_{4} \right) m^{*}_{2}n^{*}_{2} \, \omega^{*2}_{2} \bigg\} \, , \\ 
    M_{21} &=& M_{12} \, . 
\end{eqnarray}
The remaining elements of the $M$-matrix which are associated with the terms quadratic in $\Psi_{\mathbf{k}}$ and $\Psi^{*}_{\mathbf{k}}$, are given below: 
\vspace{-0.2 cm}
\begin{eqnarray}
    M_{33} &=& \left\{(8 \, g_{1}g_{2})^{-1} \left[g_{1}A_{1} \left(4g_{2}-A_{1} \right) \left(m_{1}\omega_{3} \right)^{2} - g_{1}A^{*2}_{1} \left(n^{*}_{1}\omega^{*}_{3} \right)^{2} - 2g_{1} \, m_{1}n^{*}_{1} \abs{A_{1}\omega_{3}}^{2} \right] \right . \nonumber \\ 
    && + \, (8 \, \widetilde{g}_{1}\widetilde{g}_{2})^{-1} \left[\widetilde{g}_{1}A^{*}_{2} \left(4\widetilde{g}_{2}-A^{*}_{2} \right) \left(n^{*}_{2}\omega^{*}_{4} \right)^{2} - \widetilde{g}_{1}A^{2}_{2} \left(m_{2}\omega_{4} \right)^{2} - 2\widetilde{g}_{1} \, m_{2}n^{*}_{2} \abs{A_{2}\omega_{4}}^{2} \right] \nonumber \\
    && \left . + \hspace{0.1 cm} (4 \, \gamma_{1}\gamma_{2})^{-1} \, \gamma_{2}A_{1} \left(A_{1}-2\gamma_{1} \right) \left(m_{1}\omega_{3} \right)^{2} \hspace{0.1 cm} + \hspace{0.1 cm} (4 \, \gamma_{3}\gamma_{4})^{-1} \, \gamma_{4}A^{*}_{2} \left(A^{*}_{2}-2\gamma_{3} \right) \left(n^{*}_{2}\omega^{*}_{4} \right)^{2} \right\} \, , \\ 
    M_{44} &=& \left\{(8 \, g_{1}g_{2})^{-1} \left[g_{1}A_{1} \left(4g_{2}-A_{1} \right) \left(n_{1}\omega_{3} \right)^{2} - g_{1}A^{*2}_{1} \left(m^{*}_{1}\omega^{*}_{3} \right)^{2} - 2g_{1} \, m^{*}_{1}n_{1} \abs{A_{1}\omega_{3}}^{2} \right] \right . \nonumber \\ 
    && + \, (8 \, \widetilde{g}_{1}\widetilde{g}_{2})^{-1} \left[\widetilde{g}_{1}A^{*}_{2} \left(4\widetilde{g}_{2}-A^{*}_{2} \right) \left(m^{*}_{2}\omega^{*}_{4} \right)^{2} - \widetilde{g}_{1}A^{2}_{2} \left(n_{2}\omega_{4} \right)^{2} - 2\widetilde{g}_{1} \, m^{*}_{2}n_{2} \abs{A_{2}\omega_{4}}^{2} \right] \nonumber \\
    && \left . + \hspace{0.1 cm} (4 \, \gamma_{1}\gamma_{2})^{-1} \, \gamma_{2}A_{1} \left(A_{1}-2\gamma_{1} \right) \left(n_{1}\omega_{3} \right)^{2} \hspace{0.1 cm} + \hspace{0.1 cm} (4 \, \gamma_{3}\gamma_{4})^{-1} \, \gamma_{4}A^{*}_{2} \left(A^{*}_{2}-2\gamma_{3} \right) \left(m^{*}_{2}\omega^{*}_{4} \right)^{2} \right\} \, , \\ 
    M_{34} &=& \left\{(8 \, g_{1}g_{2})^{-1} \left[g_{1}A_{1} \left(4g_{2}-A_{1} \right) m_{1}n_{1}\omega^{2}_{3} - g_{1} \, m^{*}_{1}n^{*}_{1} \left(A^{*}_{1}\omega^{*}_{3} \right)^{2} - g_{1} \abs{A_{1}\omega_{3}}^{2} \left(\abs{m_{1}}^{2} + \abs{n_{1}}^{2} \right) \right] \right . \nonumber \\ 
    && + \, (8 \, \widetilde{g}_{1}\widetilde{g}_{2})^{-1} \left[\widetilde{g}_{1}A^{*}_{2} \left(4\widetilde{g}_{2}-A^{*}_{2} \right) m^{*}_{2}n^{*}_{2}\omega^{*2}_{4} - \widetilde{g}_{1} \, m_{2}n_{2} \left(A_{2}\omega_{4} \right)^{2} - \widetilde{g}_{1} \abs{A_{2}\omega_{4}}^{2} \left(\abs{m_{2}}^{2} + \abs{n_{2}}^{2} \right) \right] \nonumber \\
    && \bigg . + \hspace{0.1 cm} (4 \, \gamma_{1}\gamma_{2})^{-1} \, \gamma_{2}A_{1} \left(A_{1}-2\gamma_{1} \right) m_{1}n_{1}\omega^{2}_{3} \hspace{0.1 cm} + \hspace{0.1 cm} (4 \, \gamma_{3}\gamma_{4})^{-1} \, \gamma_{4}A^{*}_{2} \left(A^{*}_{2}-2\gamma_{3} \right) m^{*}_{2}n^{*}_{2} \, \omega^{*2}_{4} \bigg\}  \, , \\ 
    M_{34} &=& M_{43} \, . 
\end{eqnarray}
Apart from the above-mentioned elements, all the remaining components of $M$-matrix vanish. 
\begin{eqnarray}
    M_{13} \hspace{0.1 cm} = \hspace{0.1 cm} 0 \hspace{0.3 cm} , \hspace{0.3 cm} M_{14} \hspace{0.1 cm} = \hspace{0.1 cm} 0 \hspace{0.3 cm} , \hspace{0.3 cm} M_{23} \hspace{0.1 cm} = \hspace{0.1 cm} 0 \hspace{0.3 cm} , \hspace{0.3 cm} M_{24} \hspace{0.1 cm} = \hspace{0.1 cm} 0 \, , \\ 
    M_{31} \hspace{0.1 cm} = \hspace{0.1 cm} 0 \hspace{0.3 cm} , \hspace{0.3 cm} M_{32} \hspace{0.1 cm} = \hspace{0.1 cm} 0 \hspace{0.3 cm} , \hspace{0.3 cm} M_{41} \hspace{0.1 cm} = \hspace{0.1 cm} 0 \hspace{0.3 cm} , \hspace{0.3 cm} M_{42} \hspace{0.1 cm} = \hspace{0.1 cm} 0 \, .  
\end{eqnarray}

Our next task is to provide explicit expressions for all the elements of the $K$-matrix, arrived in the third line of \ref{intermediate_4}. The $K$-matrix is constructed from those terms of the normalization factors $N_{1}$ and $N^{*}_{2}$ that explicitly depend on the complex parameters $\beta_{\mathbf{k}}$ and $\beta_{\mathbf{-k}}$. Note that, these normalization factors are chosen such that they must be associated with the respective wave functions $\psi_{r, \beta} \left(-q_{\mathbf{k_{1}}}, q_{-\mathbf{k_{2}}}, t_{1} \right)$ and $\psi^{*}_{r, \beta} \left(\widetilde{q}_{\mathbf{k_{1}}}, -\widetilde{q}_{-\mathbf{k_{2}}}, t_{2} \right)$, as given by  \ref{wave_func_integral_1} and \ref{wave_func_integral_2}. Mathematically, the $K$-matrix can be determined through the re-arrangement of the above-mentioned exponential terms present in the normalization factor:  
\begin{eqnarray}\label{K_construction}
    \exp\left(\widetilde{N}_{11} + \widetilde{N}^{*}_{22} \right) &=& \exp\bigg(\hspace{0.05 cm} K_{11} \, \delta^{2}_{\mathbf{k, R}} \, + \, K_{22} \, \delta^{2}_{\mathbf{k, I}} \, + \, K_{12} \, \delta_{\mathbf{k, R}} \, \delta_{\mathbf{k, I}} \, + \, K_{21} \, \delta_{\mathbf{k, I}} \, \delta_{\mathbf{k, R}} \bigg. \nonumber \\ 
    && \bigg. \hspace{0.45 cm} + \hspace{0.1 cm} K_{33} \, \psi^{2}_{\mathbf{k, R}} \, + \, K_{44} \, \psi^{2}_{\mathbf{k, I}} \, + \, K_{34} \, \psi_{\mathbf{k, R}} \, \psi_{\mathbf{k, I}} \, + \, K_{43} \, \psi_{\mathbf{k, I}} \, \psi_{\mathbf{k, R}} \hspace{0.05 cm} \bigg) \nonumber \\ 
    &=& \exp\bigg(\sigma^{T}_{\mathbf{k}} \, K \, \sigma_{\mathbf{k}} \bigg) \, . 
\end{eqnarray}
In the last line of \ref{K_construction}, the argument of the exponential is written in a more compact matrix notation, in terms of the $4D$ vector $\sigma_{\mathbf{k}}$ 
and the matrix $K$. The vector $\sigma_{\mathbf{k}}$ is defined as follows, 
\begin{eqnarray}
    \sigma^{T}_{\mathbf{k}} &=& 
    \begin{pmatrix}
        \hspace{0.06 cm} \delta_{\mathbf{k, R}} & \delta_{\mathbf{k, I}} & \psi_{\mathbf{k, R}} & \psi_{\mathbf{k, I}} \hspace{0.06 cm} 
    \end{pmatrix} \, . 
\end{eqnarray}

The elements of the K-matrix, associated with the terms which are quadratic in $\delta_{\mathbf{k}, \mathbf{R}}$ and $\delta_{\mathbf{k}, \mathbf{I}}$, are given below: 
\begin{eqnarray}
    K_{11} &=& \left\{- \, i \, \Im(2A_{1}-B_{1}) \left[ \, \Re(p_{1}-q_{1}) \, \right]^{2} - \, i \, \Re(p_{1}-q_{1}) \Im(p_{1}-q_{1}) \right. \nonumber \\ 
    && \left. + \, i \, \Im(2A_{2}-B_{2}) \left[ \, \Re(p_{2}-q_{2}) \, \right]^{2} + \, i \, \Re(p_{2}-q_{2}) \Im(p_{2}-q_{2}) \right\} \, , \\ 
    K_{22} &=& \left\{- \, i \, \Im(2A_{1}-B_{1}) \left[ \, \Im(p_{1}+q_{1}) \, \right]^{2} + \, i \, \Re(p_{1}+q_{1})  \Im(p_{1}+q_{1}) \right. \nonumber \\
    && \left. + \, i \, \Im(2A_{2}-B_{2}) \left[ \, \Im(p_{2}+q_{2}) \, \right]^{2} - \, i \, \Re(p_{2}+q_{2}) \Im(p_{2}+q_{2}) \right\} , \\ 
    K_{12} &=& \frac{1}{2} \bigg\{2i \Im(2A_{1}-B_{1}) \Re(p_{1}-q_{1}) \Im(p_{1}+q_{1}) + \, i \, \left[ \, \Im(p_{1}-q_{1})\Im(p_{1}+q_{1}) - \Re(p_{1}-q_{1})\Re(p_{1}+q_{1}) \, \right] \bigg. \nonumber \\ 
    && \bigg. - \, 2i \Im(2A_{2}-B_{2}) \Re(p_{2}-q_{2}) \Im(p_{2}+q_{2}) - \, i \, \left[ \, \Im(p_{2}-q_{2})\Im(p_{2}+q_{2}) - \Re(p_{2}-q_{2})\Re(p_{2}+q_{2}) \, \right] \bigg\} , \nonumber \\ 
    K_{21} &=& K_{12} \, . 
\end{eqnarray}
Likewise, the remaining elements of the $K$-matrix, being the coefficients of the terms which are quadratic in $\psi_{\mathbf{k}, \mathbf{R}}$ and $\psi_{\mathbf{k}, \mathbf{I}}$, are given below: 
\vspace{-0.2 cm}
\begin{eqnarray}
    K_{33} &=& \left\{- \, i \, \Im(2A_{1}+B_{1}) \left[ \, \Re(p_{1}+q_{1}) \, \right]^{2} - \, i \, \Re(p_{1}+q_{1})\Im(p_{1}+q_{1}) \right. \nonumber \\
    && \left. + \, i \, \Im(2A_{2}+B_{2}) \left[ \, \Re(p_{2}+q_{2}) \, \right]^{2} + \, i \, \Re(p_{2}+q_{2})\Im(p_{2}+q_{2}) \right\} , \\ 
    K_{44} &=& \left\{- \, i \Im(2A_{1}+B_{1}) \left[ \, \Im(p_{1}-q_{1}) \, \right]^{2} + \, i \, \Re(p_{1}-q_{1})\Im(p_{1}-q_{1}) \right. \nonumber \\
    && \left. + \, i \Im(2A_{2}+B_{2}) \left[ \, \Im(p_{2}-q_{2}) \, \right]^{2} - \, i \, \Re(p_{2}-q_{2})\Im(p_{2}-q_{2}) \right\} , \\ 
    K_{34} &=& \frac{1}{2} \bigg\{2i \, \Im(2A_{1}+B_{1}) \Re(p_{1}+q_{1})\Im(p_{1}-q_{1}) + \, i \, \left[ \, \Im(p_{1}-q_{1})\Im(p_{1}+q_{1}) - \Re(p_{1}-q_{1})\Re(p_{1}+q_{1}) \, \right] \bigg. \nonumber \\ 
    && \bigg. - \, 2i \, \Im(2A_{2}+B_{2}) \Re(p_{2}+q_{2})\Im(p_{2}-q_{2}) - \, i \, \left[ \, \Im(p_{2}-q_{2})\Im(p_{2}+q_{2}) - \Re(p_{2}-q_{2})\Re(p_{2}+q_{2}) \, \right] \bigg\} , \nonumber \\ 
    K_{43} &=& K_{34} \, . 
\end{eqnarray}
Likewise, all the remaining elements of $K$-matrix are found to be exactly zero. 
\begin{eqnarray}
    K_{13} \hspace{0.1 cm} = \hspace{0.1 cm} 0 \hspace{0.3 cm} , \hspace{0.3 cm} K_{14} \hspace{0.1 cm} = \hspace{0.1 cm} 0 \hspace{0.3 cm} , \hspace{0.3 cm} K_{23} \hspace{0.1 cm} = \hspace{0.1 cm} 0 \hspace{0.3 cm} , \hspace{0.3 cm} K_{24} \hspace{0.1 cm} = \hspace{0.1 cm} 0 \, , \\ 
    K_{31} \hspace{0.1 cm} = \hspace{0.1 cm} 0 \hspace{0.3 cm} , \hspace{0.3 cm} K_{32} \hspace{0.1 cm} = \hspace{0.1 cm} 0 \hspace{0.3 cm} , \hspace{0.3 cm} K_{41} \hspace{0.1 cm} = \hspace{0.1 cm} 0 \hspace{0.3 cm} , \hspace{0.3 cm} K_{42} \hspace{0.1 cm} = \hspace{0.1 cm} 0 \, .  
\end{eqnarray}

Next, we are going to concentrate on the elements of $\widetilde{T}$-matrix, appeared in the third line of \ref{intermediate_4}. Similar to the case of $K$-matrix, the $\widetilde{T}$-matrix originates from those terms of the normalization factors $N^{*}_{3}$ and $N_{4}$ that explicitly depend on the complex parameters $\alpha_{\mathbf{k}}$ and $\alpha_{\mathbf{-k}}$. Again the normalization factors are chosen such that they must be associated with the respective wave functions $\psi_{r, \alpha} \left(q_{\mathbf{k_{1}}}, q_{-\mathbf{k_{2}}}, t_{1} \right)$ and $\psi^{*}_{r, \alpha} \left(\widetilde{q}_{\mathbf{k_{1}}}, \widetilde{q}_{-\mathbf{k_{2}}}, t_{2} \right)$, as given by \ref{wave_func_integral_3} and \ref{wave_func_integral_4}. The primary difference between the matrices $K$ and $\widetilde{T} \, -$ lies in the variables whose coefficients correspond to the respective matrix elements. The elements of the $K$-matrix correspond to the coefficients of those terms quadratic in the four-dimensional vector $\mathbf{\sigma_{k}}$, whereas the elements of the $\widetilde{T}$-matrix correspond to the coefficients of those terms quadratic in the four-dimensional vector $\lambda_{\mathbf{k}}$. 

Similar to the $K$-matrix, the $\widetilde{T}$-matrix can be also determined through the re-arrangement of the above-mentioned exponential terms present in the normalization factors: 
\begin{eqnarray}\label{T_construction}
    \exp\left(\widetilde{N}^{*}_{33} + \widetilde{N}_{44} \right) &=& \exp\bigg(\hspace{0.05 cm} \widetilde{T}_{11} \, \chi^{2}_{\mathbf{k, R}} + \widetilde{T}_{22} \, \chi^{2}_{\mathbf{k, I}} + \widetilde{T}_{12} \, \chi_{\mathbf{k, R}} \, \chi_{\mathbf{k, I}} + \widetilde{T}_{21} \, \chi_{\mathbf{k, I}} \, \chi_{\mathbf{k, R}} \bigg. \nonumber \\ 
    && \bigg. \hspace{0.45 cm} + \hspace{0.1 cm} \widetilde{T}_{33} \, \Xi^{2}_{\mathbf{k, R}} + \widetilde{T}_{44} \, \Xi^{2}_{\mathbf{k, I}} + \widetilde{T}_{34} \, \Xi_{\mathbf{k, R}} \, \Xi_{\mathbf{k, I}} + \widetilde{T}_{43} \, \Xi_{\mathbf{k, I}} \, \Xi_{\mathbf{k, R}} \hspace{0.05 cm} \bigg) \nonumber \\ 
    &=& \exp\left(\lambda^{T}_{\mathbf{k}} \, \widetilde{T} \, \lambda_{\mathbf{k}} \right) \, . 
\end{eqnarray}
In the last line of \ref{T_construction}, the argument of the exponential is written in a more compact matrix form, in terms of the $4D$ vector $\lambda_{\mathbf{k}}$ 
and the matrix $\widetilde{T}$. The vector $\lambda_{\mathbf{k}}$ is defined as follows, 
\begin{eqnarray}
    \lambda^{T}_{\mathbf{k}} &=& 
    \begin{pmatrix}
        \hspace{0.06 cm} \chi_{\mathbf{k, R}} & \chi_{\mathbf{k, I}} &\Xi_{\mathbf{k, R}} & \Xi_{\mathbf{k, I}} \hspace{0.06 cm}
    \end{pmatrix} \, . 
\end{eqnarray}

A careful inspection shows that all the elements of the $\widetilde{T}$-matrix are simply the complex conjugate of the corresponding elements of $K$-matrix. The individual elements are written as follows, 
\begin{eqnarray}
    \widetilde{T}_{11} &=& K^{*}_{11} \hspace{0.4 cm} , \hspace{0.4 cm} \widetilde{T}_{22} \hspace{0.1 cm} = \hspace{0.1 cm} K^{*}_{22} \hspace{0.4 cm} , \hspace{0.4 cm} \widetilde{T}_{12} \hspace{0.1 cm} = \hspace{0.1 cm} K^{*}_{12} \hspace{0.4 cm} , \hspace{0.4 cm} \widetilde{T}_{21} \hspace{0.1 cm} = \hspace{0.1 cm} K^{*}_{21} \, , \\ 
    \widetilde{T}_{33} &=& K^{*}_{33} \hspace{0.4 cm} , \hspace{0.4 cm} \widetilde{T}_{44} \hspace{0.1 cm} = \hspace{0.1 cm} K^{*}_{44} \hspace{0.4 cm} , \hspace{0.4 cm} \widetilde{T}_{34} \hspace{0.1 cm} = \hspace{0.1 cm} K^{*}_{34} \hspace{0.4 cm} , \hspace{0.4 cm} \widetilde{T}_{43} \hspace{0.1 cm} = \hspace{0.1 cm} K^{*}_{43} \, . 
\end{eqnarray}
Following the steps of $K$-matrix, all the remaining elements of $\widetilde{T}$-matrix are found to be exactly zero. 
\begin{eqnarray}
    \widetilde{T}_{13} \hspace{0.1 cm} = \hspace{0.1 cm} 0 \hspace{0.3 cm} , \hspace{0.3 cm} \widetilde{T}_{14} \hspace{0.1 cm} = \hspace{0.1 cm} 0 \hspace{0.3 cm} , \hspace{0.3 cm} \widetilde{T}_{23} \hspace{0.1 cm} = \hspace{0.1 cm} 0 \hspace{0.3 cm} , \hspace{0.3 cm} \widetilde{T}_{24} \hspace{0.1 cm} = \hspace{0.1 cm} 0 \, , \\ 
    \widetilde{T}_{31} \hspace{0.1 cm} = \hspace{0.1 cm} 0 \hspace{0.3 cm} , \hspace{0.3 cm} \widetilde{T}_{32} \hspace{0.1 cm} = \hspace{0.1 cm} 0 \hspace{0.3 cm} , \hspace{0.3 cm} \widetilde{T}_{41} \hspace{0.1 cm} = \hspace{0.1 cm} 0 \hspace{0.3 cm} , \hspace{0.3 cm} \widetilde{T}_{42} \hspace{0.1 cm} = \hspace{0.1 cm} 0 \, .  
\end{eqnarray}

Now we proceed towards the final destination of the section. The $N$-matrix, appearing in the third line of \ref{intermediate_4}, plays a crucial role in the evaluation of the complex Gaussian integrals over the four-dimensional vector $\mathbf{\sigma_{k}}$. Similar to the $M$-matrix, the elements of the $N$-matrix are constructed from the integrand of the complex-variable integral over the parameters $(\beta_{\mathbf{k}}, \beta_{\mathbf{-k}})$. This construction is achieved by re-organizing the integrand in terms of the complex variables $\delta_{\mathbf{k, R}}, \delta_{\mathbf{k, I}}, \psi_{\mathbf{k, R}}$ and $\psi_{\mathbf{k, I}}$, as defined in \ref{coordinate_5}. Mathematically, this can be expressed as follows: 
\begin{eqnarray}\label{N_construction}
    \exp\left(\kappa_{1} + \kappa_{11} + \overline{N}_{11} + \overline{N}^{*}_{22} \right) &=& \exp\bigg(N_{11} \hspace{0.1 cm} \delta^{2}_{\mathbf{k, R}} \, + \, N_{22} \hspace{0.1 cm} \delta^{2}_{\mathbf{k, I}} \, + \, N_{12} \hspace{0.1 cm} \delta_{\mathbf{k, R}}\delta_{\mathbf{k, I}} \, + \, N_{21} \hspace{0.1 cm} \delta_{\mathbf{k, I}}\delta_{\mathbf{k, R}} \, + \, Q_{1} \hspace{0.05 cm} \delta_{\mathbf{k, R}} \, + \, Q_{2} \hspace{0.05 cm} \delta_{\mathbf{k, I}} \bigg. \nonumber \\ 
    && \bigg. + \, N_{33} \hspace{0.1 cm} \psi^{2}_{\mathbf{k, R}} \, + \, N_{44} \hspace{0.1 cm} \psi^{2}_{\mathbf{k, I}} \, + \, N_{34} \hspace{0.1 cm} \psi_{\mathbf{k, R}}\psi_{\mathbf{k, I}} \, + \, N_{43} \hspace{0.1 cm} \psi_{\mathbf{k, I}}\psi_{\mathbf{k, R}} \, + \, Q_{3} \hspace{0.05 cm} \psi_{\mathbf{k, R}} \, + \, Q_{4} \hspace{0.05 cm} \psi_{\mathbf{k, I}} \bigg) \nonumber \\ 
    &=& \exp\bigg(\sigma^{T}_{\mathbf{k}} \, N \, \sigma_{\mathbf{k}} \, + \, Q^{T} \sigma_{\mathbf{k}} \bigg) \, . 
\end{eqnarray}
In the last line of \ref{N_construction}, the argument of the exponential is written in a more compact matrix notation, involving the matrix $N$ and the four dimensional vectors $\sigma_{\mathbf{k}}$ and $Q \, -$ respectively. \\ 
The elements of $N$-matrix associated with the quadratic terms in $\delta_{\mathbf{k, R}}$ and $\delta_{\mathbf{k, I}}$, are given below: 
\begin{eqnarray}
    N_{11} &=& \frac{1}{2} \left(M_{11} + M_{22} + 2 \, M_{12} \right) \hspace{0.6 cm} , \hspace{0.6 cm} N_{12} \hspace{0.2 cm} = \hspace{0.2 cm} \frac{1}{2} \, i \left(M_{11} - M_{22} \right) \, , \\
    N_{22} &=& \frac{1}{2} \left(2 \, M_{12} - M_{11} - M_{22} \right) \hspace{0.6 cm} , \hspace{0.6 cm} N_{21} \hspace{0.2 cm} = \hspace{0.2 cm} \frac{1}{2} \, i \left(M_{11} - M_{22} \right) \, . 
\end{eqnarray}
Likewise, the elements of $N$-matrix associated with the quadratic terms of $\Psi_{\mathbf{k, R}}$ and $\Psi_{\mathbf{k, I}}$, are given below: 
\begin{eqnarray}
    N_{33} &=& \frac{1}{2} \left(M_{33} + M_{44} + 2 \, M_{34} \right) \hspace{0.6 cm} , \hspace{0.6 cm} N_{34} \hspace{0.2 cm} = \hspace{0.2 cm} \frac{1}{2} \, i \left(M_{33} - M_{44} \right) \, , \\ 
    N_{44} &=& \frac{1}{2} \left(2 \, M_{34} - M_{33} - M_{44} \right) \hspace{0.6 cm} , \hspace{0.6 cm} N_{43} \hspace{0.2 cm} = \hspace{0.2 cm} \frac{1}{2} \, i \left(M_{33} - M_{44} \right) \, . 
\end{eqnarray}
Apart from the above-mentioned elements, all the remaining components of the $(4 \cross 4)$ $N$-matrix are equal to zero. 
\begin{eqnarray}
    N_{13} \hspace{0.1 cm} = \hspace{0.1 cm} 0 \hspace{0.3 cm} , \hspace{0.3 cm} N_{14} \hspace{0.1 cm} = \hspace{0.1 cm} 0 \hspace{0.3 cm} , \hspace{0.3 cm} N_{23} \hspace{0.1 cm} = \hspace{0.1 cm} 0 \hspace{0.3 cm} , \hspace{0.3 cm} N_{24} \hspace{0.1 cm} = \hspace{0.1 cm} 0 \, , \\ 
    N_{31} \hspace{0.1 cm} = \hspace{0.1 cm} 0 \hspace{0.3 cm} , \hspace{0.3 cm} N_{32} \hspace{0.1 cm} = \hspace{0.1 cm} 0 \hspace{0.3 cm} , \hspace{0.3 cm} N_{41} \hspace{0.1 cm} = \hspace{0.1 cm} 0 \hspace{0.3 cm} , \hspace{0.3 cm} N_{42} \hspace{0.1 cm} = \hspace{0.1 cm} 0 \, .  
\end{eqnarray}
In matrix notation, the $M$ and $N$-matrix can be expressed as follows: 
\begin{eqnarray}
    M &=& 
    \begin{pmatrix}
        M_{11} & M_{12} & M_{13} & M_{14} \\ 
        M_{21} & M_{22} & M_{23} & M_{24} \\ 
        M_{31} & M_{32} & M_{33} & M_{34} \\ 
        N_{41} & M_{42} & M_{43} & M_{44} \\ 
    \end{pmatrix} 
    \hspace{0.8 cm} , \hspace{0.8 cm} 
    N \hspace{0.2 cm} = \hspace{0.2 cm}  
    \begin{pmatrix}
        N_{11} & N_{12} & N_{13} & N_{14} \\ 
        N_{21} & N_{22} & N_{23} & N_{24} \\ 
        N_{31} & N_{32} & N_{33} & N_{34} \\ 
        N_{41} & N_{42} & N_{43} & N_{44} \\ 
    \end{pmatrix} \, . 
\end{eqnarray}
It is important to note that the basis vectors used in the matrix representations of the two matrices are fundamentally different. For the $M$-matrix, the basis vector is given by $\, \Lambda^{T}_{\mathbf{k}} = \left(\delta_{\mathbf{k}} \hspace{0.15 cm} \delta^{*}_{\mathbf{k}} \hspace{0.15 cm} \Psi_{\mathbf{k}} \hspace{0.15 cm} \Psi^{*}_{\mathbf{k}} \right)$, whereas for the $N$-matrix, the basis vector is defined as $\, \sigma^{T}_{\mathbf{k}} = \left(\delta_{\mathbf{k, R}} \hspace{0.15 cm} \delta_{\mathbf{k, I}} \hspace{0.15 cm} \Psi_{\mathbf{k, R}} \hspace{0.15 cm} \Psi_{\mathbf{k, I}} \right)$. 

Similarly, in the matrix notation, the $K$ and $\widetilde{T}$-matrix can be expressed as follows: 
\begin{eqnarray}
    K &=& 
    \begin{pmatrix}
        K_{11} & K_{12} & K_{13} & K_{14} \\ 
        K_{21} & K_{22} & K_{23} & K_{24} \\ 
        K_{31} & K_{32} & K_{33} & K_{34} \\ 
        K_{41} & K_{42} & K_{43} & K_{44} 
    \end{pmatrix}
    \hspace{0.6 cm} , \hspace{0.6 cm} 
    \widetilde{T} \hspace{0.2 cm} = \hspace{0.2 cm} 
    \begin{pmatrix}
        \widetilde{T}_{11} & \widetilde{T}_{12} & \widetilde{T}_{13} & \widetilde{T}_{14} \\ 
        \widetilde{T}_{21} & \widetilde{T}_{22} & \widetilde{T}_{23} & \widetilde{T}_{24} \\ 
        \widetilde{T}_{31} & \widetilde{T}_{32} & \widetilde{T}_{33} & \widetilde{T}_{34} \\ 
        \widetilde{T}_{41} & \widetilde{T}_{42} & \widetilde{T}_{43} & \widetilde{T}_{44} 
    \end{pmatrix} \, . 
\end{eqnarray}
It is important to note that the basis vectors used in the matrix representation of the above mentioned two matrices, are fundamentally different. For the $K$-matrix, the basis vector is given be $\sigma^{T}_{\mathbf{k}} = \left(\delta_{\mathbf{k, R}} \hspace{0.15 cm} \delta_{\mathbf{k, I}} \hspace{0.15 cm} \Psi_{\mathbf{k, R}} \hspace{0.15 cm} \Psi_{\mathbf{k, I}} \right)$, whereas for the $T$-matrix, the basis vector is defined as follows $\lambda^{T}_{\mathbf{k}} = \left(\chi_{\mathbf{k, R}} \hspace{0.15 cm} \chi_{\mathbf{k, I}} \hspace{0.15 cm} \Xi_{\mathbf{k, R}} \hspace{0.15 cm} \Xi_{\mathbf{k, I}} \right)$.

\subsection{Explicit expression of $R$ and $Q \, -$ vector}

In the section, our primary focus is to provide explicit expressions for all the elements of $Q$-matrix, appearing in the third line of \ref{intermediate_4}. To simplify the expressions, we introduce a new $(4 \cross 1)$ matrix $R$, in terms of which all the elements of the $Q$-matrix can be conveniently expressed. In matrix notation, the $R$-matrix can be written as, 
\begin{eqnarray}
    R^{T} &=& 
    \begin{pmatrix}
        R_{1} & R_{2} & R_{3} & R_{4} 
    \end{pmatrix} \, .
\end{eqnarray}

The $R$-matrix is primarily constructed from the integrand of the complex-variable integral over the parameters $(\beta_{\mathbf{k}}, \beta_{\mathbf{-k}})$, as given in \ref{spin_correlation_3}. According to \ref{M_construction}, the elements of $R$-matrix are the coefficients of those terms which are linear in $\delta_{\mathbf{k}}, \delta^{*}_{\mathbf{k}}, \Psi_{\mathbf{k}}$ and $\Psi^{*}_{\mathbf{k}}$. Explicit expressions for all the components of the $R$-matrix are given below:  
\begin{eqnarray}
    R_{1} &=& \frac{1}{2 \, \gamma_{2}\gamma_{4}} \left(- A_{1}\gamma_{4}\Sigma_{4} \, m_{1}\omega_{1} - A^{*}_{2}\gamma_{2}\Sigma_{7} \, n^{*}_{2}\omega^{*}_{2} \, \right) \, , \\
    R_{2} &=& \frac{1}{2 \, \gamma_{2}\gamma_{4}} \left(A_{1}\gamma_{4}\Sigma_{4} \, n_{1}\omega_{1} + A^{*}_{2}\gamma_{2}\Sigma_{7} \, m^{*}_{2}\omega^{*}_{2} \, \right) \, , \\
    R_{3} &=& \frac{1}{2 \, \gamma_{1}\gamma_{3}} \left(-A_{1}\gamma_{3}\Sigma_{3} \, m_{1}\omega_{3} + A^{*}_{2}\gamma_{1}\Sigma_{6} \, n^{*}_{2}\omega^{*}_{4} \right) \, , \\
    R_{4} &=& \frac{1}{2 \, \gamma_{1}\gamma_{3}} \left(-A_{1}\gamma_{3}\Sigma_{3} \, n_{1}\omega_{3} + A^{*}_{2}\gamma_{1}\Sigma_{6} \, m^{*}_{2}\omega^{*}_{4} \, \right) \, . 
\end{eqnarray}
Once the elements of the $R$-matrix has been found, the remaining task is to express the components of $Q$-matrix as a linear combination to those of $R$-matrix. In matrix notation, the $Q$-matrix can be expressed as, 
\begin{eqnarray}
    Q^{T} &=& 
    \begin{pmatrix}
        Q_{1} & Q_{2} & Q_{3} & Q_{4} 
    \end{pmatrix} \, . 
\end{eqnarray}

Similar to $R$-matrix, the $Q$-matrix is also constructed from the same complex-variable integral over the parameters $(\beta_{\mathbf{k}}, \beta_{\mathbf{-k}})$, as given in \ref{spin_correlation_3}. According to \ref{N_construction}, the elements of $Q$-matrix are the coefficients of those terms which are linear in $\delta_{\mathbf{k, R}}, \delta_{\mathbf{k, I}}, \Psi_{\mathbf{k, R}}$ and $\Psi_{\mathbf{k, I}}$. Explicit expressions for the components of $Q$-matrix are given below: 
\begin{eqnarray}
    Q_{1} &=& \frac{1}{\sqrt{2}} \left(R_{1} + R_{2} \right) \hspace{0.6 cm} , \hspace{0.6 cm} Q_{2} \hspace{0.2 cm} = \hspace{0.2 cm} \frac{i}{\sqrt{2}} \left(R_{1} - R_{2} \right) \, , \\ 
    Q_{3} &=& \frac{1}{\sqrt{2}} \left(R_{3} + R_{4} \right) \hspace{0.6 cm} , \hspace{0.6 cm} Q_{4} \hspace{0.2 cm} = \hspace{0.2 cm} \frac{i}{\sqrt{2}} \left(R_{3} - R_{4} \right) \, . 
\end{eqnarray}


\printbibliography

@article{PhysRevD.23.347,
  title = {Inflationary universe: A possible solution to the horizon and flatness problems},
  author = {Guth, Alan H.},
  journal = {Phys. Rev. D},
  volume = {23},
  issue = {2},
  pages = {347--356},
  numpages = {0},
  year = {1981},
  month = {Jan},
  publisher = {American Physical Society},
  doi = {10.1103/PhysRevD.23.347},
  url = {https://link.aps.org/doi/10.1103/PhysRevD.23.347}
}

@article{Linde:1981mu,
    author = "Linde, Andrei D.",
    editor = "Fang, Li-Zhi and Ruffini, R.",
    title = "{A New Inflationary Universe Scenario: A Possible Solution of the Horizon, Flatness, Homogeneity, Isotropy and Primordial Monopole Problems}",
    reportNumber = "LEBEDEV-81-229",
    doi = {10.1016/0370-2693(82)91219-9}, 
    url = {https://www.sciencedirect.com/science/article/abs/pii/0370269382912199?via%3Dihub},
    journal = "Phys. Lett. B",
    volume = "108",
    pages = "389--393",
    year = "1982"
}

@article{Martin:2018ycu,
    author = "Martin, Jerome",
    editor = "Coccia, E. and Silk, J. and Vittorio, N.",
    title = "{The Theory of Inflation}",
    eprint = "1807.11075",
    archivePrefix = "arXiv",
    primaryClass = "astro-ph.CO",
    doi = "10.3254/ENFI200008",
    journal = "Proc. Int. Sch. Phys. Fermi",
    volume = "200",
    pages = "155--178",
    year = "2020"
}

@inproceedings{Baumann:2009ds,
    author = "Baumann, Daniel",
    title = "{Inflation}",
    booktitle = "{Theoretical Advanced Study Institute in Elementary Particle Physics}: {Physics of the Large and the Small}",
    eprint = "0907.5424",
    archivePrefix = "arXiv",
    primaryClass = "hep-th",
    reportNumber = "TASI-2009",
    doi = "10.1142/9789814327183_0010",
    pages = "523--686",
    year = "2011"
}

@article{Sriramkumar:2009kg,
    author = "Sriramkumar, L.",
    title = "{An introduction to inflation and cosmological perturbation theory}",
    eprint = "0904.4584",
    archivePrefix = "arXiv",
    primaryClass = "astro-ph.CO",
    journal = "Curr. Sci.",
    volume = "97",
    pages = "868",
    year = "2009"
}

@inproceedings{Brandenberger:1993zc,
    author = "Brandenberger, Robert H. and Feldman, H. and Mukhanov, Viatcheslav F.",
    title = "{Classical and quantum theory of perturbations in inflationary universe models}",
    booktitle = "{37th Yamada Conference: Evolution of the Universe and its Observational Quest}",
    eprint = "9307016",
    archivePrefix = "arXiv",
    primaryClass = "astro-ph", 
    reportNumber = "BROWN-HET-914",
    pages = "19--30",
    month = "7",
    year = "1993"
}

@article{Mukhanov:1990me,
    author = "Mukhanov, Viatcheslav F. and Feldman, H. A. and Brandenberger, Robert H.",
    title = "{Theory of cosmological perturbations. Part 1. Classical perturbations. Part 2. Quantum theory of perturbations. Part 3. Extensions}",
    reportNumber = "BROWN-HET-796, BROWN-HET-800, BROWN-HET-780",
    doi = "10.1016/0370-1573(92)90044-Z",
    journal = "Phys. Rept.",
    volume = "215",
    pages = "203--333",
    year = "1992"
}

@article{Martin:2004um,
    author = "Martin, Jerome",
    editor = "Amelino-Camelia, G. and Kowalski-Glikman, J.",
    title = "{Inflationary cosmological perturbations of quantum-mechanical origin}",
    eprint = "0406011",
    archivePrefix = "arXiv",
    primaryClass = "hep-th",
    doi = "10.1007/11377306_7",
    journal = "Lect. Notes Phys.",
    volume = "669",
    pages = "199--244",
    year = "2005"
}

@article{riotto2002inflation,
    title={Inflation and the theory of cosmological perturbations},
    author={Riotto, Antonio},
    eprint = "0210162", 
    archivePrefix = "arXiv", 
    primaryClass = "hep-ph", 
%   journal={arXiv preprint hep-ph/0210162},
    year={2002}
}

@article{kurki2005cosmological,
  title={Cosmological perturbation theory - part 1, 2},
  author={Kurki-Suonio, Hannu},
  year={2005},
  url = {https://www.mv.helsinki.fi/home/hkurkisu/cpt/}
}

@inproceedings{KurkiSuonio2015CosmologicalPT,
  title={Cosmological Perturbation Theory - part 2},
  author={Hannu Kurki-Suonio},
  year={2015},
  url={https://api.semanticscholar.org/CorpusID:125588012}
}

@article{Albrecht:1992kf,
    author = "Albrecht, Andreas and Ferreira, Pedro and Joyce, Michael and Prokopec, Tomislav",
    title = "{Inflation and squeezed quantum states}",
    eprint = "astro-ph/9303001",
    archivePrefix = "arXiv",
    reportNumber = "IMPERIAL-TP-92-93-21",
    doi = "10.1103/PhysRevD.50.4807",
    journal = "Phys. Rev. D",
    volume = "50",
    pages = "4807--4820",
    year = "1994"
}

@article{Martin:2012ua,
    author = "Martin, Jerome",
    editor = "Pal, Supratik and Basu, Banasri",
    title = "{The Quantum State of Inflationary Perturbations}",
    eprint = "1209.3092",
    archivePrefix = "arXiv",
    primaryClass = "hep-th",
    doi = "10.1088/1742-6596/405/1/012004",
    journal = "J. Phys. Conf. Ser.",
    volume = "405",
    pages = "012004",
    year = "2012"
}

@article{Martin:2015qta,
    author = "Martin, Jerome and Vennin, Vincent",
    title = "{Quantum Discord of Cosmic Inflation: Can we Show that CMB Anisotropies are of Quantum-Mechanical Origin?}",
    eprint = "1510.04038",
    archivePrefix = "arXiv",
    primaryClass = "astro-ph.CO",
    doi = "10.1103/PhysRevD.93.023505",
    journal = "Phys. Rev. D",
    volume = "93",
    number = "2",
    pages = "023505",
    year = "2016"
}

@article{Brahma:2020zpk,
    author = "Brahma, Suddhasattwa and Alaryani, Omar and Brandenberger, Robert",
    title = "{Entanglement entropy of cosmological perturbations}",
    eprint = "2005.09688",
    archivePrefix = "arXiv",
    primaryClass = "hep-th",
    doi = "10.1103/PhysRevD.102.043529",
    journal = "Phys. Rev. D",
    volume = "102",
    number = "4",
    pages = "043529",
    year = "2020"
}

@article{Bhattacharyya:2022rhm,
    author = "Bhattacharyya, Arpan and Hanif, Tanvir and Haque, S. Shajidul and Paul, Arpon",
    title = "{Decoherence, entanglement negativity, and circuit complexity for an open quantum system}",
    eprint = "2210.09268",
    archivePrefix = "arXiv",
    primaryClass = "hep-th",
    doi = "10.1103/PhysRevD.107.106007",
    journal = "Phys. Rev. D",
    volume = "107",
    number = "10",
    pages = "106007",
    year = "2023"
}

@article{Martin:2016tbd,
    author = "Martin, J{\'e}rome and Vennin, Vincent",
    title = "{Bell inequalities for continuous-variable systems in generic squeezed states}",
    eprint = "1605.02944",
    archivePrefix = "arXiv",
    primaryClass = "quant-ph",
    doi = "10.1103/PhysRevA.93.062117",
    journal = "Phys. Rev. A",
    volume = "93",
    number = "6",
    pages = "062117",
    year = "2016"
}

@article{Martin:2017zxs,
    author = "Martin, Jerome and Vennin, Vincent",
    title = "{Obstructions to Bell CMB Experiments}",
    eprint = "1706.05001",
    archivePrefix = "arXiv",
    primaryClass = "astro-ph.CO",
    doi = "10.1103/PhysRevD.96.063501",
    journal = "Phys. Rev. D",
    volume = "96",
    number = "6",
    pages = "063501",
    year = "2017"
}

@article{Martin:2019wta,
    author = "Martin, J{\'e}r{\^o}me",
    title = "{Cosmic Inflation, Quantum Information and the Pioneering Role of John S Bell in Cosmology}",
    eprint = "1904.00083",
    archivePrefix = "arXiv",
    primaryClass = "quant-ph",
    doi = "10.3390/universe5040092",
    journal = "Universe",
    volume = "5",
    number = "4",
    pages = "92",
    year = "2019"
}

@article{Dale:2023fnp,
    author = "Dale, Roberto and Lapiedra, Ramon and Morales-Lladosa, Juan Antonio",
    title = "{Cosmic primordial density fluctuations and Bell inequalities}",
    eprint = "2302.05125",
    archivePrefix = "arXiv",
    primaryClass = "gr-qc",
    doi = "10.1103/PhysRevD.107.023506",
    journal = "Phys. Rev. D",
    volume = "107",
    number = "2",
    pages = "023506",
    year = "2023"
}

@article{Dale:2025nhc,
    author = "Dale, Roberto and Gand{\'\i}a, Jes{\'u}s M. and Morales-Lladosa, Juan Antonio and Lapiedra, Ramon",
    title = "{Violation of Bell inequalities from Cosmic Microwave Background data}",
    eprint = "2502.13846",
    archivePrefix = "arXiv",
    primaryClass = "gr-qc",
    doi = "10.1088/1475-7516/2025/07/044",
    journal = "JCAP",
    volume = "07",
    pages = "044",
    year = "2025"
}

@article{Kanno:2017dci,
    author = "Kanno, Sugumi and Soda, Jiro",
    title = "{Infinite violation of Bell inequalities in inflation}",
    eprint = "1705.06199",
    archivePrefix = "arXiv",
    primaryClass = "hep-th",
    reportNumber = "KOBE-COSMO-17-08, KOBE-COSMO-17-08",
    doi = "10.1103/PhysRevD.96.083501",
    journal = "Phys. Rev. D",
    volume = "96",
    number = "8",
    pages = "083501",
    year = "2017"
}

@article{Choudhury:2016cso,
    author = "Choudhury, Sayantan and Panda, Sudhakar and Singh, Rajeev",
    title = "{Bell violation in the Sky}",
    eprint = "1607.00237",
    archivePrefix = "arXiv",
    primaryClass = "hep-th",
    reportNumber = "TIFR/TH/16-19, TIFR-TH-16-19",
    doi = "10.1140/epjc/s10052-016-4553-3",
    journal = "Eur. Phys. J. C",
    volume = "77",
    pages = "60",
    year = "2017"
}

@article{PhysRev.47.777,
  title = {Can Quantum-Mechanical Description of Physical Reality Be Considered Complete?},
  author = {Einstein, A. and Podolsky, B. and Rosen, N.},
  journal = {Phys. Rev.},
  volume = {47},
  issue = {10},
  pages = {777--780},
  numpages = {0},
  year = {1935},
  month = {May},
  publisher = {American Physical Society},
  doi = {10.1103/PhysRev.47.777},
  url = {https://link.aps.org/doi/10.1103/PhysRev.47.777}
}

@article{PhysicsPhysiqueFizika.1.195,
  title = {On the Einstein Podolsky Rosen paradox},
  author = {Bell, J. S.},
  journal = {Physics Physique Fizika},
  volume = {1},
  issue = {3},
  pages = {195--200},
  numpages = {6},
  year = {1964},
  month = {Nov},
  publisher = {American Physical Society},
  doi = {10.1103/PhysicsPhysiqueFizika.1.195},
  url = {https://link.aps.org/doi/10.1103/PhysicsPhysiqueFizika.1.195}
}

@article{Thearle:2018sdl,
    author = "Thearle, Oliver and others",
    title = "{Violation of Bells inequality using continuous variable measurements}",
    eprint = "1801.03194",
    archivePrefix = "arXiv",
    primaryClass = "quant-ph",
    doi = "10.1103/PhysRevLett.120.040406",
    month = "1",
    year = "2018"
}

@article{Praxmeyer:2004jsd,
    author = "Praxmeyer, L. and Englert, B. -G. and Wodkiewicz, K.",
    title = "{Violation of Bell's inequality for continuous variables}",
    eprint = "0406172",
    archivePrefix = "arXiv",
    primaryClass = "quant-ph", 
    doi = "10.1140/epjd/e2005-00021-1",
    journal = "Eur. Phys. J. D",
    volume = "32",
    pages = "227--231",
    year = "2005"
}

@article{Yarnall:2007ofr,
    author = "Yarnall, Timothy and Abouraddy, Ayman F. and Saleh, Bahaa E. A. and Teich, Malvin C.",
    title = "{Experimental Violation of Bell's Inequality in Spatial-Parity Space}",
    eprint = "0708.0653",
    archivePrefix = "arXiv",
    primaryClass = "quant-ph",
    doi = "10.1103/PhysRevLett.99.170408",
    journal = "Phys. Rev. Lett.",
    volume = "99",
    pages = "170408",
    year = "2007"
}

@article{cirel1980quantum,
  title={Quantum generalizations of Bell's inequality},
  author={Cirel'son, Boris S},
  journal={Letters in Mathematical Physics},
  volume={4},
  number={2},
  pages={93--100},
  year={1980},
  publisher={Springer}
}

@article{Ando:2020kdz,
    author = "Ando, Kenta and Vennin, Vincent",
    title = "{Bipartite temporal Bell inequalities for two-mode squeezed states}",
    eprint = "2007.00458",
    archivePrefix = "arXiv",
    primaryClass = "quant-ph",
    doi = "10.1103/PhysRevA.102.052213",
    journal = "Phys. Rev. A",
    volume = "102",
    number = "5",
    pages = "052213",
    year = "2020"
}

@article{Fritz:2010qzm,
    author = "Fritz, Tobias",
    title = "{Quantum correlations in the temporal Clauser{\textendash}Horne{\textendash}Shimony{\textendash}Holt (CHSH) scenario}",
    eprint = "1005.3421",
    archivePrefix = "arXiv",
    primaryClass = "quant-ph",
    doi = "10.1088/1367-2630/12/8/083055",
    journal = "New J. Phys.",
    volume = "12",
    number = "8",
    pages = "083055",
    year = "2010"
}

@article{Li:2011jvq,
    author = "Li, Ming and Fei, Shao-Ming and Li-Jost, Xianqing",
    title = "{Bipartite Bell Inequality and Maximal Violation}",
    eprint = "1102.5246",
    archivePrefix = "arXiv",
    primaryClass = "quant-ph",
    doi = "10.1088/0253-6102/55/3/09",
    journal = "Commun. Theor. Phys.",
    volume = "55",
    pages = "418--420",
    year = "2011"
}

@article{Martin:2016nrr,
    author = "Martin, Jerome and Vennin, Vincent",
    title = "{Leggett-Garg Inequalities for Squeezed States}",
    eprint = "1611.01785",
    archivePrefix = "arXiv",
    primaryClass = "quant-ph",
    doi = "10.1103/PhysRevA.94.052135",
    journal = "Phys. Rev. A",
    volume = "94",
    number = "5",
    pages = "052135",
    year = "2016"
}

@article{Chatterjee:2024los,
    author = "Chatterjee, Arijit and Karthik, H. S. and Mahesh, T. S. and Devi, A. R. Usha",
    title = "{Extreme Violations of Leggett-Garg Inequalities for a System Evolving under Superposition of Unitaries}",
    eprint = "2411.02301",
    archivePrefix = "arXiv",
    primaryClass = "quant-ph",
    doi = "10.1103/vydp-9qqq",
    journal = "Phys. Rev. Lett.",
    volume = "135",
    number = "22",
    pages = "220202",
    year = "2025"
}

@article{Kundu:2011sg,
    author = "Kundu, Sandipan",
    title = "{Inflation with General Initial Conditions for Scalar Perturbations}",
    eprint = "1110.4688",
    archivePrefix = "arXiv",
    primaryClass = "astro-ph.CO",
    reportNumber = "UTTG-23-11, TCC-027-11",
    doi = "10.1088/1475-7516/2012/02/005",
    journal = "JCAP",
    volume = "02",
    pages = "005",
    year = "2012"
}

@article{Ragavendra:2024qpj,
    author = "Ragavendra, H. V. and Mukherjee, Dipayan and Sethi, Shiv K.",
    title = "{Cosmological consequences of statistical inhomogeneity}",
    eprint = "2411.01331",
    archivePrefix = "arXiv",
    primaryClass = "astro-ph.CO",
    doi = "10.1103/PhysRevD.111.023541",
    journal = "Phys. Rev. D",
    volume = "111",
    number = "2",
    pages = "023541",
    year = "2025"
}

@article{Mukherjee:2025dcv,
    author = "Mukherjee, Dipayan and Ragavendra, H. V. and Sethi, Shiv K.",
    title = "{Scalar-induced gravitational waves from coherent initial states}",
    eprint = "2506.23798",
    archivePrefix = "arXiv",
    primaryClass = "astro-ph.CO",
    doi = "10.1103/qd1s-9fxl",
    journal = "Phys. Rev. D",
    volume = "113",
    number = "2",
    pages = "023533",
    year = "2026"
}

@article{Mondal:2024glo,
    author = "Mondal, Aurindam and Raveendran, Rathul Nath",
    title = "{Violation of Bell inequality from a squeezed coherent state of inflationary perturbations}",
    eprint = "2410.04608",
    archivePrefix = "arXiv",
    primaryClass = "gr-qc",
    doi = "10.1016/j.dark.2026.102218",
    journal = "Phys. Dark Univ.",
    volume = "51",
    pages = "102218",
    year = "2026"
}

@article{Bertuzzo:2026gkj,
    author = "Bertuzzo, Enrico and Salla, Gabriel M. and Tesi, Andrea",
    title = "{The effects of non Bunch-Davies initial conditions on gravitationally produced relics}",
    eprint = "2603.03430",
    archivePrefix = "arXiv",
    primaryClass = "gr-qc",
    reportNumber = "DESY-25-183",
    month = "3",
    year = "2026"
}

@article{Gour:2003wsm,
    author = "Gour, G. and Khanna, F. C. and Mann, A. and Revzen, M.",
    title = "{Optimization of Bell's Inequality Violation For Continuous Variable Systems}",
    eprint = "0308063",
    archivePrefix = "arXiv",
    primaryClass = "quant-ph", 
    doi = "10.1016/j.physleta.2004.03.018",
    month = "10",
    year = "2003"
}

@article{Revzen:2004mzw,
    author = "Revzen, M. and Mello, P. A. and Mann, A. and Johansen, L. M.",
    title = "{Bell's Inequality Violation (BIQV) with Non-Negative Wigner Function}",
    eprint = "0405100",
    archivePrefix = "arXiv",
    primaryClass = "quant-ph", 
    doi = "10.1103/PhysRevA.71.022103",
    month = "5",
    year = "2004"
}

@article{Grain:2019vnq,
    author = "Grain, Julien and Vennin, Vincent",
    title = "{Canonical transformations and squeezing formalism in cosmology}",
    eprint = "1910.01916",
    archivePrefix = "arXiv",
    primaryClass = "astro-ph.CO",
    doi = "10.1088/1475-7516/2020/02/022",
    journal = "JCAP",
    volume = "02",
    pages = "022",
    year = "2020"
}

@article{Chen:2024ckx,
    author = "Chen, Pisin and Lin, Kuan-Nan and Lin, Wei-Chen and Yeom, Dong-han",
    title = "{Possible origin of {\ensuremath{\alpha}}-vacua as the initial state of the Universe}",
    eprint = "2404.15450",
    archivePrefix = "arXiv",
    primaryClass = "gr-qc",
    doi = "10.1103/PhysRevD.111.083520",
    journal = "Phys. Rev. D",
    volume = "111",
    number = "8",
    pages = "083520",
    year = "2025"
}

@article{Wood-Saanaoui:2026bma,
    author = "Wood-Saanaoui, Melo and O. Ramos, Rudnei and Berera, Arjun",
    title = "{Choice of Quantum Vacuum for Inflation Observables}",
    eprint = "2602.22116",
    archivePrefix = "arXiv",
    primaryClass = "gr-qc",
    journal = "Symmetry",
    volume = "18",
    pages = "399",
    year = "2026"
}

@article{Yin:2023jlv,
    author = "Yin, Yuan",
    title = "{Cosmological collider signal from non-Bunch-Davies initial states}",
    eprint = "2309.05244",
    archivePrefix = "arXiv",
    primaryClass = "hep-ph",
    doi = "10.1103/PhysRevD.109.043535",
    journal = "Phys. Rev. D",
    volume = "109",
    number = "4",
    pages = "043535",
    year = "2024"
}

@article{Gessey-Jones:2021yky,
    author = "Gessey-Jones, T. and Handley, W. J.",
    title = "{Constraining quantum initial conditions before inflation}",
    eprint = "2104.03016",
    archivePrefix = "arXiv",
    primaryClass = "astro-ph.CO",
    doi = "10.1103/PhysRevD.104.063532",
    journal = "Phys. Rev. D",
    volume = "104",
    number = "6",
    pages = "063532",
    year = "2021"
}

\end{document}